\documentclass[]{spie}  

\usepackage{amsmath,amsfonts,amssymb}
\usepackage{graphicx}
\usepackage[colorlinks=true, allcolors=blue]{hyperref}

\usepackage{caption}
\usepackage{subcaption}

\usepackage{appendix}

\usepackage{wrapfig}
\usepackage{sidecap}

\title{Polarization Modeling and Predictions for DKIST Part 4: Calibration Accuracy Over Field of View, Retardance Spatial Uniformity And Achromat Design Sensitivity}

\author[a]{David M. Harrington}
\author[b]{Stacey R. Sueoka}
\affil[a]{National Solar Observatory, 8 Kiopa'a Street, Ste 201 Pukalani, HI 96768, USA}
\affil[b]{National Solar Observatory, 3665 Discovery Drive, Boulder, CO, 80303, USA}

\authorinfo{Further author information: (Send correspondence to D.M.H) \\
	D.M.H.: E-mail: dharrington@nso.edu, Telephone: 1 808 572 6888}

\begin{document} 
\maketitle

\begin{abstract}

Modern observatories and instruments require optics fabricated at larger sizes with more stringent performance requirements. The Daniel K. Inouye Solar Telescope (DKIST) will be the world's largest solar telescope at 4.0 m aperture delivering a 300 Watt beam and a 5 arc-minute field. Spatial variation of retardance is a limitation to calibration of the full field. Three polarimeters operate seven cameras simultaneously in narrow bandpasses from 380 nm to 1800 nm. The DKIST polarization calibration optics must be 120 mm in diameter at Gregorian focus to pass the beam and operate under high heat load, UV flux and environmental variability. Similar constraints apply to the three retarders for modulation within the instrument suite with large beams near focal planes at F/ 18 to F/ 62. We assess how design factors can produce more spatial and spectral errors simulating elliptical retardance caused by polishing errors. We measure over 5$^\circ$ net circular retardance and spectral oscillations over $\pm$2$^\circ$ for optics specified as strictly linear retarders. Spatial variations on scales larger than 10 mm contain 90\% of the variation. Different designs can be a factor of 2 more sensitive to polishing errors with dissimilar spatial distributions even when using identical retardance bias values and materials. The calibration for the on axis beam is not impacted once circular retardance is included. Calibration of the full field is limited by spatial retardance variation unless techniques account for this variation. We show calibration retarder variation at amplitudes of 1$^\circ$ retardance for field angles greater than roughly one arc minute for both quartz and MgF$_2$ retarders at visible wavelengths with significant variation between the three DKIST calibration retarders. We present polishing error maps to inform new calibration techniques attempting to deliver absolute accuracy of system calibration below effective cross-talk levels of 1$^\circ$ retardance. 

\end{abstract}

\keywords{Instrumentation, Polarization, Mueller matrix, DKIST}

\section{Introduction}
\label{sec:introduction}

The Daniel K. Inouye Solar Telescope (DKIST) on Haleakal\={a}, Maui, Hawai'i is planning on science operations beginning in 2020 with commissioning beginning very soon. The off-axis altitude azimuth telescope has a 4.0 m diameter F/ 2 primary mirror. A suite of polarimetric instrumentation is located in the coud\'{e} laboratory \cite{2014SPIE.9145E..25M, Keil:2011wj, Rimmele:2004ew}. When calibrating instruments that can deviate substantially from the telescope optical bore sight, an understanding of the polarimetric calibration issues across the field of view is critical. We must understand the mirror polarization response across the field and also the spatial inhomogeneity of all optics not at a pupil plane. In this study, we focus on the calibration retarders and modulators as they are a primary source of polarization variation when calibrating the full field of view. Modulating retarders for these instruments work in beams with focal ratios varying from F/ 8 to F/ 62 with the retarders mounted inside the instruments.  The instruments all can scan the telescope field of view either with steering mirrors or by stepping a spectrograph slit. Thus modulators may have a constant beam footprint but sample the variable polarization across the telescope field.

DKIST uses seven mirrors to feed the beam to the rotating coud\'{e} platform \cite{Marino:2016ks, McMullin:2016hm,Johnson:2016he,2014SPIE.9147E..0FE, 2014SPIE.9147E..07E, 2014SPIE.9145E..25M}. Operations involve four polarimetric instruments spanning the 380 nm to 5000 nm wavelength range. At present design, three different retarders are in fabrication for use in calibration near the Gregorian focus \cite{2014SPIE.9147E..0FE,Sueoka:2014cm,Sueoka:2016vo}. These calibration retarders see a beam with 300 Watts of optical power, a focal ratio F/ 13 with an extremely large clear aperture of 105 mm. Five more mirrors deliver the beam to the coud\'{e} laboratory.  A train of dichroic beam splitters in the collimated coud\'{e} beam after the adaptive optics (AO) deformable mirror (DM) allows for rapid changing of instrument configurations. Different wavelengths can be observed simultaneously by three polarimetric instruments covering 380 nm to 1800 nm all using the AO system \cite{2014SPIE.9147E..0FE, 2014SPIE.9147E..07E, 2014SPIE.9147E..0ES, SocasNavarro:2005bq}. Another instrument (CryoNIRSP) can receive all wavelengths using an all-reflective beam to 5000 nm wavelength but without adaptive optics.

Complex polarization modulation and calibration strategies are required for such a multi-instrument system \cite{2014SPIE.9147E..0FE,2014SPIE.9147E..07E, Sueoka:2014cm, 2015SPIE.9369E..0NS, deWijn:2012dd, 2010SPIE.7735E..4AD}. The planned 4 m European Solar Telescope (EST), though using an on-axis primary mirror, will also require similar calibration considerations \cite{SanchezCapuchino:2010gy, Bettonvil:2011wj,Bettonvil:2010cj,Collados:2010bh}.  Many solar and night-time telescopes have performed polarization calibration of complex optical pathways \cite{DeJuanOvelar:2014bq, Joos:2008dg, Keller:2009vj,Keller:2010ig, Keller:2003bo, Rodenhuis:2012du, Roelfsema:2010ca, 1994A&A...292..713S, 1992A&A...260..543S,  1991SoPh..134....1A, Schmidt:2003tz, Snik:2012jw,  Snik:2008fh, Snik:2006iw, SocasNavarro:2011gn, SocasNavarro:2005jl, SocasNavarro:2005gv, Spano:2004ge, Strassmeier:2008ho, Strassmeier:2003gt, Tinbergen:2007fd, 2005A&A...443.1047B, 2005A&A...437.1159B}.  We refer the reader to recent papers outlining the various capabilities of the first-light instruments \cite{McMullin:2016hm, 2014SPIE.9147E..07E, 2014SPIE.9145E..25M, 2014SPIE.9147E..0FE, Rimmele:2004ew}. 

For the DKIST system, we have been pursuing a detailed campaign of system-level simulations and performance predictions for polarimetry.  In our first paper \cite{Harrington:2017dj}, we showed the variation of polarization properties for powered mirrors, the impact of mirror coatings and polarization variation with wavelength and field of view. The F/ 2 primary mirror and F/ 13 secondary only introduced depolarization at amplitudes below 0.2\%. The field dependence was at magnitudes of 0.02 in a Mueller matrix element across the 5 arcminute field. In the next two papers \cite{Harrington:2018cx,Harrington:2017jh}, we explored polarization fringes caused by multi-crystal retarders and their dependence on retarder design, beam F/ number,  thermal variation and thermal loading. The fringes are present at high spectral frequency, clearly resolvable by our instruments and with significant fringes in retardance, diattenuation and transmission. There is strong dependence on beam F/ number through an average over the beam footprint.  For DKIST, fringes in the F/ 13 beam of the calibration optics can be factors of few to $>$30 reduced compared to the collimated case, strongly dependent on wavelength.

\begin{table}[htbp]
\vspace{-2mm}
\caption{Retarder Names and Design Style}
\label{table:retarder_types}
\centering
\begin{tabular}{l | l l l}
\hline \hline
ID	& Common Name				& N		& Design Description		\\
\hline \hline
1a	& Zero-order retarder			& 1		& Single retarder polished to net retardance $<$1 wave		\\
1b	& Multi-order retarder			& 1		& Single retarder polished to net retardance $>$1 wave		\\
2a	& Compound retarder			& 2		& Two retarders same material bonded in subtraction \\
	&							& 		& with fast axes oriented at 90$^\circ$ \\
2b	& Compound zero order retarder 	& 2		& Same as 2a but with net retardance $<$ 1 wave \\
2c	& Bi-crystalline compound retarder	& 2		& Same as 2a but with dissimilar materials for 1 and 2 \\
3	& Pancharatnam achromat		& 3		& 3 retarders total, the outer two A style with matching \\
	&							&		& retardance and orientation, mounted in A-B-A order. \\
4	& Pancharatnam superachromat	& 6		& 6 retarders total used as 3 pairs of compound retarder \\
	&							&		& (bi-crystalline or not) in same A-B-A ordering \\
5	& 5-layer achromat				& 5		& Common polycarbonate design A-B-C-B-A style \\
	&							& 		& with A \& B pairs at the same orientation	\\
\hline \hline		
\end{tabular}
\vspace{-2mm}
\end{table}

In this paper, we analyze several kinds of retarder and their spatial variation of retardance. We utilize standard terminology for various styles of single and multi-element achromatic retarders. Super-achromatic retarders are often made of several components, themselves combinations of achromatic retarders. Table \ref{table:retarder_types} shows several types of retarders and the terminology used in this paper.  Other definitions of these terms are in use so we just describe how we use the terms here in Table \ref{table:retarder_types}.  

A retarder design was introduced by Pancharatnam \cite{Pancharatnam:1955iw} to make an achromatic retarder using a combination of three retarders. The two outer retarders were labeled A and had the same retardance magnitude and orientation.  The inner retarder, labeled B, was at some alternate retardance magnitude and orientation.  The magnitude of A, magnitude of B and orientation of B were optimized for design. Later designs for super-achromats used three compound or bi-crystalline achromats in place of A and B for six total crystals \cite{2014SPIE.9147E..0FE}. This increased the wavelength range when requiring achromatic linear retardance of various specifications to achieve high efficiency of modulation or calibration.

There are many degrees of freedom with six crystals one allows different materials, retardance values and orientations. The Pancharatnam designs are usually simplified by choosing just two materials and making the outer two bi-crystalline or compound retarders identical. This simple design uses an A-B-A type alignment where the two outer crystal pairs are mounted with their fast axes aligned. Provided the bi-crystalline or compound pairs are treated as perfect linear retarders, there is a simple theoretical formula for the linear retardance of such an A-B-A design. 

\begin{wrapfigure}{r}{0.45\textwidth}
\centering
\vspace{-5mm}
\begin{equation}
\label{pan_stack1}
\cos \frac{\Delta}{2}  =  \cos \frac{\delta_B}{2}  \cos \delta_A  - \sin \frac{\delta_B}{2}  \sin \delta_A  \cos 2\theta  
\end{equation}
\begin{equation}
\cot 2\Theta =  \frac{ \sin \delta_A  \cot \frac{\delta_B}{2}  +  \cos \delta_A \cos  2\theta }   {  \sin 2\theta }
\label{pan_stack2}
\end{equation}
\vspace{-7mm}
\end{wrapfigure}

If we take the retardance of the A crystals as $\delta_A$ and the B crystals as $\delta_B$, and the relative orientation between the A and B crystal pairs as $\theta$, we can write the formula for the resulting superachromatic optic retardance ($\Delta$) and fast axis orientation ($\Theta$) as in Equations \ref{pan_stack1} and \ref{pan_stack2} \cite{Pancharatnam:1955iw}. We also analyze super-achromatic polycarbonate retarders that have a design style of A-B-C-B-A where the magnitudes of A, B and C as well as the orientations of the B pair and the individual C retarder are all optimized. Though achromatic performance is not required for modulation or calibration, many modern instrument designs still use this configuration. \cite{2010SPIE.7735E..4AD, Tomczyk:2010wta, Wijn:2011wt}  We note that these same retarder design strategies can be used to make efficient modulators as elliptical retarders. If we relax the constraint of requiring retarder pairs to remain parallel, we create elliptical retardance and can optimize for quite large wavelength ranges, as in our DKIST instrument use cases. We use the standard axis-angle formalism for the $QUV$ to $QUV$ elements of the elliptical retarder as a rotation matrix and outline details in Appendix \ref{sec:appendix_eliret}. 

\begin{table}[htbp]
\vspace{-0mm}
\caption{DKIST Crystal Retarder Design Properties: Wavelengths, Bias, Magnitude, Orientations}
\label{table:dkist_retarders}
\centering
\begin{tabular}{l l l l l l l}
\hline \hline
Name		& Linear		& Wavelngth		& Crys.		& Bias		& Design Magnitude			& Design A-B-A			\\
			& or			& range			& Mat.		& Net		& Net waves retardance		& Fast Axis		\\
			& Elliptical		& nm			 	&			& Waves		& at $\lambda$=633nm		& Orientation		\\
\hline \hline	
ViSP SAR		& Linear		& 380-1100		& SiO$_2$	&  30$\pm$1	& 0.328 - 0.476 - 0.328		& 0$^\circ$ - 70.25$^\circ$ - 0$^\circ$		\\
ViSP PCM	& Elliptic		& 380-1100		& SiO$_2$	 & 30$\pm$1	& 0.476 - 0.328 - 0.476		& 0$^\circ$ - 41.28$^\circ$ - 148.23$^\circ$ 	\\
\hline
DL SAR		& Linear		& 900-2500		& SiO$_2$	 & 30$\pm$1	& 0.683 - 1.000 - 0.683		& 0$^\circ$ - 65.00$^\circ$ - 0$^\circ$		\\
DL PCM		& Elliptic		& 900-2500		& SiO$_2$	 & 30$\pm$1	& 1.000 - 0.683 - 1.000		& 0$^\circ$ - 42.20$^\circ$ - 152.51$^\circ$	\\
\hline
Cryo SAR		& Linear		& 2500-5000		& MgF$_2$	 & 40$\pm$1	& 2.230 - 3.346 - 2.230		& 0$^\circ$ - 107.75$^\circ$ - 0$^\circ$		\\
Cryo PCM		& Elliptic		& 1000-5000		& MgF$_2$	 & 40$\pm$1	& 1.893 - 1.282 - 1.893		& 0$^\circ$ - 71.86$^\circ$ - 30.39$^\circ$		\\
\hline \hline		
\end{tabular}
\vspace{-1mm}
\end{table}

The wavelengths for our retarder optimization are set by the DKIST spectropolarimetric instrumentation as shown in Table \ref{table:dkist_retarders}. In naming the optics, the project chose Super Achromatic Retarder (SAR) for the calibration retarders. They are named according to specific instruments, but we note that any SAR can be used to with any instrument to calibrate. The project named the elliptical retarders used as modulators to be Poly-Chromatic Modulators (PCM). Three spectropolarimetric instruments on DKIST simultaneously use the adaptive optics system: the Visible Spectro-Polarimeter (ViSP), the Visible Tunable Filter (VTF) and the Diffraction Limited Near-InfraRed Spectro-Polarimeter (DL-NIRSP).  The retarders operating at wavelengths shorter than 2500 nm used crystal quartz (SiO$_2$) which were polished to roughly 2 mm physical thickness giving a net bias around 30 waves retardance at the 633 nm metrology wavelength.  The longer wavelength designs used MgF$_2$ crystals also around 2 mm physical thickness, giving 40 waves net retardance bias at 633 nm for this higher birefringence material.

Another first light instrument was designed to include infrared capabilities at wavelengths as long as 5000 nm.  This instrument, the Cryogenic Near-Infrared Spectro-Polarimeter (CryoNIRSP) does not use the adaptive optics system and has a separate optical path. The ViSP covers visible wavelengths 380 nm to 900 nm with VTF being similar at 520 nm to 870 nm.  The DL-NIRSP nominally covers 500 nm to 1800 nm though at the time of the retarder designs the wavelength range was 900 nm to 2500 nm. The Cryo-NIRSP presently covers wavelengths from 1000 nm to 5000 nm with potential capabilities at shorter wavelengths. The optics designated for ViSP have a wavelength range for efficient modulation and calibration from 380 nm to 1100 nm. The optics we designate for the DL-NIRSP have a wavelength range for reasonably efficient modulation from 500 nm to 2500 nm with the calibration retarder covering 900 nm to 2500 nm. The optics we designate for Cryo-NIRSP cover 2500 nm to 5000 nm for the calibration retarder but the modulator has a wider range of 1000 nm to 5000 nm.  

For the DKIST retarder designs, each retarder only uses a single material. Crystal quartz was used for the retarders working at wavelengths shorter than 2500 nm. Crystal MgF$_2$ was used for the retarders nominally operating at longer wavelengths to 5000 nm. The individual compound MgF$_2$ crystal retarders are less than one wave net retardance at 5000 nm wavelength when the two crystals are bonded in subtraction, but are roughly five and seven waves each at shorter visible wavelengths. Each two-crystal achromat have their axes oriented at 90$^\circ$ with respect to each other. We had to provide elliptical retarders that act as efficient modulators within each instrument in addition to calibration retarders giving us six retarder designs. Each A and B retarder would thus be two crystals of the same material instead of bi-crystalline. This has benefit for reducing polarization fringes as well as cost in simplicity of manufacture. 

\begin{wrapfigure}{l}{0.60\textwidth}
\centering
\vspace{-4mm}
\begin{tabular}{c} 
\hbox{
\hspace{-1.1em}
\includegraphics[height=6.7cm, angle=0]{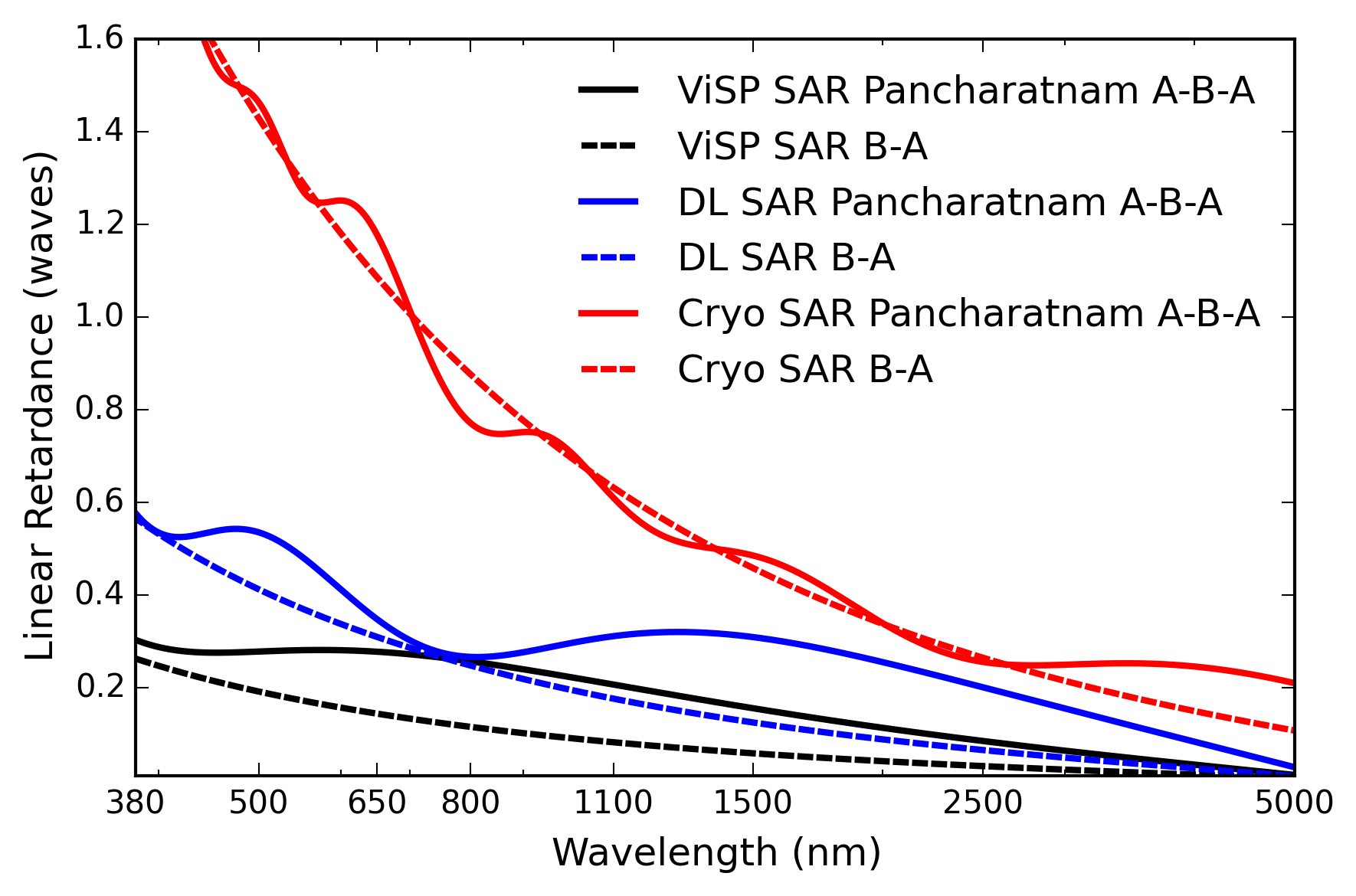}
}
\end{tabular}
\caption[] 
{ \label{fig:dkist_retarder_design_pancharatnam} The Pancharatnam linear retardance magnitude following Equation \ref{pan_stack2} as solid lines. The difference between two-crystal achromats A and B are shown for reference as the dashed lines.  Each color represents a different design wavelength.}
\vspace{-3mm}
\end{wrapfigure}

Figure \ref{fig:dkist_retarder_design_pancharatnam} shows the net retardance for each of these two-crystal achromats.  The black and blue curves show the crystal SiO$_2$ designs while red shows the crystal MgF$_2$. We also show the difference between the two-crystal A and B achromats in Figure \ref{fig:dkist_retarder_design_pancharatnam} as the dashed lines as these properties generally set the range for achromatic performance. The solid lines show that the linear retardance magnitude from Equation \ref{pan_stack2} does indeed oscillate about the B minus A difference, and the magnitude may be a few times larger than the difference between the A-B two-crystal achromats. Note that each of the crystal pairs are a fraction of a wave different in the respective bandpasses of operation. As the crystal MgF$_2$ is designed for wavelengths longer than 1000 nm, the two crystal pairs are over two waves net retardance at visible wavelengths.  With this study, we show how various designs show increased sensitivity to polishing and crystal alignment errors and to provide the predictions of the impact to the calibration and modulation processes.


In Section \ref{sec:example_metrology_cryo_sar}, we will examine vendor metrology, DKIST metrology and the associated elliptical retarder parameter fits.  We will show that the circular retardance component is present at magnitudes of several degrees and that this can easily be included in an elliptical retarder model for the optic. The presence and magnitude is consistent with our metrology. In Section \ref{sec:spatial_polish}, we then show how mounting these retarders near focal planes in converging beams couples spatial variation of retardance in to errors in system calibration. We show spatial maps of retardance magnitude across the clear aperture of individual two-crystal compound retarders. In Section \ref{sec:spatial_ER_variation}, these individual crystal pair maps are then stacked together to create a model for a six-crystal design tolerance analysis used in estimating orientation error and thickness error sensitivities. We will show that different designs can vary in sensitivity by at least a factor of 2 and that the wavelength dependence of these errors is strongly variable. We show statistics of spatial variation that will introduce both depolarization and calibration errors. In Section \ref{sec:as_built_spatial_maps} we summarize the metrology of the suite of DKIST six-crystal super achromatic retarders with examples of dependence on spatial location and wavelength.  Section \ref{sec:impact_to_fov} describes the impact to DKIST calibration across the field of view showing measured retardance variation across DKIST footprints and also depolarization caused by spatial variation within a footprint.  Section \ref{sec:field_dependend_demod} first shows the impact to modulation within DKIST instruments and compares with depolarization from a continuously rotating retarder.  We then show how to create a short but efficient calibration sequence that can use one optic to calibrate all DKIST post-AO instruments simultaneously. We use a simple model for the calibration process that creates demodulation errors as functions of field angle included in the process. In Appendix \ref{sec:appendix_eliret} we summarize some mathematical properties of elliptical retardance.  In Appendix \ref{sec:appendix_spatial_maps}, we show much more complete metrology of all DKIST retarders for reference. In Appendix \ref{sec:appendix_uniformity_polycarb_FLC}, we put this work in context of the heat loads, UV flux levels and field of view requirements of other telescopes such as the GREGOR solar telescope, the Dunn Solar Telescope and the Goode Solar Telescope.  We show retardance spatial non-uniformity of polycarbonate and ferro-electric liquid crystal type retarders.

\clearpage
\section{Validation of Fabricated Six-Crystal Calibration Retarders}
\label{sec:example_metrology_cryo_sar}

When developing specifications for large aperture precision retarders, the spectral and spatial variation of retardance properties can impact the ultimate instrument performance through multiple pathways. Uncalibrated spatial variation can lead to improper system Mueller matrices or demodulation matrices. Spectral oscillations from crystal orientation errors and polishing errors can introduce circular retardance, often complicating calibration with multiple additional variables and / or constraining spectral calibration models. Our super-achromatic designs are effectively an imperfect stack of six rotation matrices (retarders) with errors fully elliptical after multiple successive imperfect optics. DKIST performed extensive metrology to ensure the calibration optics are built within tight design tolerances and their properties are known in detail.

\subsection{Physical Thickness \& Individual Crystal Measurements at Meadowlark Optics}
\label{sec:cryosar_meadowlark_metrology}

In this section, we show a sample of metrology performed by Meadowlark Optics as part of acceptance testing as well as internal DKIST metrology. We used a wide variety of metrology tools to verify the as-built retarders achieve the required performance. This ensures our subsequent performance models are valid as we have accurate knowledge of the components inside each assembly. We examine one of our six retarder optics and then in later sections show the impact of some manufacturing errors on the DKIST instrument performance. We denote the Pancharatnam theoretical A-B-A design using the DKIST optic labeling convention for the CryoNIRSP SAR components: G-H-G.  The numbering of each pair was sequential so the optic contains six crystals in the ordering 2G - 3H - 4G. Within each compound zero order retarder, the two crystals are oriented with fast axes within $\pm$0.3$^\circ$ of crossed.  The G retarder is nominally two crystal retarders roughly 2 mm thick each representing 40 waves net retardance in each crystal. DKIST used the convention where the plate with the larger desired net retardance was called the subtraction crystal while the plate with the lower net retardance was called the {\it bias} crystal. For example, the G compound zero order retarder has a net retardance of 2.2 waves at 633 nm wavelength after the two crystals are assembled in subtraction with fast axes crossed. This design creates a compound true zero-order retarder at wavelengths longer than 2000 nm. For the H pair, the two retarders subtract to 3.46 waves net retardance at 633 nm wavelength.

Meadowlark Optics provided DKIST with several types of measurements to show the crystals meet manufacturing tolerances.   The physical thickness was measured by a Heidenheim MT 60M metrology system with $\sim$0.5 $\mu$m thickness accuracy.  The individual crystals that make up the compound retarders were also tested spectrally to derive a net retardance for the assembled crystal pairs. This measurement was derived by fitting broad band spectra collected with a Varian spectrophotometer and mounting the crystal between crossed polarizers.  Once the individual crystals were measured, they were assembled into compound retarders.  These compound retarders were measured for net retardance at one wavelength. 

\begin{wraptable}{l}{0.60\textwidth}
\vspace{-2mm}
\caption{Measured Cryo-NIRSP SAR Crystal Properties}
\label{table:CNsar}
\centering
\begin{tabular}{l | l l | l l | l l | l}
\hline \hline
N		& Des.	& Meas	& Ret		& Meas	& Des		& Meas			& Ori				\\
		& mm	& mm	& Bias		& Waves	& Pair		&$\delta\pm$0.01	& $\pm0.3^\circ$	\\
\hline \hline
2G$_s$	& 2.27	& 2.309	& 42$\pm$1	& 42.909	& 			& 				& 0$^\circ$		\\
2G$_b$	& 2.15	& 2.190	& 40$\pm$1	& 40.683	& 2.230		& 2.2319			& 90$^\circ$		\\
\hline
3H$_s$	& 2.33	& 2.373	& 43$\pm$1	& 44.084	& 			& 				& 107.75$^\circ$	\\
3H$_b$	& 2.15	& 2.192	& 40$\pm$1	& 40.736	& 3.346		& 3.3521			& 197.75$^\circ$	\\
\hline
4G$_s$	& 2.27	& 2.278	& 42$\pm$1	& 42.336	& 			& 				& 0$^\circ$		\\
4G$_b$	& 2.15	& 2.159	& 40$\pm$1	& 40.108	& 2.230		& 2.2306			& 90$^\circ$		\\
\hline \hline		
\end{tabular}
\vspace{-2mm}
\end{wraptable}

In Table \ref{table:CNsar} we show a comparison of the design and measured parameters for the six crystals in this optic measured at the center of the clear aperture.  The first column shows the name of the crystal, whether G or H and whether bias or subtraction.  The second column shows the nominal physical thickness in millimeters with all crystals. The third column shows the physical thickness measured with the Heidenheim tooling at accuracies of better than $\pm$0.001 millimeters. The fourth column shows the design net retardance in waves.

The manufacturing tolerance was $\pm$1 wave net for each individual crystal but with a much stricter requirement after the compound zero-order G and H pairs were assembled.  The fifth column shows the measured net retardance of the crystal at 633 nm wavelengths. It was also supported by later measurements retardance with a spatial scanning system using a 599 nm central wavelength filter having a 10 nm FWHM bandpass.   The sixth column of Table \ref{table:CNsar} shows the desired net retardance of the crystal pairs when mounted in subtraction.  The G pairs should have 2.230 waves net retardance while the H pair should be 3.346.  The polishing tolerance was specified at $\pm$0.01 waves net.  The seventh column shows the actual measured net retardance.  The last column shows the orientation (clocking) of each crystal.

\begin{table}[htbp]
\vspace{-3mm}
\caption{Compound Zero Order Achromat Spatial Polishing Metrology Results}
\label{table:polish_error_measured_plates_as_built}
\centering
\begin{tabular}{r c c c c c c c c c }
\hline\hline
Bi-Crystalline	& Mean 		& Peak-Peak  	& RMS		& Measured 	& Meas. 	& Meas. 	& Meas.	& Meas.		\\
Achromat		& Retardance	& Ret. Error 	& Ret. Error	& Retardance	& Ret. 	& Ret.	& Ret.	& Ret.		\\
Name		& waves		& waves x100   & waves x100	& Center		& 1		& 2		& 3		& 4			\\
\hline
\hline
ViSP SAR B2 	& 0.32774	& 0.43		& 0.15	& 0.3256 	& 0.3278 	& 0.3267 	& 0.3299 	& 0.3287		\\
ViSP SAR A3 	& 0.47224	& 0.31		& 0.11	& 0.4703 	& 0.4724 	& 0.4730 	& 0.4721 	& 0.4734		\\
ViSP SAR B4 	& 0.32512	& 0.31		& 0.10	& 0.3238 	& 0.3248 	& 0.3252 	& 0.3249 	& 0.3269		\\
\hline
ViSP PCM  A2 & 0.4732	& 0.31  		& 0.11	& 0.4751 	& 0.4725 	& 0.4720 	& 0.4736 	& 0.4728		\\
ViSP PCM  B3 & 0.3239	& 0.22  		& 0.08	& 0.3236 	& 0.3233 	& 0.3255 	& 0.3236 	& 0.3236		\\
ViSP PCM  A4 & 0.4758 	& 0.51  		& 0.10	& 0.4740 	& 0.4774 	& 0.4767 	& 0.4779 	& 0.4728		\\
\hline
DL SAR D2 	& 0.6909 	& 0.26		& 0.11 	& 0.6920 	& 0.6897 	& 0.6914 	& 0.6921 	& 0.6895		\\
DL SAR C3 	& 1.0008	& 0.44		& 0.17	& 0.9996 	& 1.0023 	& 1.0001 	& 1.0032 	& 0.9988		\\
DL SAR D4 	& 0.6868	& 0.81		& 0.33	& 0.6875 	& 0.6908 	& 0.6834 	& 0.6897 	& 0.6826		\\
\hline
DL PCM C2 	& 0.9991	& 0.42		& 0.14	& 1.0016 	& 0.9984 	& 0.9974 	& 0.9989 	& 0.9994		\\
DL PCM D3 	& 0.6836	& 0.19		& 0.08	& 0.6843 	& 0.6841 	& 0.6824 	& 0.6842 	& 0.6829		\\
DL PCM C4 	& 1.0033	& 0.05 		& 0.01	& 1.0032 	& 1.0033 	& 1.0031 	& 1.0034 	& 1.0036		\\
\hline
Cryo SAR  G2 	& 2.2319 	& 2.36		& 1.00 	& 2.2404 	& 2.2201 	& 2.2437 	& 2.2201 	& 2.2351		\\
Cryo SAR  H3 	& 3.3555	& 1.11		& 0.40	& 3.3536 	& 3.5556 	& 3.3529 	& 3.3632 	& 3.3521		\\
Cryo SAR  G4 	& 2.2282	& 1.91		& 0.67	& 2.2400 	& 2.2244 	& 2.2209 	& 2.2251 	& 2.2306		\\
\hline
Cryo PCM E2 	& 1.8873 	& 2.69		& 1.13	& 1.9007 	& 1.8738 	& 1.8986 	& 1.8750 	& 1.8882		\\
Cryo PCM F3 	& 1.2921	& 1.35		& 0.46	& 1.2992 	& 1.2938 	& 1.2857 	& 1.2890 	& 1.2929		\\
Cryo PCM E4 	& 1.8841	& 0.98		& 0.36	& 1.8820 	& 1.8813 	& 1.8835 	& 1.8826 	& 1.8911		\\
\hline
\hline
\end{tabular}
The acceptance measurements from Meadowlark Optics. SAR denotes the Super Achromatic calibration Retarders. PCM denotes Poly-Chromatic Modulator type retarders.  Crystal pairs A, B, C and D are quartz. Crystal pairs E, F, G and H are MgF$_2$.  One measurement was performed at the center of the aperture while another four measurements were done spatially de-centered by 30 mm uniformly across the aperture. 
\vspace{-3mm}
\end{table}

\subsection{Meadowlark Optics Compound Retarder Metrology at 5 Aperture Locations}
\label{sec:spatial_acceptance_data}

In this section we show how similar crystals can have substantially different uniformity when two crystals were mounted in subtraction to form compound zero-order retarders at their specific design bandpasses. We also show the polishing and mounting process for the pairs of SiO$_2$ crystals had very different retardance uniformity than for the pairs of MgF$_2$ crystals. We were provided high accuracy linear retardance measurements at five spatial locations for the two-crystal subtraction pairs manufactured by Meadowlark Optics. One point was at the center of the aperture while four other points were evenly distributed at a radius of 30 mm from optical center representing the de-centered edge of the 2.8 arc minute field beam near Gregorian focus. These five-location tests were done after oiling and aligning of the crystal pairs but before final assembly and bonding of the six crystal stacks. Table \ref{table:polish_error_measured_plates_as_built} summarizes the metrology results for the final as-built parts when a 3 mm diameter test beam spot-checked the retardance.  

As part of final acceptance of the optics in 2017, we received Meadowlark uniformity data for the ViSP quartz crystal subtraction pairs. The retarder pairs had nominal targets of A = 0.476 waves retardance at 633.443 nm wavelength. The B-type crystal pairs were designed as 0.328 waves net retardance. The super-achromatic calibration retarder (SAR) for ViSP was configured B-A-B while the modulator was A-B-A. Similarly, the DL-NIRSP retarders were designed as D = 1.000 waves retardance and C = 0.683 waves retardance. The SAR was configured with retarders in D-C-D ordering while the modulator was designed as C-D-C. For the CryoNIRSP retarders made of MgF$_2$ crystal, the design did not utilize identical crystal thicknesses for calibration and modulation retarders. The modulator used 1.893 waves retardance for E and 1.282 waves retardance for F in an E-F-E style design.  The calibration retarder used 2.230 waves retardance for G and 3.346 waves retardance for H in a G-H-G style design. We summarize the polishing non-uniformity for all six DKIST retarder designs in Table \ref{table:polish_error_measured_plates_as_built_stack_error}.  The first column shows the optic name.  The second column shows the sum of the peak to peak variations from Table \ref{table:polish_error_measured_plates_as_built} above.  This sum is the maximum possible polish error should the bi-crystalline plates have the errors linearly stacked.  The third column shows the sum of the RMS variations in Table \ref{table:polish_error_measured_plates_as_built}. 

\begin{wraptable}{l}{0.27\textwidth}
	\vspace{-4mm}
	\caption{As Built Summary}
	\label{table:polish_error_measured_plates_as_built_stack_error}
	\centering
	\begin{tabular}{l c c }
	\hline
	\hline
	Optic			& P-P	& RMS 	\\
	Name		& waves	& waves 	\\
				& x0.01	& x0.01 	\\
	\hline \hline
	ViSP SAR		& 1.05	& 0.36	\\
	ViSP PCM	& 1.04	& 0.29	\\
	\hline
	DL SAR		& 1.51	& 0.61	\\
	DL PCM		& 0.66	& 0.23	\\
	\hline
	Cryo SAR		& 5.38	& 2.07	\\
	Cryo PCM		& 5.02	& 1.95	\\
	\hline
	\hline
	\end{tabular}
	\vspace{-4mm}
\end{wraptable}

Table \ref{table:polish_error_measured_plates_as_built_stack_error} shows the peak-to-peak (P-P) and RMS polishing errors in hundredths of a wave retardance. For instance, the ViSP SAR has 0.0105 waves retardance variation peak-to-peak and is listed as 1.05 in Table \ref{table:polish_error_measured_plates_as_built_stack_error}. Significant variations in delivered polish between the various optic designs. The ViSP designs came in with very similar peak to peak as well as RMS values. For the DL-NIRSP, the calibration retarder (SAR) has more than double the peak to peak variation and closer to triple the RMS variation. Both the ViSP and DL-NIRSP designs use SiO$_2$ crystals at 2.1 mm physical thickness but we see spatial variation properties varying significantly between the as built optics. The Cryo-NIRSP designs used MgF$_2$ crystals at roughly 2.2 mm physical thickness. A different polishing process was used on the softer MgF$_2$ crystals which resulted in polishing performance roughly three to eight times worse when summing the individual P-P and RMS crystal pair measurements. 

We also need to ensure that the crystals were cut with the crystal axes parallel and perpendicular to the optical propagation direction. Crystal axis tilt errors were measured to be less than 0.06$^\circ$.  Given the polishing errors can create spatially variable retardance, we neglect this slight spatial offset in crystal axis.

For the DKIST retarders, the SiO$_2$ crystal optics were a factor of roughly 5 less spatially variable in both peak-to-peak and RMS retardance variation than the MgF$_2$ crystals.  In the next section, we take these magnitudes of spatial retardance variation and compute simulations of polishing error sensitivity in some DKIST retarder designs and compare with example metrology results for spatial retardance mapping.

\subsection{Mueller matrix measurements with NSO Lab Spectropolarimeter (NLSP)}
\label{sec:cryosar_nlsp_mm}

The National Solar Observatory Laboratory Spectro-Polarimeter (NLSP) uses two spectrographs to measure polarized spectra with rotating polarization optics a wire grid polarizing beam splitter as an analyzer. We use a fiber-coupled collimated light source stopped to a circular beam of 4 mm diameter using laser cut metal masks.  A polarization state generator consists of a rotating wire grid polarizer and rotating third-wave achromatic linear retarder mounted upstream of the sample location. After the sample, a rotating third-wave linear retarder is mounted as a modulator. The final optic is a fixed orientation analyzing wire grid polarizer also used as a polarizing beam splitter.  

As detectors, we use visible and near-infrared spectrographs from Avantes. The visible spectrograph covers 380 nm to 1200 nm wavelength while the NIR spectrograph covers 900 nm to 1650 nm wavelength. The beam transmitted through the wire grid polarizer feeds the visible spectrograph via filters, aperture stops and a lens. At the lens focus, a fiber couples light to the spectrograph. The beam reflected off the wire grid polarizer is passed through a separate set of filter, aperture stop and lens optics into the near infrared (NIR) spectrograph. This NIR arm has an additional polarizer with wires parallel to the analyzer to remove the fresnel reflection component off the glass and maintain high contrast. We achieve reasonable signal to noise ratio from roughly 400 nm to 1600 nm wavelength with a single exposure. We achieve simultaneous measurements in both visible and near-infrared systems in the 950 nm to 1100 nm bandpass with reasonable signal to noise ratio. This allows us to estimate systematic effects and ensure stable results.

\begin{figure}[htbp]
\begin{center}
\vspace{-2mm}
\hbox{
\hspace{-1.0em}
\includegraphics[height=12.6cm, angle=0]{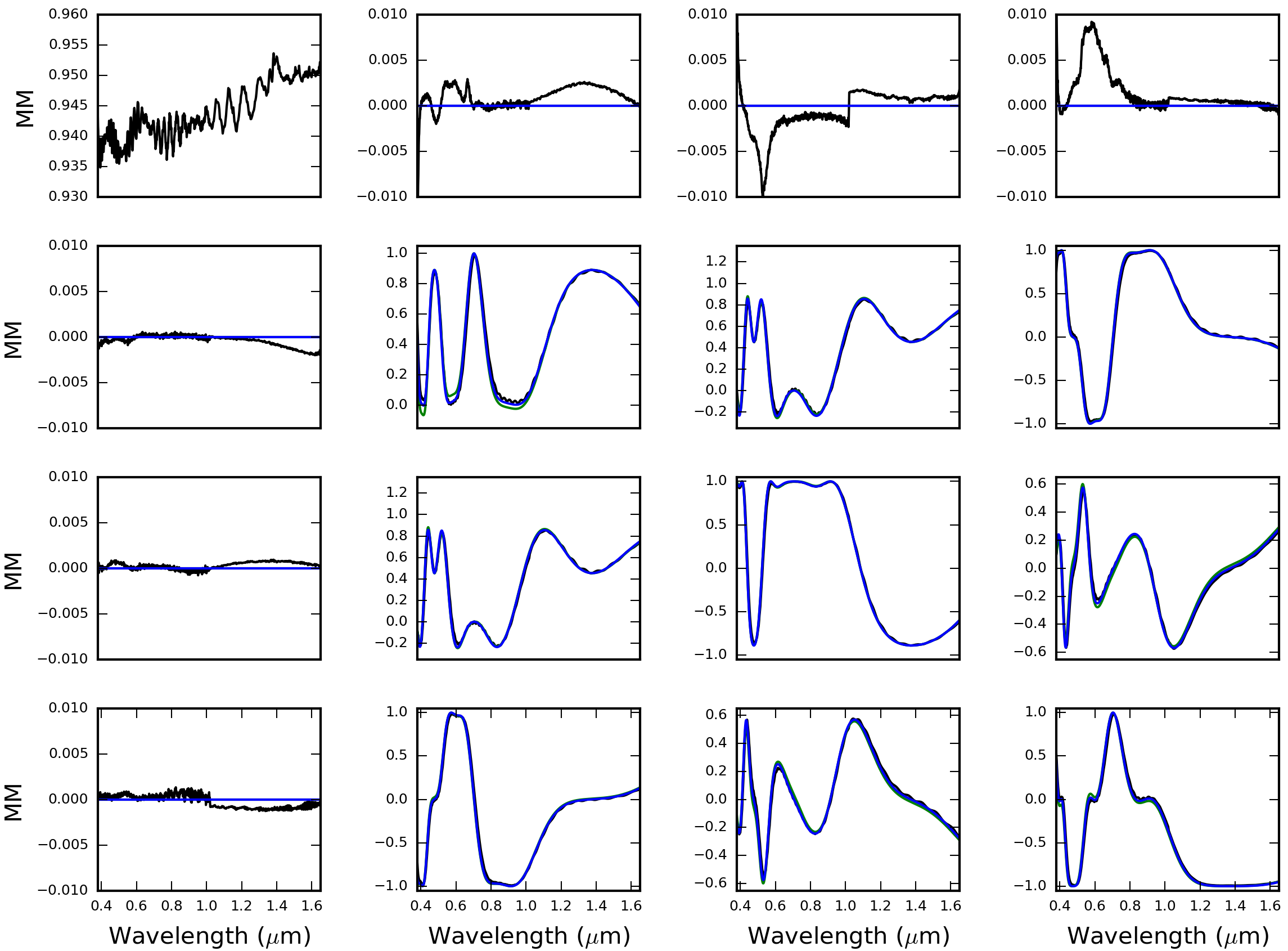}
}
\caption[] 
{\label{fig:measure_nlsp_cryo_sar} The NLSP measured CryoNIRSP SAR Mueller matrix and associated fits. The VIS and NIR spectrograph measurements have been spliced together at 1020 nm wavelength. The transmission is roughly 94\% to 95\% and all other Mueller matrix elements were normalized by this $II$ term. The black curve shows the NLSP data. The green curve shows the theoretical Mueller matrix computed as a stack of six ideal linear retarders rotated to match the measured NLSP data set using the as-built crystal thicknesses. The blue curve shows the same theoretical calculation but with the design crystal physical thicknesses}
\vspace{-6mm} 
\end{center}
\end{figure}

Figure \ref{fig:measure_nlsp_cryo_sar} shows the NLSP measurements of the Cryo-NIRSP SAR along with the theoretical elliptical retarder model. The measurements have a spectral resolving power of roughly 290 at 380 nm wavelength rising to 800 at 1050 nm wavelength on the visible spectrograph. With the NIR spectrograph, we obtain resolving power of 200 at 950 nm wavelength rising to 460 at 1530 nm wavelength. With the as-built thicknesses of Table \ref{table:CNsar}, we compare the theoretical predicted Mueller matrix shown in blue to the measurements shown in black. The optic was mounted in a rotary stage and the best-fit orientation of the theoretical model to the measurements gave an orientation of 155.5$^\circ$.

Figure \ref{fig:measure_nlsp_cryo_sar} shows excellent agreement between the measurements in black and theoretical as-built Mueller matrix from six ideal linear retarders in blue curve. The first row and column of the Mueller matrix show some small artifacts and depolarization at levels below 1\%. The transmission is above 93.5\% as expected for the refractive index matching oil between the MgF$_2$ crystals and the Fresnel losses at the air interfaces. The retardance shows strong variation in the lower 3x3 matrix elements. The Mueller matrix approaches an identity matrix representing full wave integer retardance magnitude for wavelengths around 700 nm per both the design and measured crystal thicknesses. With the theoretical Pancharatnam equations and the as-built crystal thicknesses, we can construct a bounded fit for elliptical retardance parameters. We use the standard axis-angle formalism for retarders as rotation matrices. In this representation, the magnitude of the retardance is the root sum square (RSS) of the three individual retardance components. Each component of retardance represents a rotation on the Poincar\'{e} sphere about a $QUV$ coordinate axis. We outline details of this standard retarder-as-rotation-matrix model Appendix \ref{sec:appendix_eliret}. 

\begin{wrapfigure}{l}{0.62\textwidth}
\centering
\vspace{-4mm}
\begin{tabular}{c} 
\hbox{
\hspace{-1.1em}
\includegraphics[height=7.7cm, angle=0]{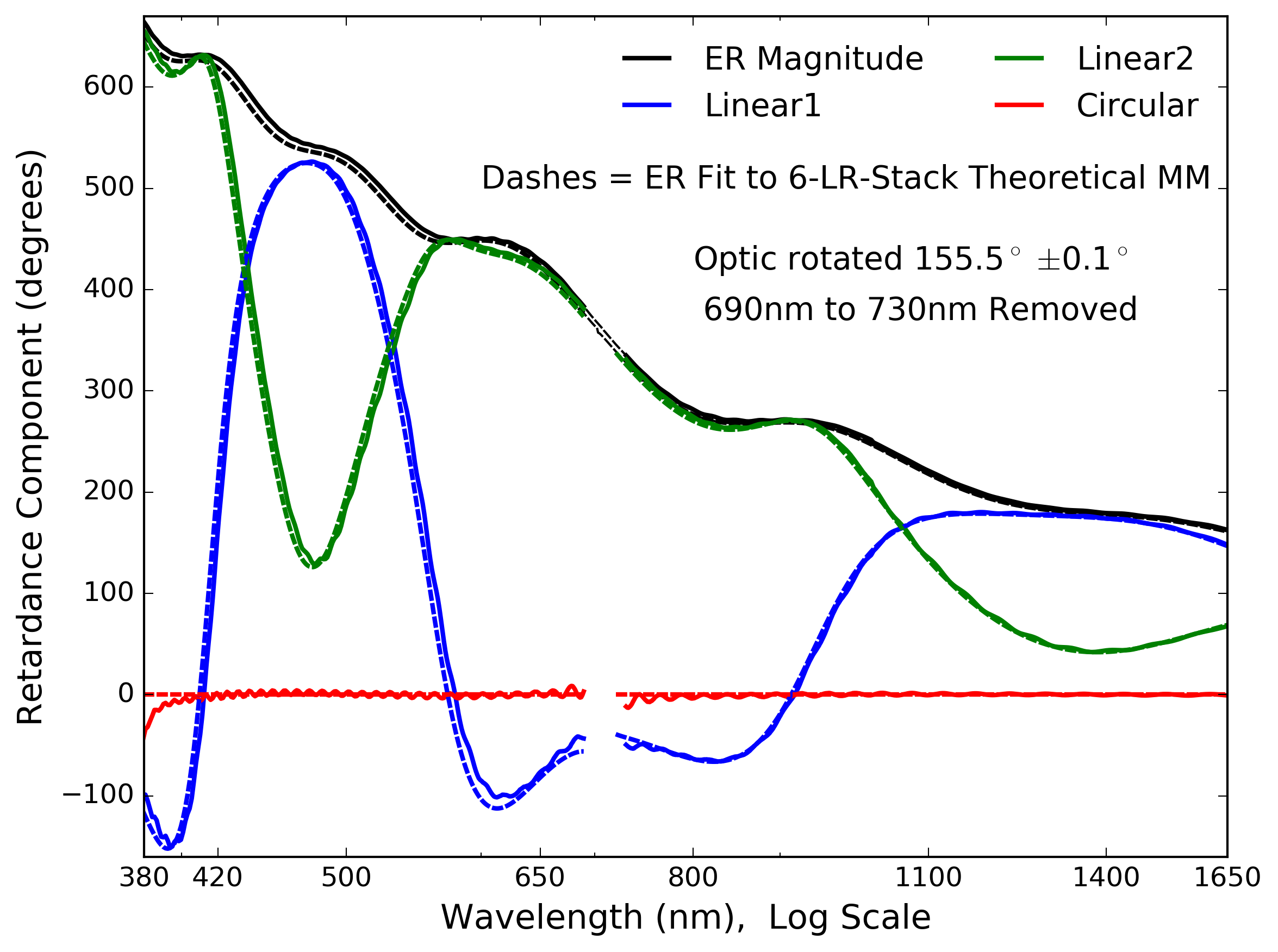}
}
\end{tabular}
\caption[] 
{ \label{fig:fit_nominal_ER_nlsp_cryo_sar} The elliptical retardance model fit to the NLSP measured Cryo SAR Mueller matrix.  The black curve shows the elliptical retardance magnitude as the root-sum-squared of all three retardance components. The blue and green curves shows the first and second components of linear retardance. The arctangent of the two components represents the fast axis orientation with strong spectral changes visible in the data. The red curve shows circular retardance. We included as dashed lines an elliptical retarder fit to the theoretical Mueller matrix computed from the birefringence equations for the six crystals using the design physical thickness.  We rotated the optic by 155.5$^\circ\pm$0.1$^\circ$ as the least-squares-fit to the orientation of the optic in the NLSP setup.}
\vspace{-5mm}
\end{wrapfigure}

Figure \ref{fig:fit_nominal_ER_nlsp_cryo_sar} shows fits of the elliptical retarder models to NLSP measurements. We can fit the design Mueller matrix with a retardance model whose magnitude is near 2 waves magnitude at 380 nm wavelength dropping to 1 wave magnitude around 700 nm with a decrease to quarter-wave at the design wavelength range 2500 nm to 5000 nm. The black curve shows the elliptical retardance magnitude as the blue and green curves show the two components of linear retardance. The spectral variation of the blue and green curves represent the orientation of the fast axis of linear retardance. The red curve shows circular retardance. 

As this optic was designed to be a linear retarder, this term is near zero with visible ripples from clocking errors. The gap near 700 nm wavelength represents the degeneracy in this particular elliptical retarder solution as a full wave of rotation can have any arbitrary pole orientation. Fits diverge rapidly for solutions close to integer multiples of full wave magnitude. If an alternate model is chosen which is driven through zero retardance at 700 nm, no such ambiguities are seen.  This is explored in more detail in Appendix \ref{sec:appendix_eliret}.

When we assess the impacts of spatial uniformity variation on DKIST system calibration, we must know the complexity of the calibration model from the standpoint of fitting errors and model degneracy as well as the expected magnitude of the various terms. The model of the optic may introduce unexpected complexities as we make attempts to simultaneously fit for time dependence of a thermal model and spatial variation of a retardance model while trying to minimize the number of exposures when observing at many simultaneous wavelengths with reduced calibration efficiencies.

As an example of some trade-offs, this Cryo-NIRSP SAR has very beneficial thermal properties when illuminated with the DKIST 300 Watt optical beam \cite{Harrington:2018cx}. This optic sees no heat when used downstream of the calibration polarizer and sees a factor of several less than our other two quartz-based calibration retarders. The thermal drifts of retardance can be neglected in calibration fitting routines for a much longer time for this retarder under illumination. The polarization fringes in this optic are also greatly reduced and temporally stabilized through the thermal behavior \cite{Harrington:2017jh, Harrington:2018cx}.  However, these benefits must be traded off against other error sources such as the exacerbated spatial variation shown later in this paper and / or complexity of the model through the larger magnitudes of circular retardance. Other considerations include the difficulty in obtaining large MgF$_2$ crystals compared to readily available and inexpensive SiO$_2$ crystals. When manufacturing, changing the crystal thickness often imposes unrealistically tight tolerances on other parameters such as alignment requirements, wave front errors or methods of bonding  (optical contact, epoxies or oils). 

\begin{wrapfigure}{l}{0.61\textwidth}
\centering
\vspace{-4mm}
\begin{tabular}{c} 
\hbox{
\hspace{-1.1em}
\includegraphics[height=7.5cm, angle=0]{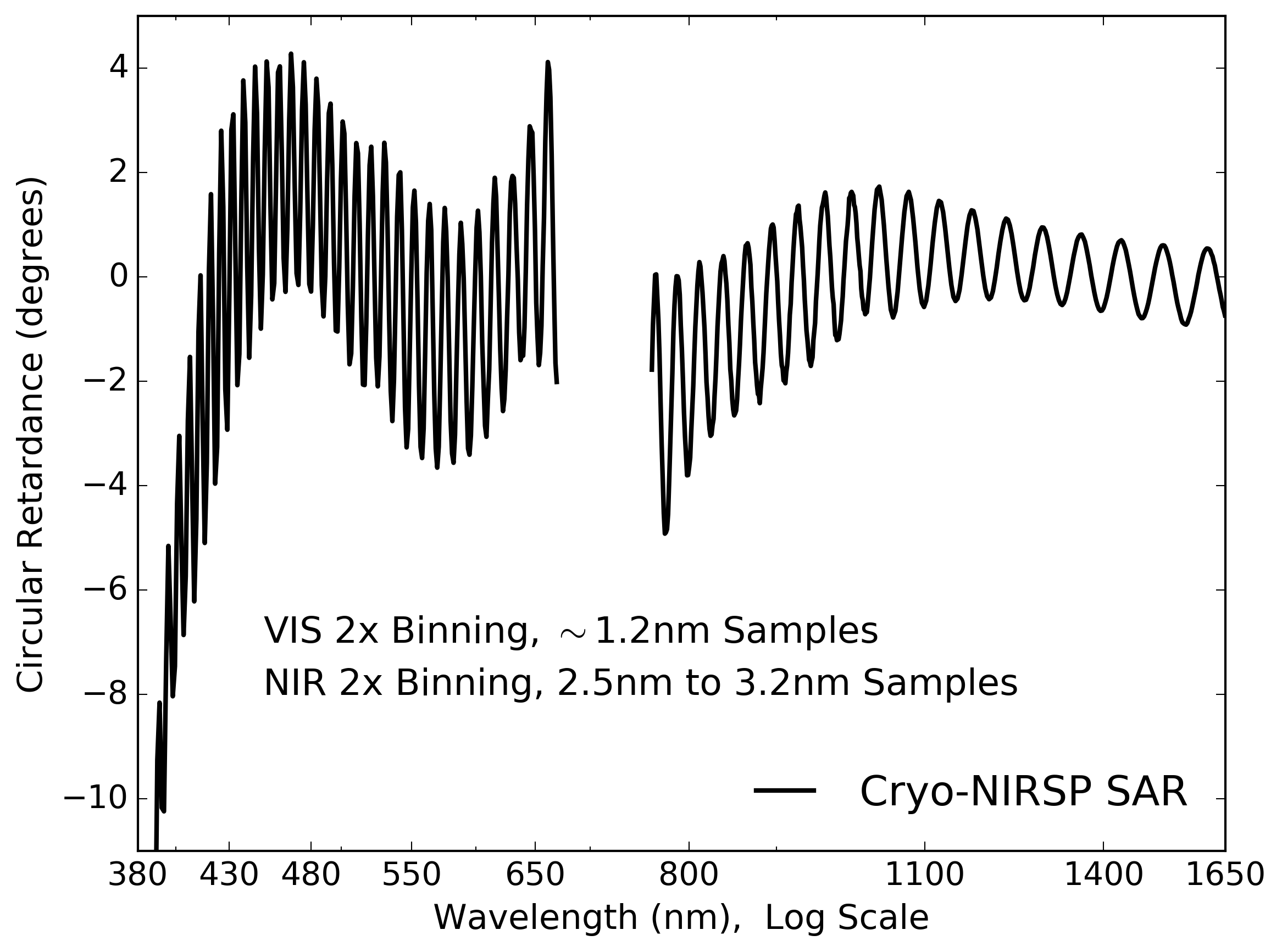}
}
\end{tabular}
\caption[] 
{ \label{fig:fit_nominal_ER_nlsp_cryo_sar_CIRCULAR} The circular retardance component from fits to the NLSP measured Cryo SAR Mueller matrix. Clocking errors cause obvious spectral ripples.  The polishing and plate alignment errors both combine to create net circular retardance.}
\vspace{-4mm}
\end{wrapfigure}

Many calibration procedures assume or fit only for linear retardance components to reduce the number of variables. But for some optics ignoring circular retardance is one of the bigger errors. In Figure \ref{fig:fit_nominal_ER_nlsp_cryo_sar_CIRCULAR} we show the circular retardance component measured in the Cryo-NIRSP SAR. We again do not show fits at wavelengths near 700 nm due to the degeneracies in the elliptical retarder model. This circular component is several degrees in magnitude at visible wavelengths. The retardance spectrally oscillates from the rotational misalignments between the six crystals with a magnitude of 1$^\circ$ to 4$^\circ$ and this oscillation becomes spectrally faster at shorter wavelengths. The magnitude of circular retardance also increases very strongly for the shortest DKIST wavelength of 393 nm with a magnitude of 10$^\circ$. We show this metrology here to motivate further measurements of elliptical retardance for all DKIST retarders and to show the necessity of including this term in DKIST calibrations.

Clocking errors between the two-crystal achromatic pairs creates ellipticity. Additionally, clocking errors between individual crystals add elliptical retardance oscillations at spectral periods of order nanometers. These spectral ripples behave very differently than polishing errors. Spectral periods of these oscillations are determined by the birefringence of each bias plate instead of the refractive index for a given crystal thickness. As such, these oscillations are usually two orders of magnitude slower. For a retarder to function as a calibrator, the optical properties must be known significantly better than the desired calibration accuracy.

The NLSP has been checked for accuracy using several techniques. We verified transmission measurements using uncoated bare glass and crystal substrates. We generally find agreement to simple Fresnel equation calculations to better than 0.3\% transmission for normal incidence. We similarly verified diattenuation by tilting thin uncoated glass substrates and achieve agreement with Fresnel equations within less than 0.3\% for tilt angles less than 30$^\circ$ and substrates thinner than 0.5 mm to minimize beam deviation. When comparing our retardance measurements with other vendor measurements, we often achieve retardance agreement better than 1$^\circ$ for single layer true zero order polycarbonate parts at normal incidence.  A forthcoming paper is in preparation with a detailed analysis of NLSP.  As DKIST instruments will directly derive their own fits to optic retardance, these measurements provide acceptance testing and high quality estimates of system performance.

We now have physical thickness of each individual crystal in every optic.  Meadowlark Optics provided retardance measurements of individual crystals with spectral fitting techniques as well as assessment of the retardance for each compound retarder.  The compound retarders were also measured at 5 locations across the clear aperture. Using NLSP, the DKIST team has also measured the full assembly Mueller matrix and derived elliptical retardance models for every optic.   With this metrology package, we can now begin the measurement and simulation process for the parts across the clear aperture. In the next section we show the impact of spatial variation when retarders are placed near focal planes, as is common in astronomical calibration systems. With measurements of spatial and spectral variation of the elliptical retardance, we can then show impacts to DKIST calibration as a function of wavelength, field of view and describe consequences of optical choices such as the retarder material, clear aperture and bias crystal thickness.

\section{Polishing Errors \& Spatial Uniformity of Retarders}
\label{sec:spatial_polish}

One of the more challenging aspects of manufacturing thin, large crystal retarders is achieving the required uniformity along with meeting all other performance requirements. Typically, the calibration optics must be mounted as far up the optical path as possible. The retardance properties must be known as accurately as possible. A known input polarization must be created as early in the optical path as possible to avoid multiple sources of unknown polarization impact. 

Prime focus or Gregorian focus are ideal locations but these come with severe constraints when using large solar telescopes. The beam diameters are often very large as are the thermal loads imposed by absorption of the high irradiance beam. Material choices often must be carefully considered for wavelength coverage as well as material hardness when achieving polishing uniformity tolerances. 

\begin{wraptable}{l}{0.39\textwidth}
\vspace{-2mm}
\caption{Retarder Beam Properties}
\label{table:dkist_retarder_F_d}
\centering
\begin{tabular}{l l l l l}
\hline \hline
Name		& F/ 		& FP			& CA		& FoV	\\
			&		& mm		& mm	& 		\\
\hline \hline	
ViSP SAR		& 13		& 26.6		& 98		& 5$'$		\\
DL SAR		& 13		& 26.6		& 98		& 5$'$		\\
Cryo SAR		& 13		& 26.6		& 98		& 5$'$		\\
\hline
ViSP PCM	& 32		& 8.1			& 82		& 120$''$	 \\
DL PCM a		& 24		& 22.0		& 11		& 28$''$ 	\\
DL PCM b		& 24		& 6.7			& 11		& 6$''$	\\
DL PCM c		& 62		& 2.6			& 7		& 3$''$	\\
Cryo PCM		& 18		& 37.0		& 105	& 180$''$	\\
\hline \hline		
\end{tabular}
\vspace{-2mm}
\end{wraptable}

Often, the retarder design becomes a case-by-case compromise between many competing design factors. Spatial variation of retardance can complicate the calibration task. Minimizing the beam size can reduce uniformity issues but heat loads and incidence angle issues increase. For modulators after many telescope optics, the more demagnified the pupil, the greater the field dependence becomes. Spatial variation of retarders at pupil planes can be neglected (introducing some small depolarization \cite{Sueoka:2014cm,Chipman:2014ta,2012ApOpt..51..735N,2012OExpr..20...17N,Noble:2011wx,Chipman:2006iu}) but each field propagates through the crystals at a higher angle of incidence, introducing other errors with field angle. Issues arise with allowable wavefront error and beam deflection for mounting near pupil planes or mounting early in the beam path as these errors impact diffraction limited imaging performance, beam wobble and adaptive optics system performance. When mounted near a pupil plane in a collimated beam, fringes are maximized and depolarization from a non-uniform footprint is minimized \cite{Harrington:2017jh,Harrington:2018cx}. All fields illuminate the same spatial region of the retarder and have the same average \cite{Harrington:2017dj, Sueoka:2016vo, Sueoka:2014cm, 2014SPIE.9147E..0FE}. In Appendix \ref{sec:appendix_uniformity_polycarb_FLC}, we outline calibration optics for several modern solar and night time observatories from the perspective of retardance properties and manufacturing decisions.

\begin{wrapfigure}{r}{0.53\textwidth}
\centering
\vspace{-1mm}
\begin{tabular}{c} 
\hbox{
\hspace{-1.1em}
\includegraphics[height=7.8cm, angle=0]{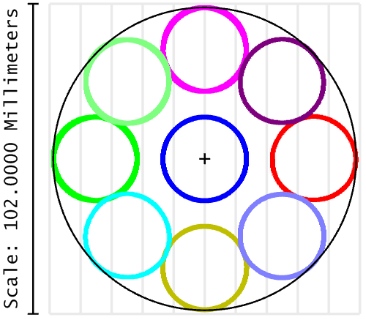}
}
\end{tabular}
\caption[Footprints] 
{ \label{fig:footprint_GOS}  The footprints of the on-axis beam and 2.5 arc minute field angle beam on the DKIST calibration retarder during an eight-orientation calibration sequence computed using the system Zemax model. Blue shows all eight footprints overlapping for the beam on-axis at field center. The other colors show a 2.5 arc minute radius field point as the crystal rotates. The footprint diameter is 26.6 mm on a 120 mm diameter optic with a clear aperture of 105 mm. }
\vspace{-3mm}
\end{wrapfigure}

For DKIST, a location 300 mm above Gregorian focus in a converging F/ 13 beam was chosen as a combination between clear aperture, irradiance, uniformity, fringes, mechanical packaging and other factors. The polarization fringes are quite large for crystal optics at slow F/ numbers and can dominate polarization errors.  Other observatories chose polycarbonate solutions but our beam contains significant UV flux levels and optics must simultaneously cover a very wide wavelength range at high heat loading.  We show in Table \ref{table:dkist_retarder_F_d} the illumination properties for each DKIST retarder. The second column shows the beam focal ratio (F/ number). The third column shows the footprint diameter in millimeters of an individual field point.  The fourth column shows the clear aperture (CA) in millimeters for the relevant instrument field. 

As an example of the beam footprints that sample the calibration retarder at DKIST, Figure \ref{fig:footprint_GOS} shows the eight individual 26.6 mm regions of the calibration retarders illuminated during a typical calibration sequence. The calibration retarder is rotated in steps of 45$^\circ$ to create a diverse set of polarization inputs. The need for spatial uniformity is apparent as most techniques assume the calibration retarder is strictly constant over these eight footprints. For the on-axis beam, the spatial variation between retarder orientations is negligible and varies only by slight offsets from spatial alignment tolerances. For the edge of the FoV however, we see eight completely independent realizations of retardance. Common techniques assuming constant calibration retardance can significantly change the assumed modulation matrix and derived telescope properties at these field angles.

\subsection{Mapping Retardance Uniformity Across The MgF$_2$ Two-Crystal Achromats}
\label{sec:spatial_mgf2_data_cryosar}

Following the initial 5-point retarder metrology and acceptance testing efforts, we improved our retardance analysis codes. We required high accuracy retardance spatial mapping measurements across the full clear aperture to match new simulations of the impact of polishing error. In January of 2017, Meadowlark Optics created this mapping capability on their AE 4 automated retardance measuring system. The system maps elliptical retardance magnitude across the clear aperture of a retarder using computer controlled translation stages. In this section we show and analyze new spatial retardance maps.

\begin{figure}[htbp]
\begin{center}
\vspace{-0mm}
\hbox{
\includegraphics[height=4.4cm, angle=0]{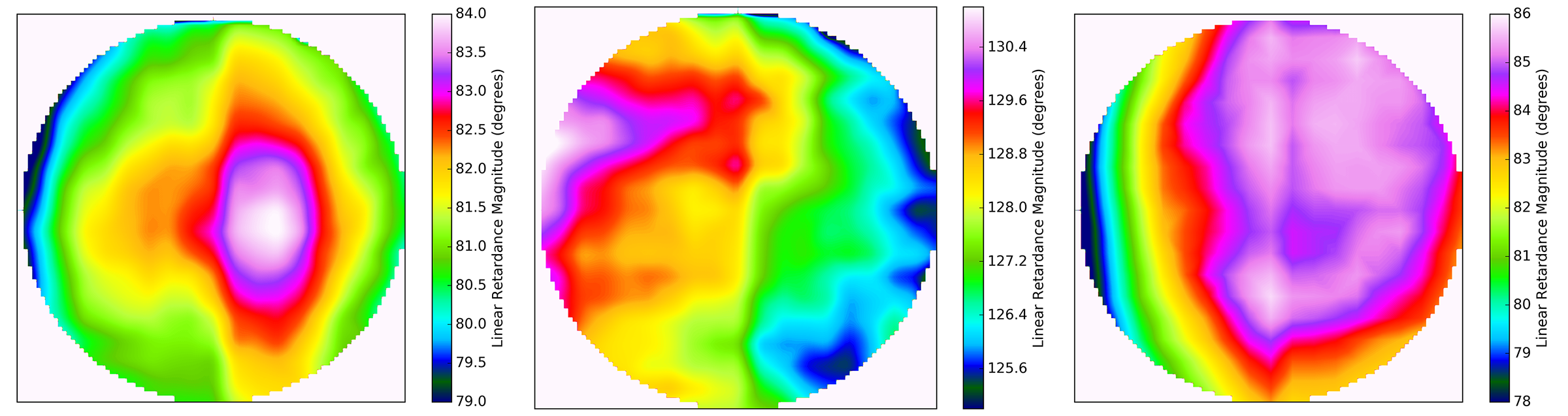}
}
\caption[] 
{\label{fig:cryo_uniformity_mlo_GHG_crystal_maps}  The measured CryoNIRSP SAR linear retardance uniformity maps for individual crystal pairs.  The A-B-A design for this calibration retarder used MgF$_2$ crystal subtraction plates of style names G2-H3-G4.  These two-crystal plates were measured at 5 mm spatial separation between points across a 100 mm aperture. The system used a 3 mm diameter probe beam.  The first two G-style pairs were designed to have 2.23 waves net retardance at 633 nm wavelength when combined in subtraction. The middle crystal pair is H-style and was designed to have 3.346 waves net retardance.  The color scale shows uniformity variation of roughly $\pm$0.02 waves from nominal design retardance across each 2-crystal stack. }
\vspace{-6mm}
\end{center}
\end{figure}

\begin{wrapfigure}{r}{0.55\textwidth}
\centering
\vspace{-3mm}
\begin{tabular}{c} 
\hbox{
\hspace{-1.0em}
\includegraphics[height=6.2cm, angle=0]{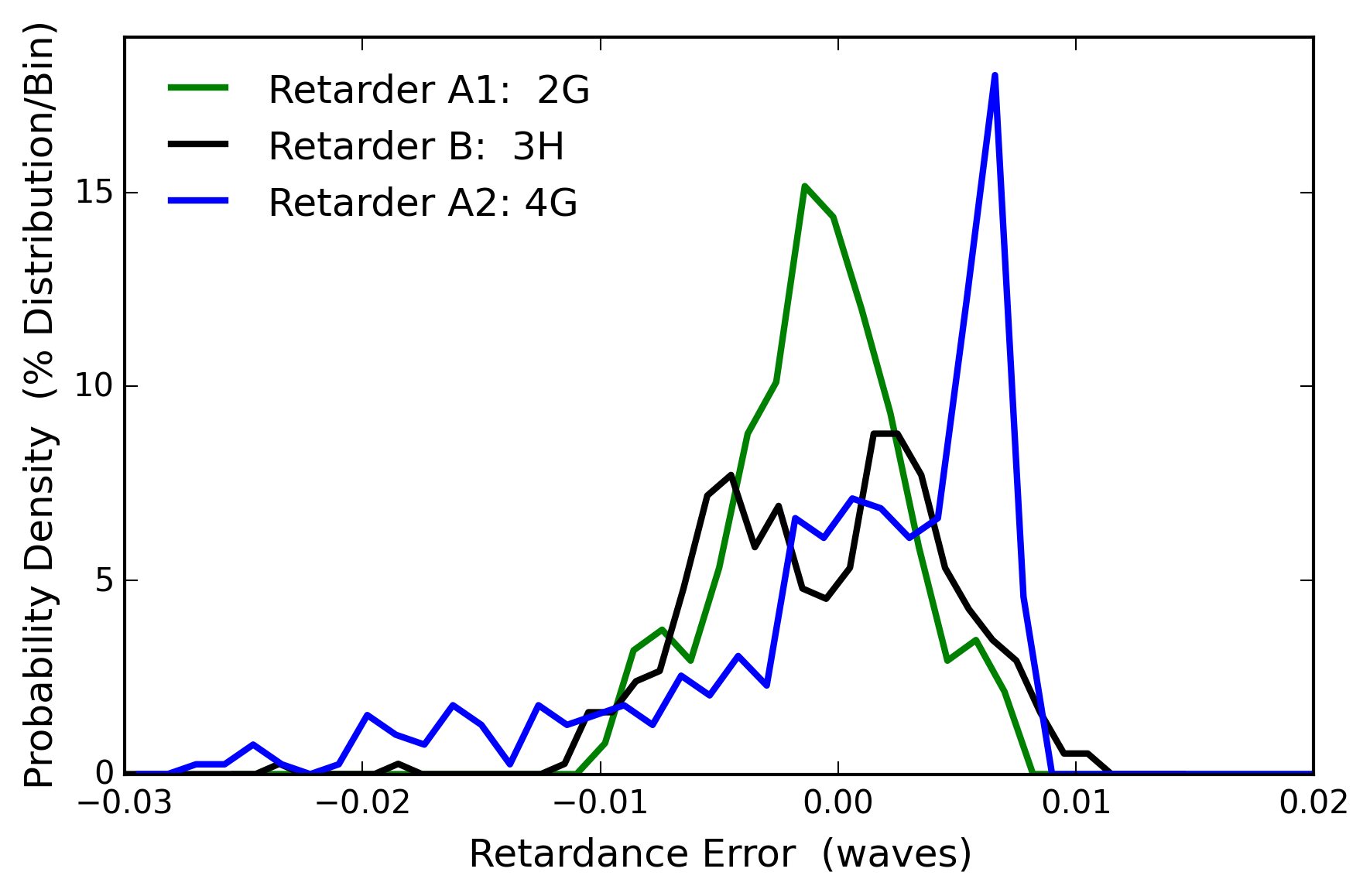}
}
\end{tabular}
\caption[] 
{\label{fig:cryo_uniformity_polish_histogram}  The histogram of retardance polishing nonuniformity values in the CryoNIRSP SAR retarder pairs. The A-B-A design for this calibration retarder used plates of styles G2-H3-G4 in the manufacturing documentation. The G plates were designed at 2.23 waves net retardance when the two MgF$_2$ crystals subtract.  The H plates were designed at 3.346 waves net retardance.  The 4G pair of retarder uniformity was dominated by a large patch of above average retardance in the spatial maps skewing the distribution significantly.   }
\vspace{-4mm}
 \end{wrapfigure}

In our initial testing, only the full elliptical retardance magnitude was measured. We used the Cryo-NIRSP SAR pairs of MgF$_2$ crystal retarders (G2-H3-G4) to demonstrate the capability and to simulate the impact of polishing errors in our six-crystal retarder designs. After initial testing, the capability to output the orientation of the linear retardance fast axis was added. We measured retardance at 415 points on a grid spacing of 5 mm with a test beam diameter of 3 mm. Machine time for this measurement is about 7 hours. In this initial setup, the clear aperture had to be scanned in two halves, leading to a slight discontinuity in the data as the part was rotated and centered. 

For the G-style crystal pairs, the design provides a net retardance of 2.30 waves at a measurement wavelength of 633.443 nm. The Meadowlark Optics AE4 system uses filters for wavelength selection and the closest available filter to that wavelength is centered at 634.0 nm and has a bandpass full width at half maximum of 10.5 nm. The ideal design retardance for this waveplate pair in subtraction at this wavelength is 2.228 waves. Similarly for the H style retarder pairs, we anticipate 3.346 waves net retardance for the two MgF$_2$ crystals in subtraction. The fast and slow axes of the crystals are along the x and y axes in these plots. For reference, the contracted five point measurements for uniformity acceptance of this crystal pair previously supplied to DKIST were taken at the center and at 45 degrees to these axis directions.

Figure \ref{fig:cryo_uniformity_mlo_GHG_crystal_maps} shows the retardance uniformity across a 90 mm clear aperture sampled on a 5 mm spatial grid with the 3 mm probe beam. From these data sets, we can derive distributions of polishing errors as shown in Figure \ref{fig:cryo_uniformity_polish_histogram}. The first pair of plates tested are called G2 and showed a reasonably Gaussian distribution centered about the nominal retardance values.  The second pair of plates called 4G were polished to the same nominal design but with a retardance uniformity distribution that was skewed and showed polishing errors within the clear aperture over 0.02 waves. The 3H pair of plates showed a non-Gaussian profile but with polishing errors mostly within 0.01 waves retardance variation. 

We also performed a Zernike polynomial decomposition to the polishing uniformity data. There was substantial amplitude for the first five terms where the retardance non-uniformity resembles tilt, power and astigmatism. The higher order terms do not substantially improve the fit when using the Zernike basis.  We also computed a simple 2D Fourier analysis of the retardance maps within a 60 mm rectangle sampled within the clear aperture. The two lowest spatial frequencies account for 90\% of the variation.  This shows the polishing error is dominated by variation at the largest spatial scales.

\begin{wrapfigure}{r}{0.63\textwidth}
\centering
\vspace{-3mm}
\begin{tabular}{c} 
\hbox{
\hspace{-1.0em}
\includegraphics[height=6.8cm, angle=0]{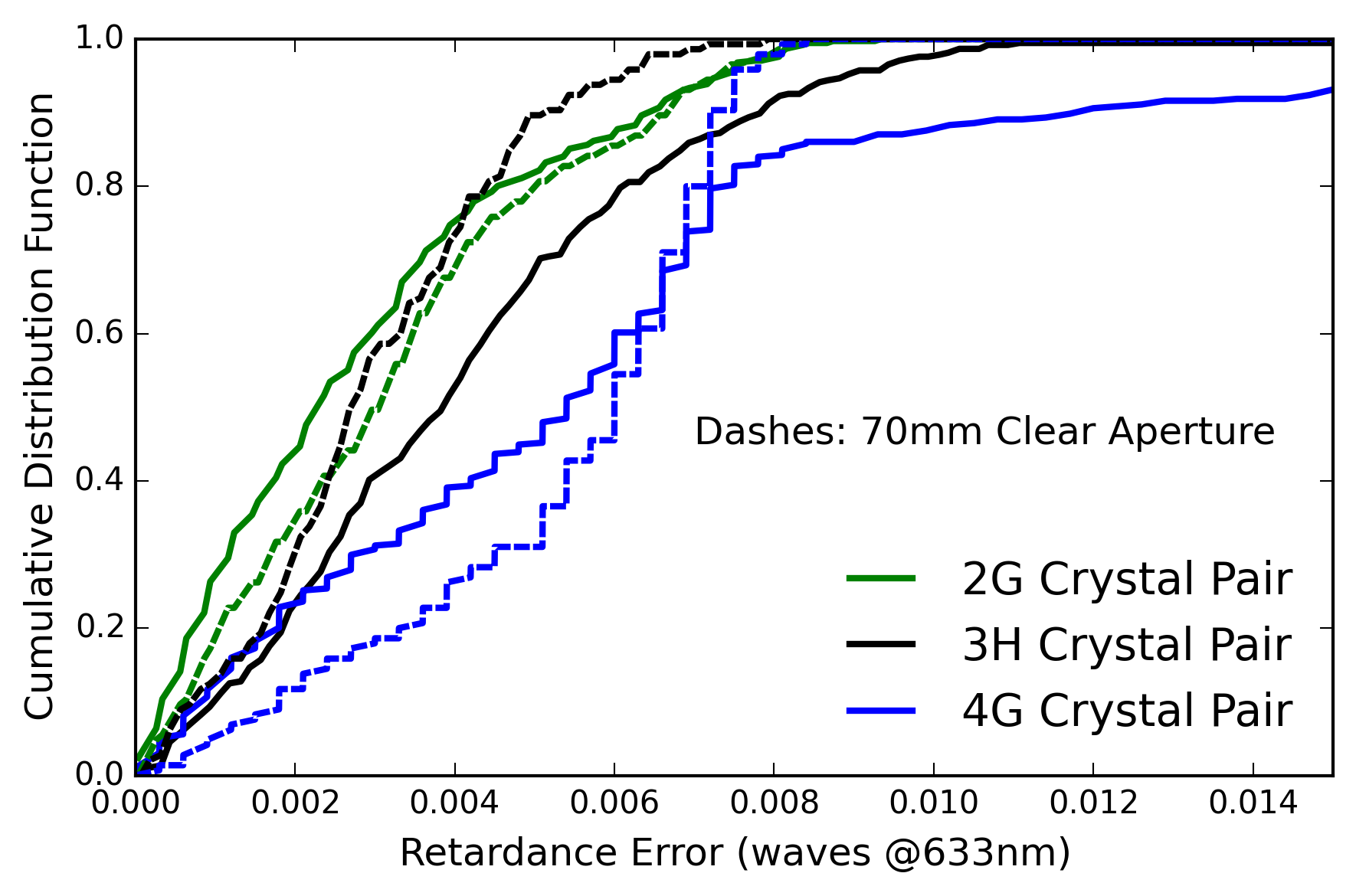}
}
\end{tabular}
\caption[] 
{ \label{fig:cryo_uniformity_polish_CDF} 
The cumulative distribution of retardance polishing nonuniformity values in the CryoNIRSP SAR retarder pairs  G-H-G.  The solid lines show a 95 mm clear aperture while the dashed lines show a reduced 70 mm clear aperture to avoid edge artifacts. }
\vspace{-6mm}
 \end{wrapfigure}

The cumulative distribution function for the polishing errors shows what amplitude of error to expect across the clear aperture. Figure \ref{fig:cryo_uniformity_polish_CDF} shows the cumulative distribution function (CDF) of retardance spatial nonuniformity for the 2G - 3H - 4G bicrystalline retarder pairs that make up the CryoNIRSP SAR.  Solid lines shows the polishing error distribution across the full 95 mm test aperture while the dashed lines show the distributions considering only points within an 70 mm clear aperture corresponding roughly to 3 arc minutes DKIST FOV at Gregorian focus.  

The 2G pair had a reasonably Gaussian-shaped distribution in the histogram of errors.  Correspondingly, the green curve in Figure \ref{fig:cryo_uniformity_polish_CDF} shows a CDF where roughly 80\% of the clear aperture sees spatial non-uniformity of 0.005 waves retardance following a reasonably Gaussian-style curve of growth.  More than 95\% of the clear aperture has retardance deviations less than the nominal polishing specification of 0.01 waves retardance.  

However, the 4G retarder pair had a section near the edge of the clear aperture with stronger deviations from the average.  The distribution of errors above in Figure \ref{fig:cryo_uniformity_polish_histogram} showed a skewed distribution with many points lying up to 0.02 waves retardance below the average retardance value. The CDF for this 4G retarder pair shows that only 40\% of the clear aperture will see retardance errors less than 0.005 waves while 85\% of the clear aperture will see polishing errors below the 0.01 waves retardance specification.  {\it Note that this was outside the 5-point test location specified in the fabrication contract}.   As the spatial maps show this 4G retardance error is highly concentrated near the edge of the optic, DKIST instruments using a reduced footprint would see substantially less spatial non-uniformity. To demonstrate this, histograms and CDFs were recomputed considering a 70mm clear aperture.  The skewed distribution on the 4G plate is still present but over 99\% of the points within the 70mm clear aperture have less than the 0.01 waves retardance error spec.  

For DKIST, observing modes and instruments using the extreme edge of the clear aperture will be impacted by this error. These modes and instruments include the ViSP PCM which uses a full 100 mm clear aperture, wide field calibrations with most instruments, and scanning modes were steering mirrors repoint the instrument to the edge of the calibration retarders at Gregorian focus.

\section{Simulating Six-Crystal Retardance Spatial Variation Design Sensitivity Using Individual Compound Retarder Maps }
\label{sec:spatial_ER_variation}

Next, we use retardance spatial uniformity maps of the individual compound retarders as representative of polishing errors for these two-crystal subtraction pairs. We then simulate the full design to derive the elliptical retardance design sensitivities of our six crystal retarders. We take the spatial retardance maps for the MgF$_2$ G-H-G retarders of Figure \ref{fig:cryo_uniformity_mlo_GHG_crystal_maps} and translate the retardance maps into physical thickness polishing error maps for use in simulations. To simulate the sensitivity of each calibration retarder design to polishing errors, we apply a linear scaling of these physical thickness errors to the various metrology results for the SiO$_2$ and MgF$_2$ crystal pairs. We compute models for all six DKIST retarders with magnitudes drawn from Tables \ref{table:polish_error_measured_plates_as_built} and \ref{table:polish_error_measured_plates_as_built_stack_error}. We compute the Mueller matrix for all wavelengths and spatial locations across the optic after applying the derived physical thickness polishing errors to the relevant crystals in the stack.  We will later compare this with measurements of the as-built six-crystal retarders. 

\begin{figure}[htbp]
\begin{center}
\vspace{-0mm}
\hbox{
\includegraphics[height=6.6cm, angle=0]{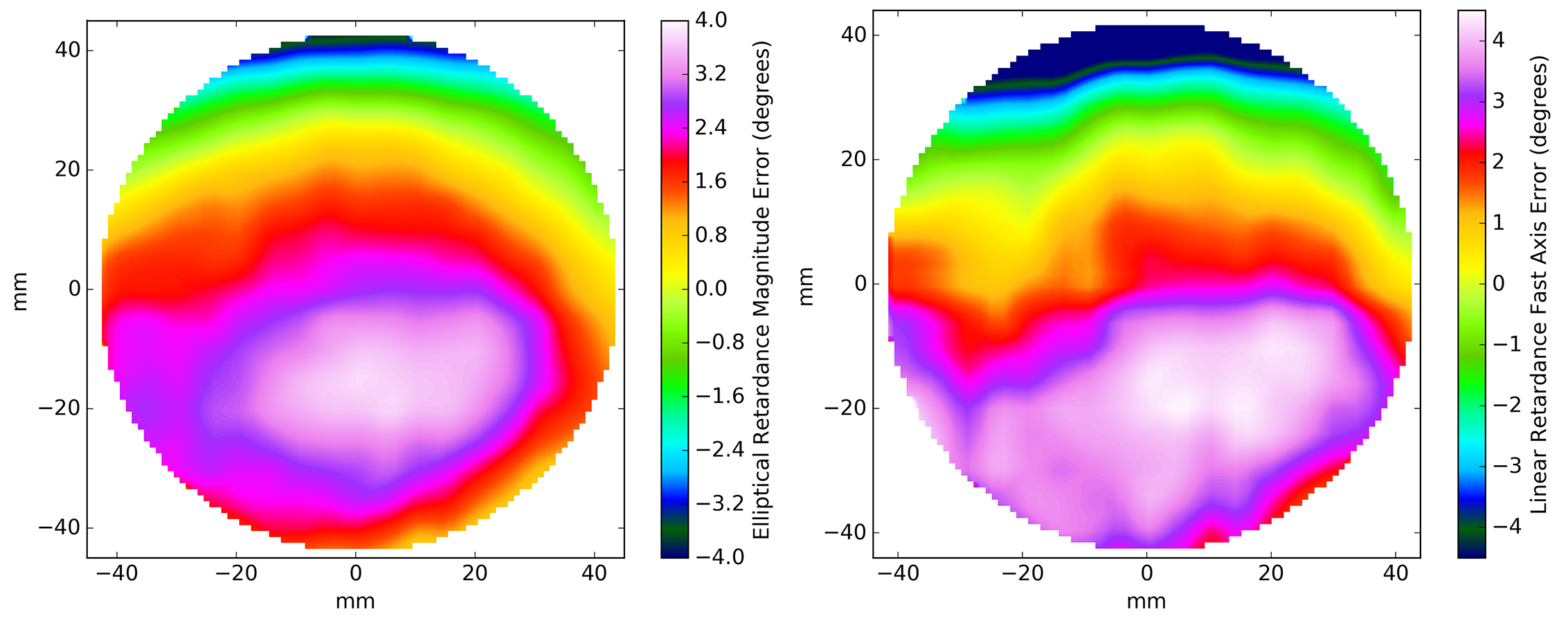}
}
\caption[] 
{\label{fig:cryo_uniformity_mlo_ERmap_656} The predicted Cryo-NIRSP SAR elliptical retardance parameter uniformity maps. A wavelength of 656 nm is used assuming each crystal pair is polished per the errors derived from Figure \ref{fig:cryo_uniformity_mlo_GHG_crystal_maps}. The elliptical retardance magnitude spatial error is shown in the right plot with a scale of $\pm$4$^\circ$. The linear retardance fast axis orientation error is shown in the left plot.  At this wavelength, polishing errors mostly change the magnitude of linear retardance (not fast axis orientation) and introduce mild circular retardance.}
\vspace{-6mm}
\end{center}
\end{figure}

For the simulations in this paper, the quartz retarder substrates were adjusted to 0.005 waves retardance variation at 633 nm wavelength. This is a crystal thickness variation of 0.349 $\mu$m. For the Cryo SAR we simulated 0.01 waves of polishing error for the MgF$_2$ crystals, corresponding to 0.538  $\mu$m of physical thickness. For the Cryo PCM, we modeled 0.02 waves for 1.076  $\mu$m crystal thickness variation. For the models here, we simulate polishing error by changing only the second crystal in each pair. This reflects polishing tolerance levels near the measured levels of 0.05 waves per pair for quartz crystals and 0.01 to 0.02 waves RMS retardance variation per pair for MgF$_2$ crystals. Note, the choice of which plate to change in simulated polish error is arbitrary as seen in our tolerancing analysis. The only impact is sign changes and slight design perturbations, not the magnitude or spatial distribution of polishing non-uniformity.

With the full Mueller matrices computed for each optic, we can perform an elliptical retarder fit at every spatial location across the clear aperture. Figure \ref{fig:cryo_uniformity_mlo_ERmap_656} shows an example for the MgF$_2$ Cryo-NIRSP calibration retarder. At a wavelength of 656 nm, we can do a simple bounded elliptical retarder fit, restricting the domain to be less than one wave net retardance to avoid any ambiguities. The nominal design fit gives elliptical retardance parameters of (-36$^\circ$, 45$^\circ$ and 0$^\circ$) for the two linear and the circular retardance components in the axis-angle formalism. This is equivalent to a 60$^\circ$ linear retarder with a fast axis orientation of 127$^\circ$ and no circular retardance. We see spatial variation of the retardance components about these average values. As seen in the spatial error maps computed in Figure \ref{fig:cryo_uniformity_mlo_ERmap_656}, the linear retardance magnitude varies by $\pm$4$^\circ$.  In the individual elliptical retarder parameter fits, the first component changes from -27$^\circ$ to -42$^\circ$ and the second component changes as well to give the fast axis variation seen in the right hand graphic of Figure \ref{fig:cryo_uniformity_mlo_ERmap_656}.  Roughly 3$^\circ$ circular retardance variation is seen as well. This level of spatial variation translates to 8$^\circ$ degrees variation in the magnitude of linear retardance (54$^\circ$ to 62$^\circ$) as well as a similar variation of the orientation of the fast axis of linear retardance.  As we show in Appendix \ref{sec:degenerate_solutions}, when we use the unsrestricted $>$1 wave solution, the predicted elliptical retarder component errors are larger than when using bounded type rotation matrices that always assume the retardance is less than one wave $quv$ rotation.  However, the computed Mueller matrices are always the same. We note that the spatial variation has strong wavelength dependence. In the same model at 396 nm wavelength, we see up to 20$^\circ$ peak to peak of circular retardance as well as over 30$^\circ$ degrees retardance variation for both linear retardance components. The linear retardance magnitude ranges from 258$^\circ$ to 275$^\circ$ spatially.  The fast axis similarly wanders by 10$^\circ$ across the optic.

\subsection{Sensitivity to Error Spatial Distributions: Order and Orientation Dependence}
\label{sec:spatial_sensitivity_ordering}

The spatial variance of the retardance errors illustrates how each design can be more or less sensitive to polishing variation. Using the spatial maps for each simulation, we take a uniformly weighted standard deviation for each elliptical retardance parameter across the clear aperture.

\begin{figure}[htbp]
\begin{center}
\vspace{-0mm}
\hbox{
\includegraphics[height=5.4cm, angle=0]{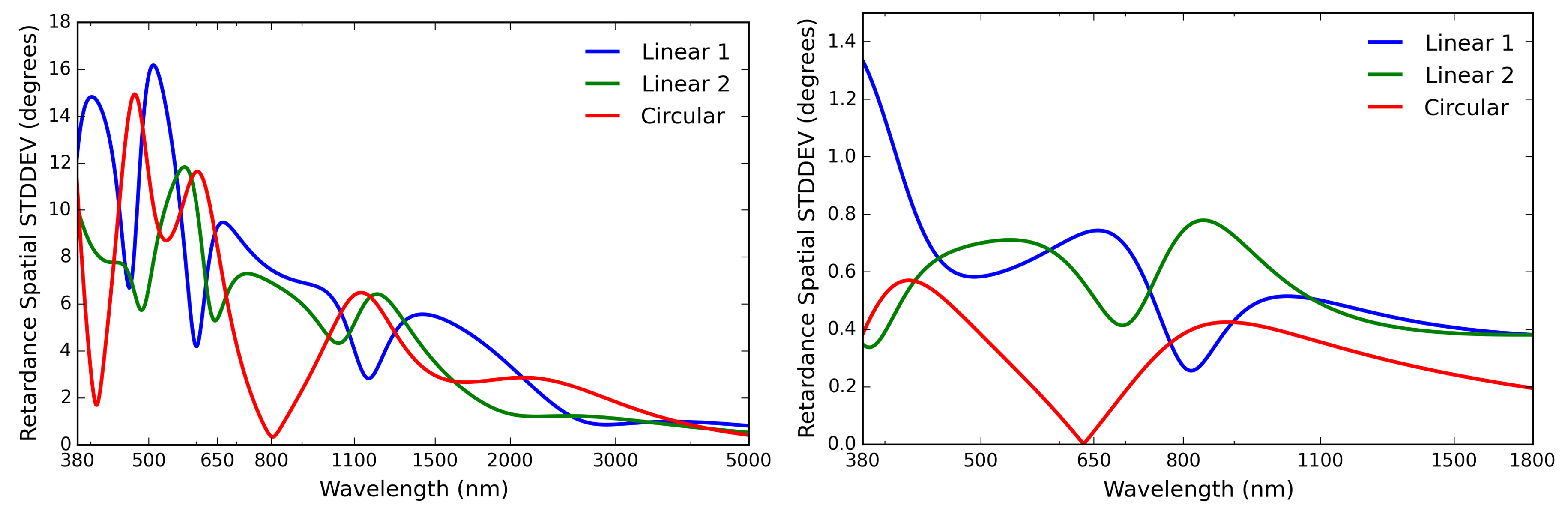}
}
\caption[] 
{\label{fig:sar_retardance_stddev_spatially} The modeled standard deviation of the three elliptical retarder components spatial variation across the clear aperture for two different designs. The left shows the CryoNIRSP PCM with polishing error scaled to 0.02 waves error while the right shows the DLNIRSP SAR with the polishing error was scaled to 0.005 waves retardance at 633 nm.  Each component of retardance can have very different wavelength dependence and sensitivity for different designs. The CryoNIRSP PCM is a designed for wavelengths longer than 1000 nm and shows low sensitivity in that bandpass.  At short wavelengths, the sensitivity is an order of magnitude higher. }
\vspace{-6mm}
\end{center}
\end{figure}

Figure \ref{fig:sar_retardance_stddev_spatially} shows this spatial variation of retardance for two of our six retarder designs. The left hand side of Figure \ref{fig:sar_retardance_stddev_spatially} shows the Cryo-NIRSP modulator (PCM) while the right side shows the DL-NIRSP calibration retarder. There is significant differences between the wavelength dependencies for each the three elliptical retardance components. The Cryo-NIRSP modulator is an elliptical retarder by design so there is significant circular retardance in nominal bandpass of 1000 nm to 5000 nm. Given the net retardance of each crystal is multiple waves for visible wavelengths, the spatial variation is up to 16$^\circ$ in each elliptical retardance component at the shortest wavelengths.  

The DL-NIRSP calibration retarder on the right has more than an order of magnitude less spatial variation even though the retardance uniformity tolerance was only four times tighter. The CryoNIRSP PCM design includes plates of 1.9 : 1.3 : 1.9 waves net retardance while the DL SAR is 0.7 : 1.0 : 0.7 at 633 nm wavelength, which gives a factor of roughly two in increased sensitivity if you only scale by compound retarder net magnitude. Each retarder design is sensitive not only to the spatial statistics of the polishing error, but there is also sensitivity to the order in which the errors in each crystal stack. With polishing error maps, we can apply errors with different magnitudes, orientations and applied in different orders to each pf the six crystals in each optic to simulate the impact on different designs. We created a model with different spatial errors as follows: The simulated error distribution on retarder pair 1 is the spatial transpose of the polishing error measured on retarder pair 2. The simulated error on retarder pair 2 becomes the error measured on retarder pair 1 rotated by 90$^\circ$. The simulated error on retarder pair 3 becomes the error measured on retarder pair 3 spatially reversed (flipped) left to right. 

\begin{figure}[htbp]
\begin{center}
\vspace{-2mm}
\hbox{
\includegraphics[height=5.5cm, angle=0]{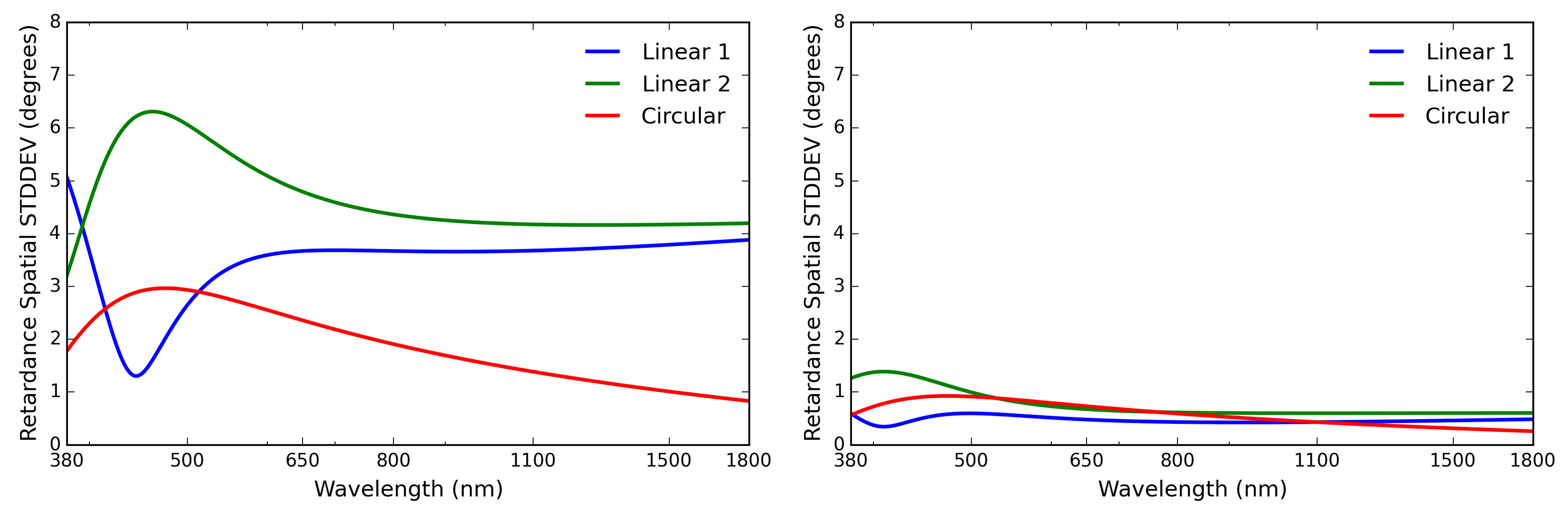}
}
\caption[] 
{\label{fig:visp_sar_retardance_stddev_spatially} The standard deviation of spatial variation in elliptical retardance components for the ViSP SAR. We used nominal polishing error spatial distributions on the left and used the spatially modified polishing errors on the right. Derived spatial sensitivity is strongly tied to the distribution, orientation and order of which crystals have which errors. Both cases used 0.005 waves net retardance magnitude error amplitude at 633 nm wavelength. Blue and green show the first and second components of linear retardance respectively.  Red shows circular retardance.}
\vspace{-8mm}
\end{center}
\end{figure}

\begin{wrapfigure}{l}{0.60\textwidth}
\centering
\vspace{-5mm}
\begin{tabular}{c} 
\hbox{
\hspace{-1.0em}
\includegraphics[height=7.6cm, angle=0]{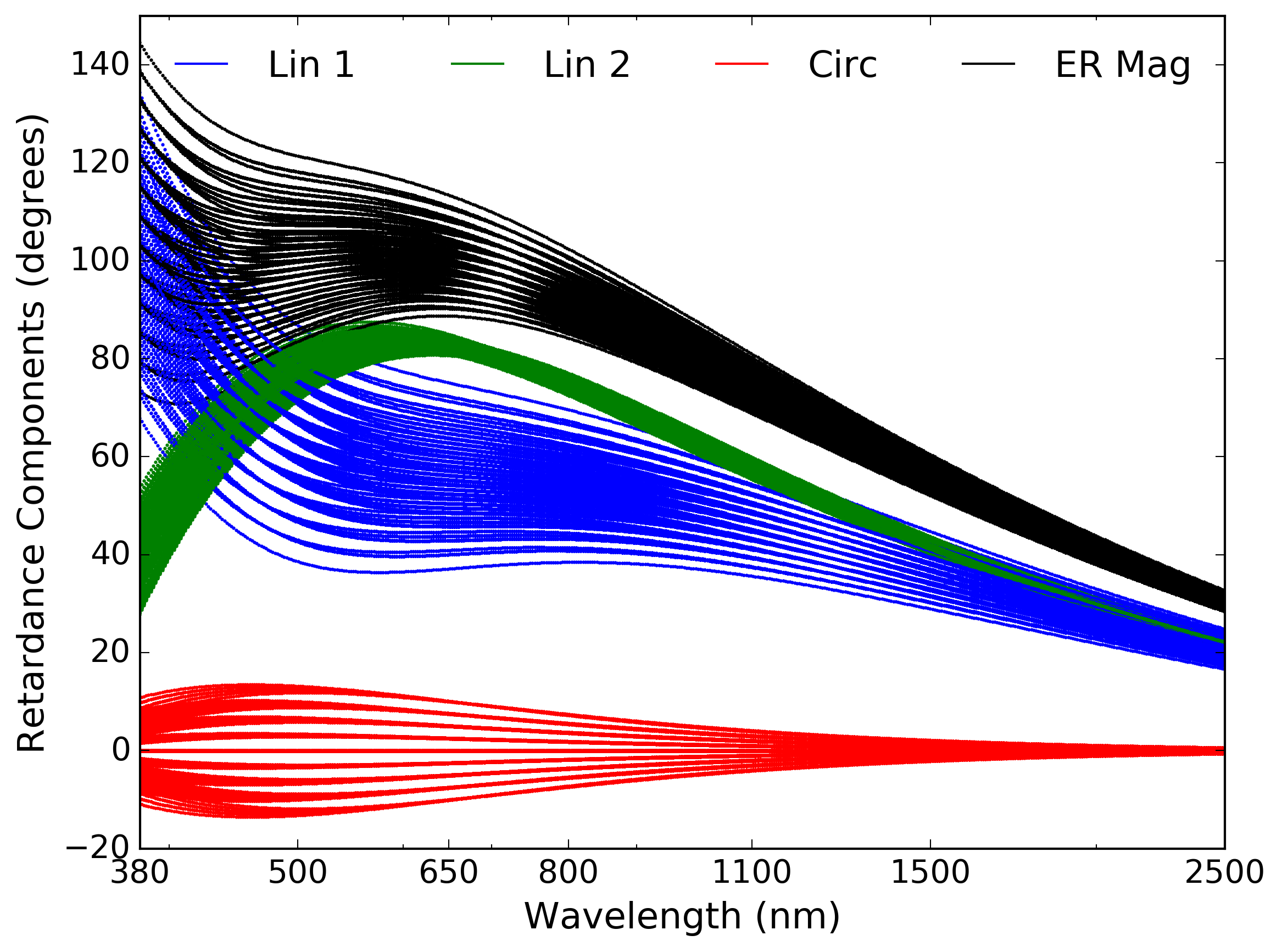}
}
\end{tabular}
\caption[] 
{\label{fig:visp_polish_tolerance_ER_fit}  The elliptical retarder model fits to polishing error simulations applied to the ViSP SAR design. Blue and green show the first and second components of linear retardance. Red shows circular retardance. Black shows the elliptical retardance magnitude as the root-sum-square of all three components. See text for details.}
\vspace{-5mm}
 \end{wrapfigure}

For the various designs, the modeled spatial variation could be factors of a few greater depending on the specific orientation of the polishing errors. We recomputed models for all six DKIST retarders using the same polishing error magnitudes but with this alternate spatial variation. The spatial distributions are similar to previous results, but with different magnitudes and spectral behavior. The ViSP SAR design wavelength range of 380 nm to 1000 nm. The design has several times less deviation as the DL-NIRSP SAR design at amplitudes in the range of 0.25$^\circ$ to 1.0$^\circ$ made of the same SiO$_2$ material when simulated with the same magnitude polishing errors.  The ViSP SAR is also a factor of few less in net retardance so the rough scaling of the polishing error sensitivity is expected.   

Figure \ref{fig:visp_sar_retardance_stddev_spatially} compares the spatial variation of the three elliptical retardance components across the clear aperture for the ViSP calibration retarder (SAR) optical model with variable orientation of polishing errors. In both retarder models, the magnitude of the polishing thickness errors was the same 0.005 waves net retardance at 633 nm wavelength applied the SiO$_2$ crystals. The nominal polishing error spatial distributions lead to standard deviations shown in the left plot of Figure \ref{fig:visp_sar_retardance_stddev_spatially}.  
 
Spatial standard deviations are in the range of 1.5$^\circ$ to over 6$^\circ$. The circular retardance shown in red has peak spatial variance occurs much short wavelengths while the linear retardance variation is significantly higher. When using the polishing errors at the alternate rotated and transposed orientations, the sensitivity of the ViSP SAR design is always less than 1.5$^\circ$ as seen in the right plot of Figure \ref{fig:visp_sar_retardance_stddev_spatially}. The circular retardance component variation is greater than the linear retardance terms at some wavelengths. The only difference between the two models is the order of application and spatial orientation of the polishing errors. 

To assess the sensitivity of each DKIST retarder design to polishing errors more generally, we compute a large grid of spectral models. The Mueller matrix of each optic was computed with polishing errors of the appropriate retardance error at 633 nm wavelength applied to each crystal independently against each other crystal. For completeness, we applied positive, negative and no error for all crystals. This resulted in 3$^6$ models for each DKIST six crystal retarder design.  An example of elliptical retardance fits is shown in Figure \ref{fig:visp_polish_tolerance_ER_fit} for the 729 Mueller matrix models for the ViSP calibration retarder (SAR) design. The blue and green curves show the two linear retardance components.  Their wavelength variation shows rotation of the fast axis of linear retardance. This retarder is nominally a quarter-wave net retardation in the 380 nm to 1000 nm bandpass with zero circular retardance. The red curve shows the circular retardance component has greater sensitivity to polishing errors around 400 nm and 900 nm with maximal sensitivity around 500 nm wavelength.  The black curve of Figure \ref{fig:visp_polish_tolerance_ER_fit} shows the total elliptical retardance magnitude as the root sum square of the three retardance components in the axis-angle rotation matrix formalism.

\section{Spatial Measurements of As-Built Six-Crystal Retarders}
\label{sec:as_built_spatial_maps}

In this section we will show an example of the measured spatial retardance variation for the as-built six-crystal DKIST retarders. The individual compound retarder maps and simulations in the last section will be compared with the metrology of the six crystal stacks. The spatial distributions vary between SiO$_2$ and MgF$_2$ materials which were polished using different techniques and to different tolerances, which drives calibration strategy. Meadowlark Optics provided several spatial maps of retardance across the clear aperture of the as-built retarder stacks as part of our metrology efforts. We tested our retarders at wavelengths where the system has good efficiency, where retardance values are more than 30$^\circ$ away from multiples of 0$^\circ$ or 180$^\circ$.  

\begin{wraptable}{l}{0.44\textwidth}
	\vspace{-2mm}
	\caption{Retardance Spatial Metrology}
	\label{table:mlo_spatial_map_parameters}
	\centering
	\begin{tabular}{l c c c c}
		\hline
		\hline
		\hline
		Optic				& $\lambda$ 	& Spc 	& $\delta$	& $\delta_{circ}$ \\
		Name			& nm			& mm	& wave	& wave	\\
		\hline
		\hline
		Cryo SAR G2		& 633 		& 5 		& 2.23	& 0		\\
		Cryo SAR H3		& 633 		& 5 		& 3.35	& 0		\\
		Cryo SAR G4		& 633 		& 5 		& 2.23	& 0		\\
		\hline
		\hline
		Cryo SAR 		& 600 		& 3 		& 0.25	& 0		\\
		\hline
		ViSP SAR 		& 420 		& 3 		& 0.25	& 0		\\
		ViSP SAR 		& 633 		& 2 		& 0.25	& 0		\\		
		ViSP PCM 		& 420 		& 3 		& 0.33	& 0.07	\\
		Cryo PCM 		& 600 		& 5 		& 0.33	& 0		\\
		DLNIRSP SAR 		& 666 		& 3 		& 0.33	& 0		\\		
		DLNIRSP PCM 	& 600 		& 3 		& 0.67	& 0.05	\\
		\hline
		\hline
	\end{tabular}
	\vspace{-3mm}
\end{wraptable}

Table \ref{table:mlo_spatial_map_parameters} shows the retarder metrology wavelengths and spatial sampling. The first column shows the name of the optic under test. The first three optics are the two-crystal achromats used in the Cryo-NIRSP calibration retarder (SAR).  The rest of the optics are the as-built six-crystal retarders.  The second column shows the wavelength of the interference filter.  All filters had a $\sim$10 nm FWHM.  The third column shows the spatial sampling (Spc) ranging from 2 mm to 5 mm. All testing was done with a 3 mm diameter probe beam.  The fourth column shows the design retardance magnitude of the assembly.  The final column shows the circular retardance present in the design $\delta_{circ}$. We used wavelengths where the as-built six crystal retarders were either quarter-wave or third-wave linear retarders.  For the DL-NIRSP and ViSP modulators (PCMs) there were no wavelenghts without circular retardance so we chose short-wavelength bandpasses with minimal circular retardance.

In this section, we compare maps at multiple wavelengths for the shorter wavelength ViSP calibration retarder made from SiO$_2$ crystals. We added the capability to additionally map the fast axis orientation of linear retardance.  The spatial distributions, error magnitudes and wavelength dependencies are substantially different.  The Meadowlark Optics spatial metrology system mechanics were upgraded in summer 2017 to allow for scanning of parts over a full 110mm clear aperture continuously without manual removal and remounting of the optic.  With this upgrade, Meadowlark also updated the software capability to output the best fit linear retardance fast-axis orientation in addition to elliptical retardance magnitude for each spatial position measured. Additional fitting parameters were output but the fundamental procedure was unchanged, measuring elliptical retardance magnitude and linear fast-axis orientation.

This upgrade removed the requirement to remount our large parts during metrology. The ViSP SAR was first measured with a $\sim$3 mm footprint beam, a 2 mm spatial step size giving 2290 individual spatial measurements.  The wavelength is set by a 632.4 nm interference filter with 10 nm FWHM. At this wavelength, the optic was nominally a $\sim$100$^\circ$ to $\sim$110$^\circ$ linear retarder.  The second set of measurements was done at 420 nm wavelength using a 10 nm FWHM interference filter.

\begin{wrapfigure}{r}{0.55\textwidth}
\centering
\vspace{-3mm}
\begin{tabular}{c} 
\hbox{
\hspace{-1.0em}
\includegraphics[height=6.1cm, angle=0]{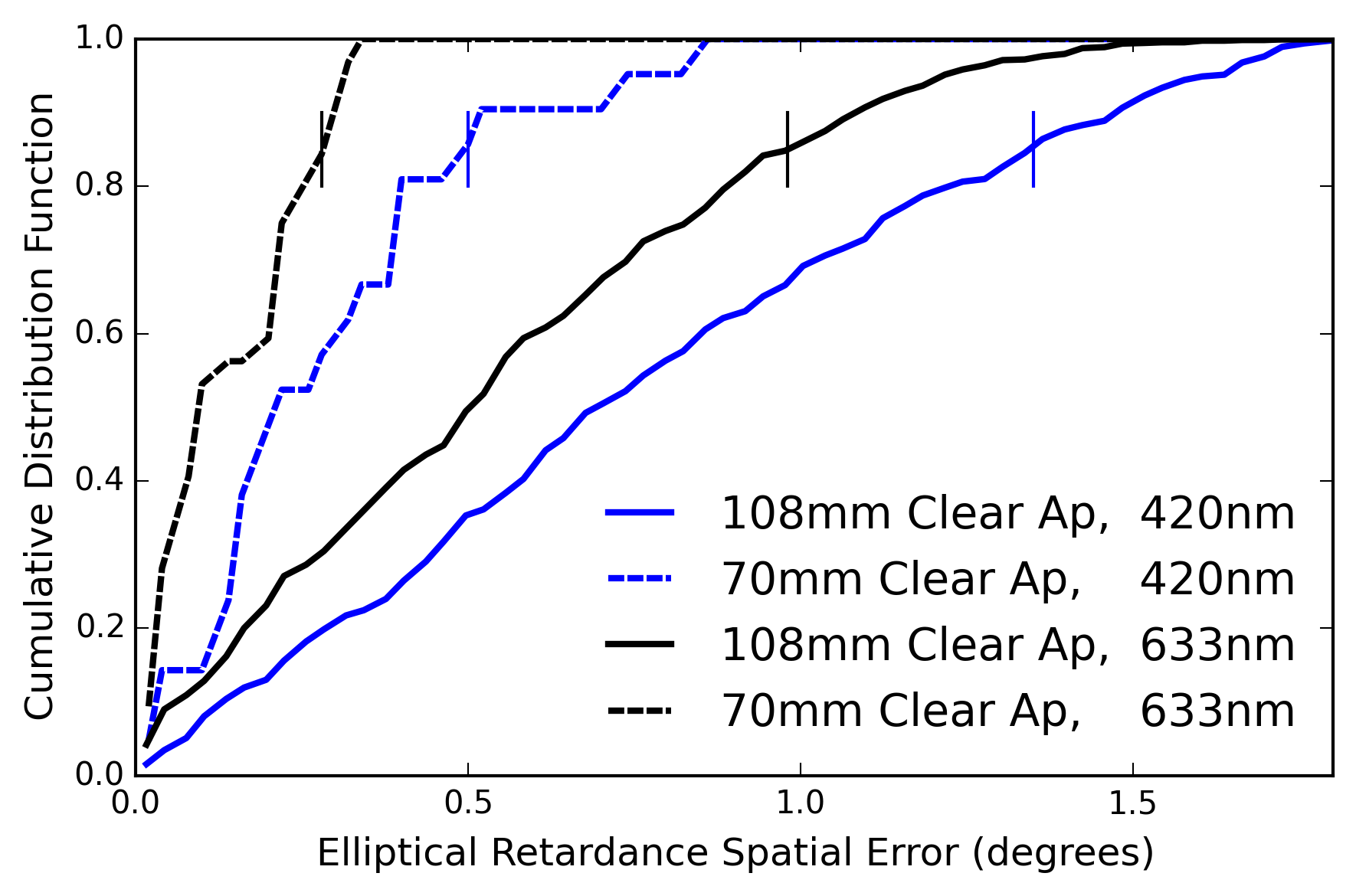}
}
\end{tabular}
\caption[] 
{ \label{fig:visp_sar_uniformity_polish_CDF} 
The cumulative distribution function of measured elliptical retardance spatial variation in the SiO$_2$ ViSP calibration retarder.  The solid lines show a 108mm clear aperture while the dashed lines show a reduced 70mm clear aperture.  Blue shows 420 nm wavelength while black shows 633 nm.}
\vspace{-4mm}
 \end{wrapfigure}

Figure \ref{fig:visp_sar_uniformity_polish_CDF} shows the cumulative distribution function for spatial retardance errors in this quartz calibration retarder at both wavelengths. The quartz retarder crystals were double-side polished and as such have a few times better spatial uniformity in thickness. This improvement is reflected in the cumulative distribution functions compared to Cryo-NIRSP modulator above. For a 108 mm clear aperture, 80\% of the elliptical retardance magnitudes measured are less than 1.0$^\circ$ spatial variation at 633 nm wavelength and less than 1.35$^\circ$ at 420 nm wavelength.

\begin{figure}[htbp]
\begin{center}
\vspace{-0mm}
\hbox{
\includegraphics[height=13.0cm, angle=0]{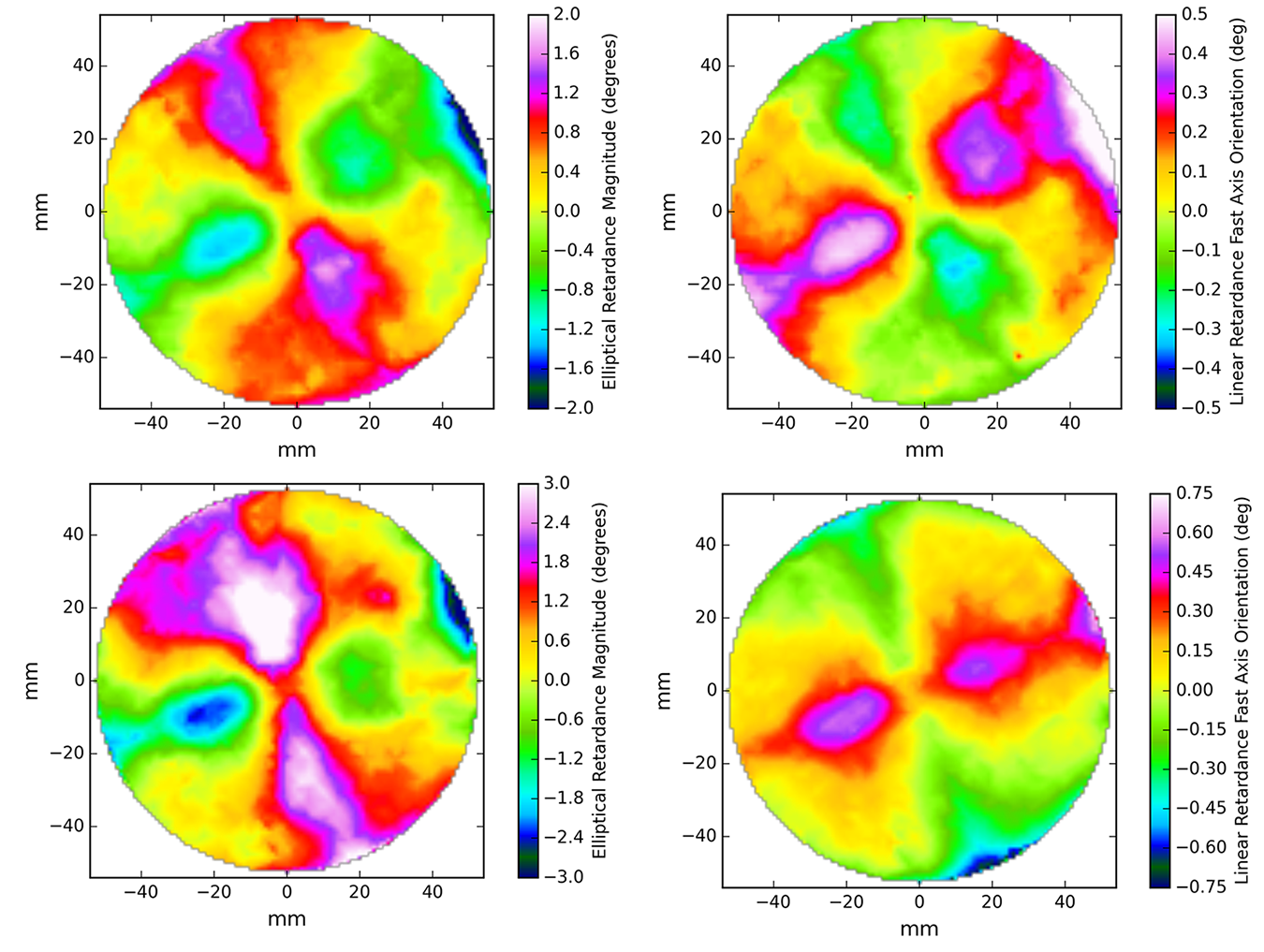}
}
\caption[] 
{\label{fig:mlo_map_visp_sar_632nm_and_420nm}  The spatial measurements of retardance properties for the ViSP SAR as-built.  The top graphics show maps measured at 632.4 nm using a $\sim$10 nm bandpass narrow-band filter and spatial sampling of $\sim$2 mm. The top right panel shows the fast axis of linear retardance spatial variation with all points were within a range of $\pm$0.5$^\circ$. The bottom graphics show maps measured at 420 nm wavelength using a $\sim$10 nm bandpass narrow-band filter and spatial sampling of 3 mm. The bottom graphics had the Y axis rang multiplied by the wavelength ratio of 1.5.  Elliptical retardance magnitude spatial variation is 7$^\circ$ peak to peak in the lower left had plot. Fast axis variation is only slightly larger at 420 nm than 633 nm. }
\vspace{-6mm}
\end{center}
\end{figure}

Figure \ref{fig:mlo_map_visp_sar_632nm_and_420nm} shows the spatial maps of elliptical retardance magnitude as well as the linear retardance fast axis orientation in the top two graphics. The spatial maps show that low spatial frequency variation dominates the distribution. A spatial FFT analysis showed that over 90\% of the retardance spatial variation was contained in the lowest few spatial frequencies (spatial periods $>$ 10 mm).

Figure \ref{fig:mlo_map_visp_sar_632nm_and_420nm} shows the spatial maps of elliptical retardance magnitude as well as the linear retardance fast axis orientation in the bottom two graphics for this shorter wavelength. The spatial sampling was reduced to 3 mm using the same 3 mm diameter beam giving 1010 independent spatial measurements at 420 nm wavelength. The peak-to-peak variation of elliptical retardance magnitude is increased to 7$^\circ$ at 420 nm wavelength, from 4$^\circ$ at 633 nm wavelength. Fast axis orientation spatial variation is also increased at shorter wavelength.

\begin{wraptable}{l}{0.60\textwidth}
\vspace{-4mm}
\caption{Measured ViSP SAR Crystal Properties}
\label{table:ViSPsar}
\centering
\begin{tabular}{l | l l | l l | l l | l}
\hline \hline
N		& Des.	& Meas	& Ret		& Meas	& Des		& Meas			& Ori				\\
		& mm	& mm	& Bias		& Waves	& Pair		&$\delta\pm$0.01	& $\pm0.3^\circ$	\\
\hline \hline
2B$_s$	& 2.12	& 2.154	& 31$\pm$1	& 30.78	& 			& 				& 0$^\circ$		\\
2B$_b$	& 2.10	& 2.132	& 30$\pm$1	& 30.46	& 0.328		& 0.3277			& 90$^\circ$		\\
\hline
3A$_s$	& 2.13	& 2.187	& 31$\pm$1	& 30.78	& 			& 				& 70.25$^\circ$	\\
3A$_b$	& 2.10	& 2.153	& 30$\pm$1	& 30.26	& 0.476		& 0.4722			& 160.25$^\circ$	\\
\hline
4B$_s$	& 2.12	& 2.155	& 31$\pm$1	& 30.79	& 			& 				& 0$^\circ$		\\
4B$_b$	& 2.10	& 2.132	& 30$\pm$1	& 30.46	& 0.328		& 0.3251			& 90$^\circ$		\\
\hline \hline		
\end{tabular}
\vspace{-2mm}
\end{wraptable}

When the clear aperture is reduced to 70 mm as appropriate for narrower field of view instruments, the results are roughly three times better. Figure \ref{fig:visp_sar_uniformity_polish_CDF} shows that 80\% of the measurements fall within 0.3$^\circ$ retardance of the average and nearly all points are within 0.4$^\circ$ for a wavelength of 633 nm.  At 420 nm wavelength, the distribution shifts to less than 0.5$^\circ$ retardance error for 80\% spatial coverage with all points below 0.9$^\circ$.  Similar results are seen for fast axis orientation. This 70 mm clear aperture represents the footprint used for this SAR to calibrate a $<$3 arc-minute DKIST field of view. When DKIST instruments set their scanning / pointing settings to match the optical boresight, the spatial variation in linear retardance magnitude should be significantly smaller.

\begin{figure}[htbp]
\begin{center}
\vspace{-0mm}
\hbox{
\includegraphics[height=5.4cm, angle=0]{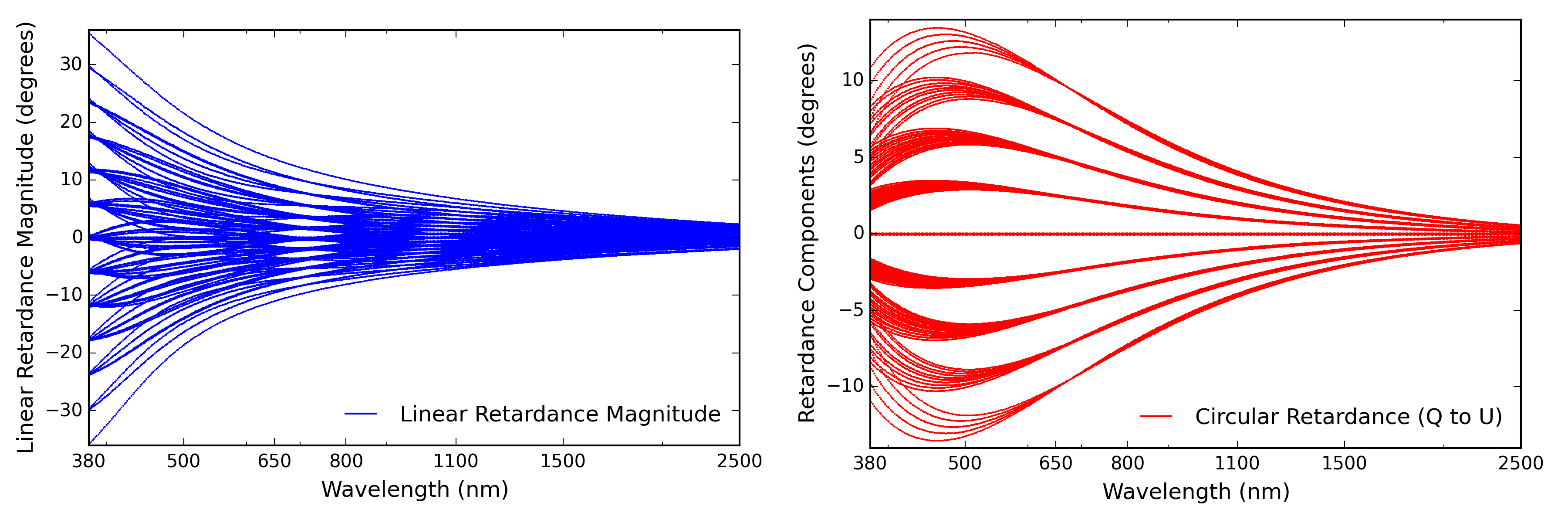}
}
\caption[] 
{\label{fig:polish_residual_ER_components_vispSAR} Left shows variation of the linear retardance magnitude computed as the RSS of the two linear retardance components for the 729 polishing error simulations on the ViSP SAR. Right shows the circular retardance.}
\vspace{-6mm}
\end{center}
\end{figure}

The models of polishing non-uniformity applied to each of the crystals results in significant circular retardance at shorter wavelengths. Per Table \ref{table:polish_error_measured_plates_as_built}, the three individual ViSP SAR compound retarder pairs B2-A3-B4 all had very similar polishing errors, unlike the DL-NIRSP components. The peak to peak retardance errors were 0.003 to 0.004 waves with RMS errors around 0.001 waves. When each crystal pair in the design has less than 0.4$^\circ$ RMS retardance variation, we can anticipate small magnitudes of spatial variation. The left graph of Figure \ref{fig:polish_residual_ER_components_vispSAR} shows the linear retardance magnitude difference from nominal design for the elliptical retardance model in our 729 trial polishing error simulations.

At short wavelengths the retardance magnitudes can change very substantially. Similar results are seen in the right hand graphic which shows the non-zero circular retardance introduced in the same models. We would expect circular retardance up to 10$^\circ$ magnitudes for some of the worst case simulations.   We know from the Meadowlark metrology that these magnitudes for polishing errors are reasonable.  Table \ref{table:ViSPsar} shows the metrology of the individual components making up the ViSP SAR optic.  The net retardance of each compound retarder B-A-B is within a few thousandths of a wave.

The predicted Mueller matrix shows spectrally variable sensitivity to physical thickness errors in the SiO$_2$ crystals through the changes in linear retardance magnitude, fast axis orientation and circular retardance. As these simulations are derived from applying a series of polishing errors to each plate, the final spatial distribution of errors represents a variable combination of all these potential outcomes. 

We note that these simulations used 0.005 waves net retardance variation per crystal pair, with the metrology suggesting the delivered parts were 20\% to 50\% better than this error.  The measured peak to peak elliptical magnitude errors of 7$^\circ$ at 420 nm wavelength and 4$^\circ$ at 633 nm wavelength are in line with the lower models of Figure \ref{fig:polish_residual_ER_components_vispSAR}. The linear retardance fast axis orientation variation is similarly quite small, consistent with the design sensitivity seen in Figure \ref{fig:polish_residual_ER_components_vispSAR}. As our simulation stacks up worst-case errors against all other worst case errors, these polishing simulations will be significantly worse than the combination of three plates with random spatial distrubutions, subject to the varying design sensitivities discussed in the previous section. 

\begin{figure}[htbp]
\begin{center}
\vspace{-0mm}
\hbox{
\includegraphics[height=6.1cm, angle=0]{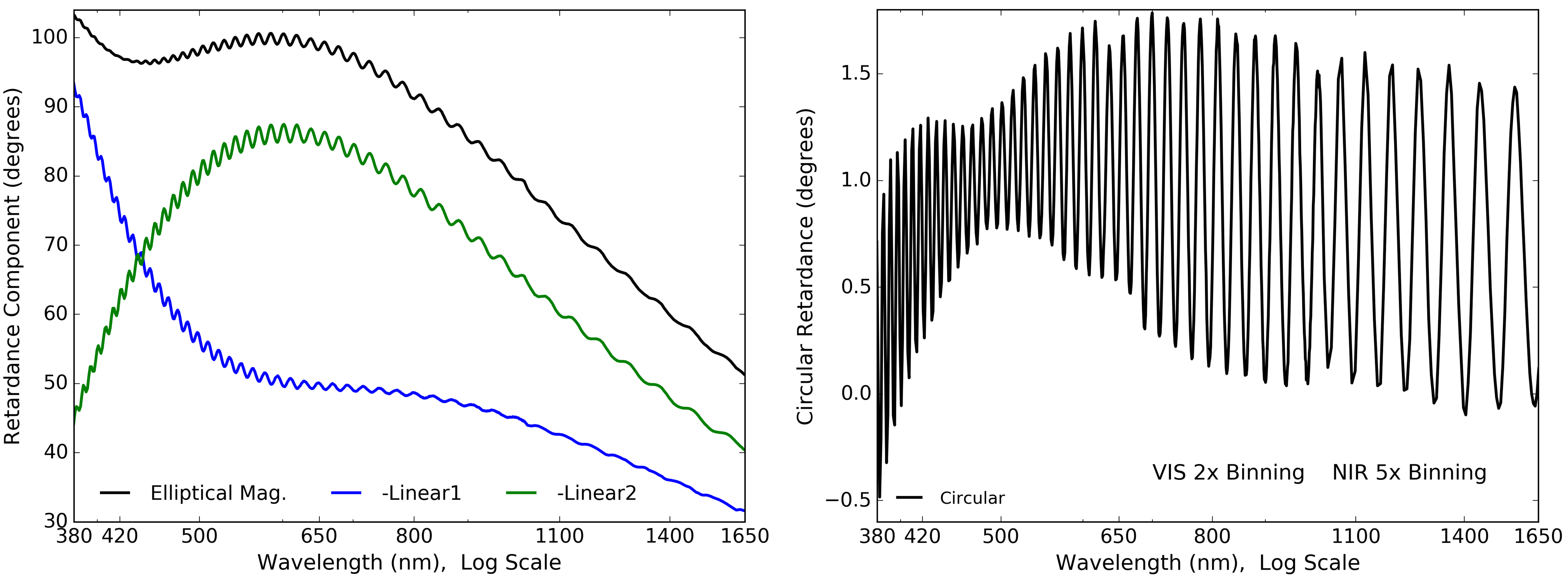}
}
\caption[] 
{\label{fig:vispSAR_nlsp_ER_Fit} The elliptical retardance fits to the NLSP Mueller matrix measurements of the ViSP SAR. The left plot shows the linear retardance components 1 and 2 in blue and green respectively. The sign of the first component has been reversed to keep the plots consistent. Black shows the elliptical retardance magnitude. The right graphic shows the circular retardance with magnitudes from -0.5$^\circ$ to +1.6$^\circ$.}
\vspace{-6mm}
\end{center}
\end{figure}

The ViSP calibration retarder has fairly small net circular retardance measured with our NLSP spectropolarimeter. Figure \ref{fig:vispSAR_nlsp_ER_Fit} shows elliptical retarder parameter fits to the center of the optic with a 4 mm beam footprint using our higher spectral resolving power configuration. We see that the retardance magnitude is very close to the design with retardance nominally around 95$^\circ$ to 102$^\circ$ over the entire ViSP instrument wavelength range.  The optic has over 55$^\circ$ retardance magnitude at the longest DL-NIRSP wavelength of 1565 nm, so this optic is capable of calibrating every DKIST instrument using the adaptive optics corrected beam:  ViSP, VTF and DL-NIRSP.  The optic does however have a few degrees of spectral oscillation from rotational misalignments between the crystals (clocking errors) and similar behavior in circular retardance.

\begin{wrapfigure}{l}{0.60\textwidth}
\centering
\vspace{-0mm}
\begin{tabular}{c} 
\hbox{
\hspace{-1.2em}
\includegraphics[height=7.1cm, angle=0]{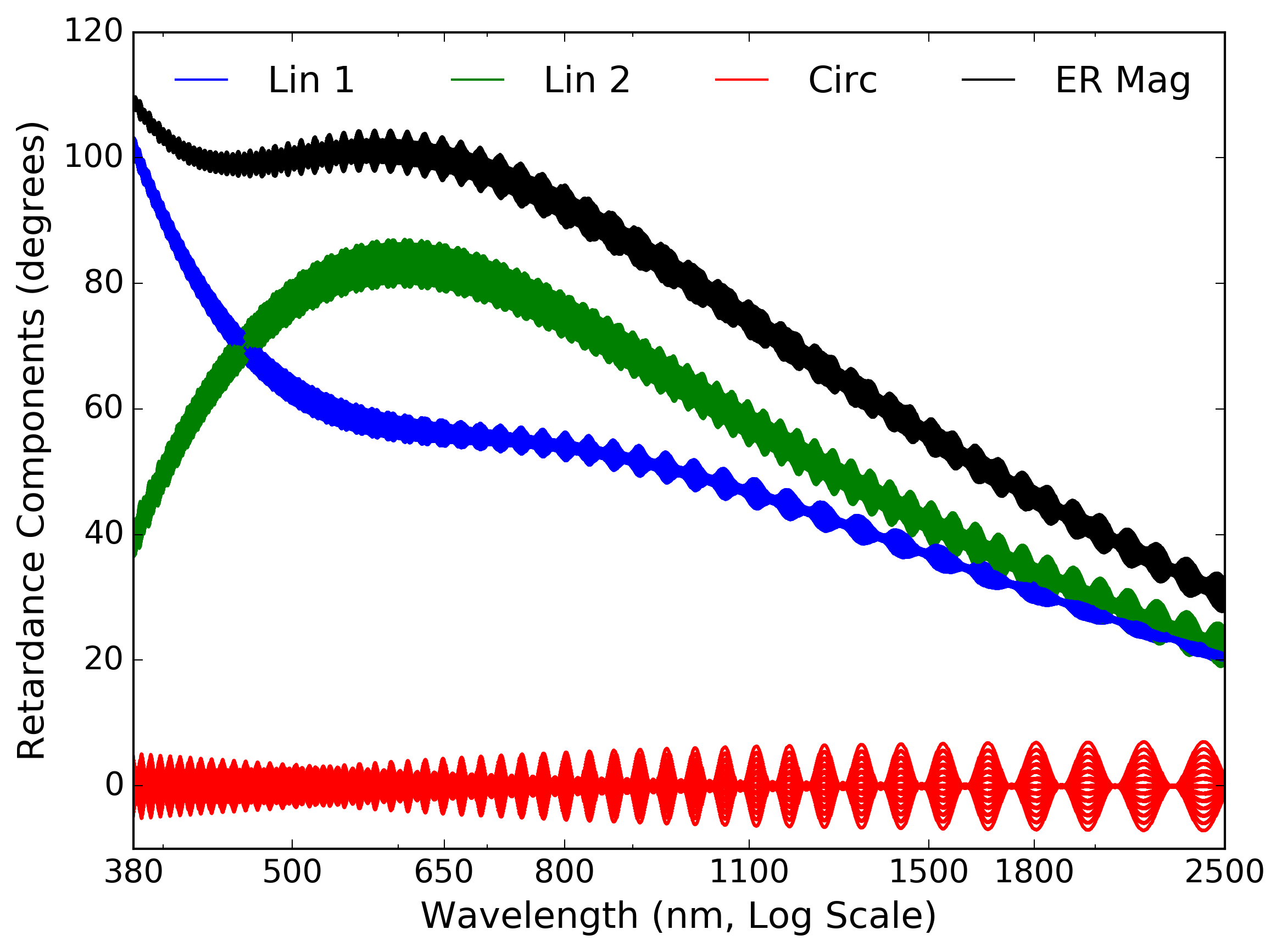}
}
\end{tabular}
\caption[] 
{\label{fig:visp_clock_tolerance_ER_fit}  The elliptical retarder model fits to clocking error simulations applied to the ViSP SAR design. Blue and green show the first and second components of linear retardance. Red shows circular retardance. Black shows the elliptical retardance magnitude.}
\vspace{-4mm}
 \end{wrapfigure}

We can compare the polishing errors of Figure \ref{fig:visp_polish_tolerance_ER_fit} to a clocking model simulation with 0.3$^\circ$ retardance error. Figure \ref{fig:visp_clock_tolerance_ER_fit} shows the 3$^6$ models where every crystal is mis-aligned with respect to every other crystal in a grid of models.  Blue shows the first component of linear retardance (a rotation about the Q axis rotating U into V).  Green shows the second components of linear retardance (a rotation about the U axis rotating Q into V). Red shows circular retardance (a rotation about the V axis rotating Q into U). Black shows the elliptical retardance magnitude as the root sum square magnitude of all three retardance components. For convenient plotting, we have reversed the signs of the linear retardance components, the same as in Figure \ref{fig:visp_polish_tolerance_ER_fit}. The clocking errors show spectrally fast oscillations at short wavelengths with significant slowing of the ripple period at near infrared wavelengths.

\clearpage
\section{Impact to DKIST Calibration: Field of View Variation}
\label{sec:impact_to_fov}

There is significant impact to the DKIST calibration process and ultimate performance from spatial variation of retardance. The calibration process in typical astronomical instruments makes the assumption that the calibration retarder can be fit with just one set of parameters for all field angles. Often times an additional variable is added to account for a temporal trend in retardance magnitude. High signal-to-noise ratios can be achieved by averaging data to lower spatial sampling. However, the longer a calibration sequence takes under thermal loading the more polarization fringes drift and the net retardance changes. With significant polishing error, the net retardance depends on which part of the optic the instruments are observing. Observational efficiency is also lost.

\begin{wrapfigure}{r}{0.62\textwidth}
\centering
\vspace{-6mm}
\begin{tabular}{c} 
\hbox{
\hspace{-1.1em}
\includegraphics[height=8.0cm, angle=0]{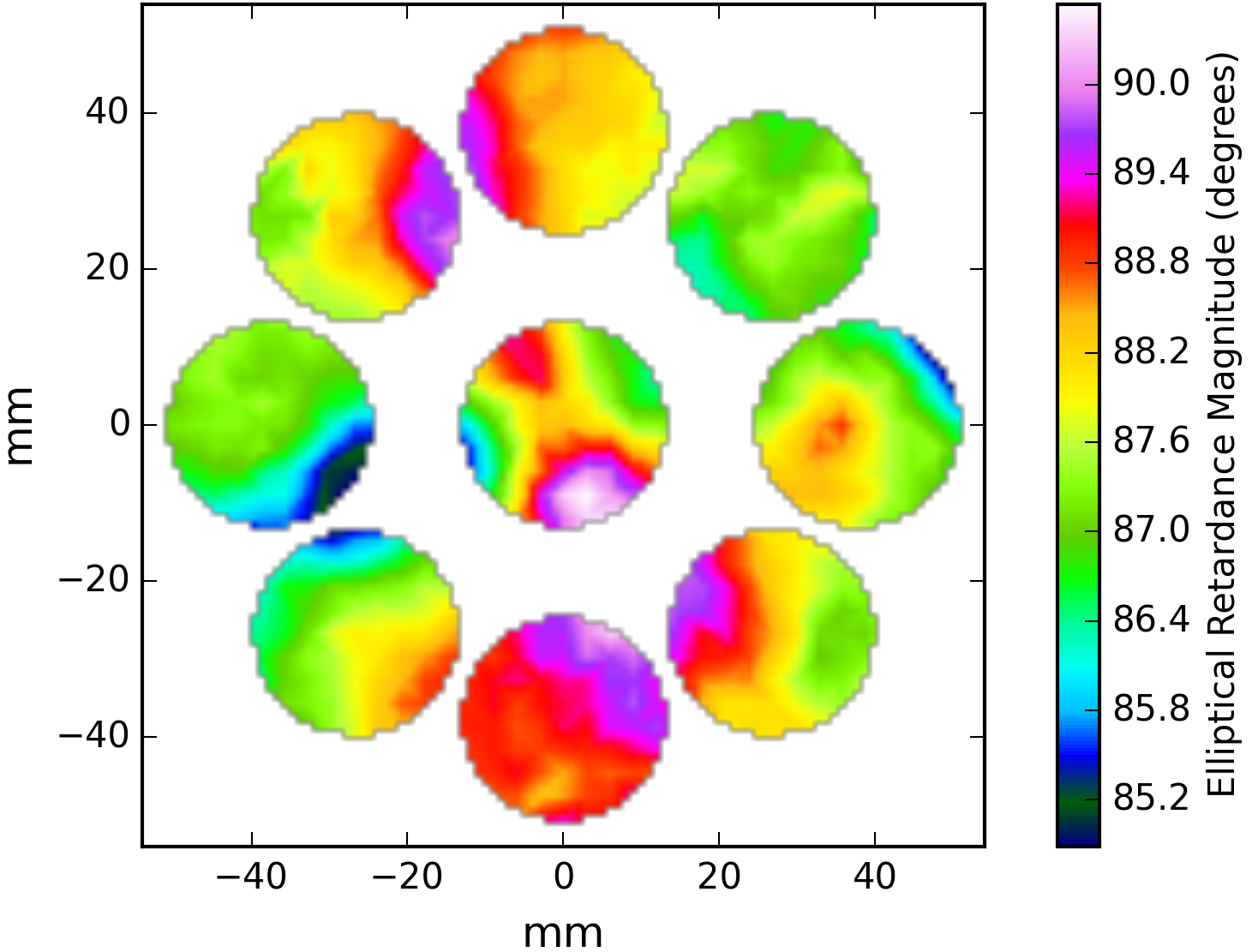}
}
\end{tabular}
\caption[] 
{ \label{fig:footprint_ViSPsar_AsBuilt_420nm} 
The retardance variation within footprints of the beam on the retarder during an eight-orientation calibration sequence at 420 nm wavelength. The beam at field center (field angle = 0) sees a constant footprint but with changing orientation. The edge of the field of view (field angle = 2.5') sees eight independent illuminations of the retarder spatial variation.}
\vspace{-5mm}
\end{wrapfigure}

In future calibrations, field dependent optical system Mueller matrices may require independent fits. The problem for this field dependent calibration process is that modulation states and system polarization models need to account for spatial variation of mirror induced polarization as well as spatially dependent modulation across the modulating retarder.  There are degeneracies when attempting to add spatial variability free parameters to a calibration sequence when also fitting for time dependence (from thermal changes).  

Figure \ref{fig:footprint_ViSPsar_AsBuilt_420nm} shows the spatial variation of retardance for the eight footprints on the ViSP calibration retarder at 420 nm wavelength as this optic rotates during a typical calibration sequence. The center footprint remains constant for the on-axis beam and would see the same average retardance for all retarder orientations. For the footprint corresponding to the 5 arc-minute field edge, clear spatial separation is seen between footprints along with significant retardance variation between steps. 

\begin{wrapfigure}{l}{0.61\textwidth}
\centering
\vspace{-2mm}
\begin{tabular}{c} 
\hbox{
\hspace{-1.1em}
\includegraphics[height=6.7cm, angle=0]{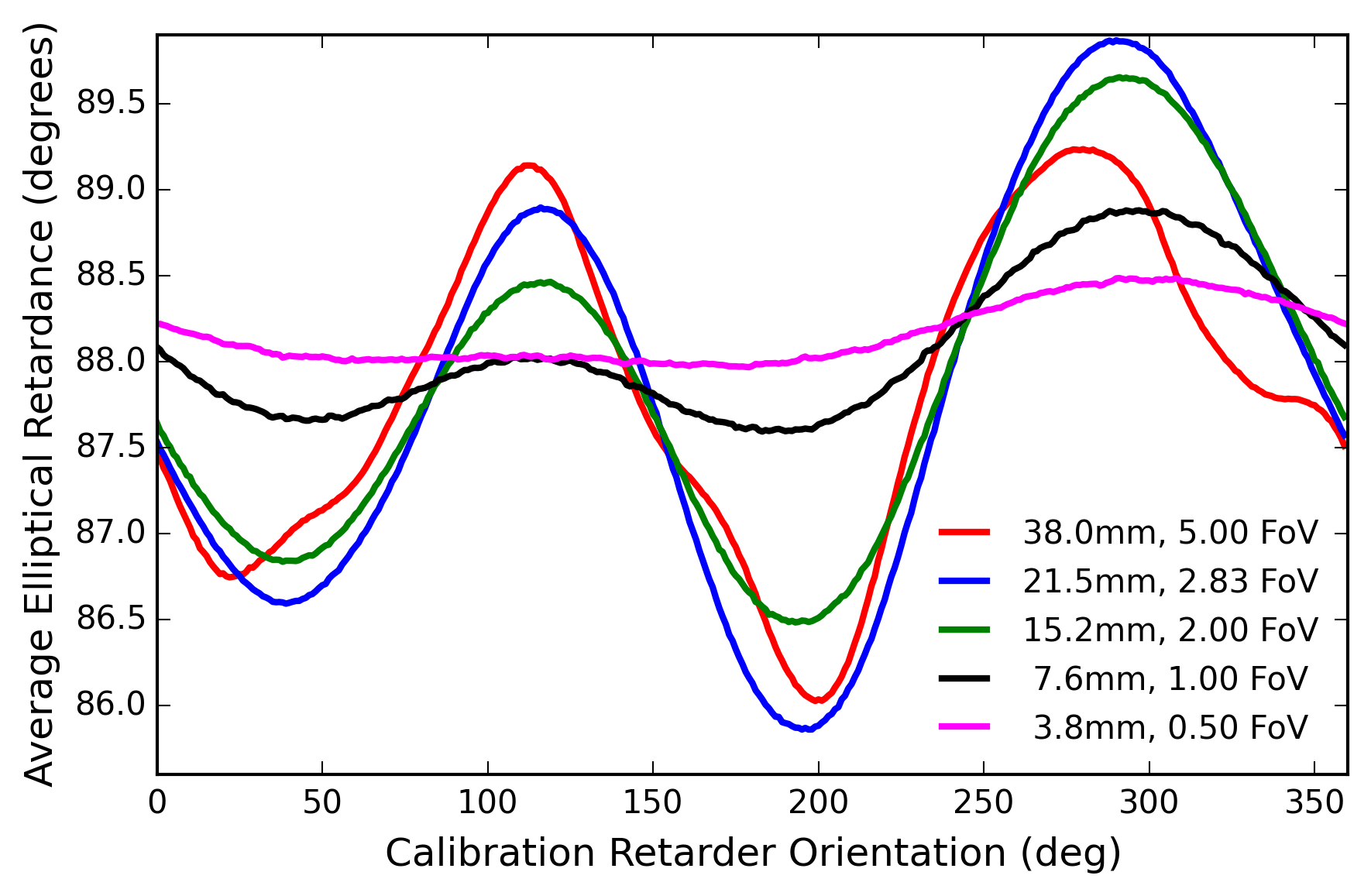}
}
\end{tabular}
\caption[] 
{ \label{fig:spatial_retardance_variation_ViSPsar_AsBuilt_420nm} 
The retardance variation computed as an average over the footprint as the ViSP calibration retarder spins during a calibration sequence at 420 nm wavelength. Different colors show the beam de-centered to match the edge of the different field angles.}
\vspace{-2mm}
\end{wrapfigure}

We can quantify the retardance variability during calibration by taking spatial averages over the footprint for varying FoV.  Figure \ref{fig:spatial_retardance_variation_ViSPsar_AsBuilt_420nm} shows an example of the average retardance across a footprint as the calibration retarder rotates. The various curves correspond to select field angles using the ViSP calibration retarder at 420 nm wavelength.  Each individual footprint is 26.6 mm at this optical location in the converging F/ 13 beam. Magenta shows a very small field of view at 0.5 arc minutes diameter with a beam decenter of 3.8 mm and 0.5$^\circ$ peak to peak spatial variation.  Black shows a 1 arc minute field of view corresponding to a beam center 7.6 mm away from the retarder center and 1$^\circ$ peak to peak spatial variation.  Green shows a 2.0 arc minute field of view with 15.2 mm of decenter.  Blue shows the edge of the 2.83 arc minute field of view and a 21.5 mm beam decenter.   Red corresponds to the maximal DKIST field of 5.0 arc minutes and a 38.0 mm beam decenter. As wavelength increases, the magnitude of polishing spatial variation and retardance non-uniformity decreases.  

At 633 nm wavelength, the maximal retardance spatial non-uniformity is 2$^\circ$ at 5.0 arc minute field of view corresponding to the edge of the clear aperture. This peak-to-peak variation is about half the magnitude as at 420 nm wavelength though the wavelength is only 150\% longer. Within a smaller 0.5 arc minute diameter field, the retardance variation is 0.2$^\circ$ rand 0.6$^\circ$ for a 1.0 arc minute field. We note that for these two particular wavelengths, the ViSP SAR maintained similar spatial patterns to the retardance non-uniformity.  

There is another small but important distinction about using a calibration retarder in a converging beam near a focal plane.  Several aperture-averaging effects lead to depolarization and a change in net retardance behavior. Similar to estimates from Sueoka \cite{Sueoka:2016vo,Sueoka:2014cm} for the DKIST retarder in a converging F/ 13 beam \cite{Sueoka:2016vo}, by averaging over a bundle of rays with spatially varying retardance properties, we encounter a change in elliptical retardance as well as depolarization. \cite{Chipman:2006iu,Anonymous:2004wl,2004OExpr..12.4941D,Chipman:2007ey,Anonymous:s4Rv1_wo,2005ApOpt..44.2490C,Chipman:2010tn,Chipman:2014ta} We showed the effect of averaging over the M1 and M2 beam introduces some small 0.1\% to 0.2\% diagonal depolarization \cite{Harrington:2017dj}. The DKIST specification for the calibration polarizer requires a contrast greater than 1000 to produce pure input Stokes vectors with less than 0.1\% depolarization.  However, similar to Sueoka \cite{Sueoka:2016vo,Sueoka:2014cm}, we find the retarder in the converging beam is a depolarizer of 5 to 10 times larger than this specification.

\begin{wrapfigure}{r}{0.61\textwidth}
\centering
\vspace{-5mm}
\begin{tabular}{c} 
\hbox{
\hspace{-1.1em}
\includegraphics[height=6.7cm, angle=0]{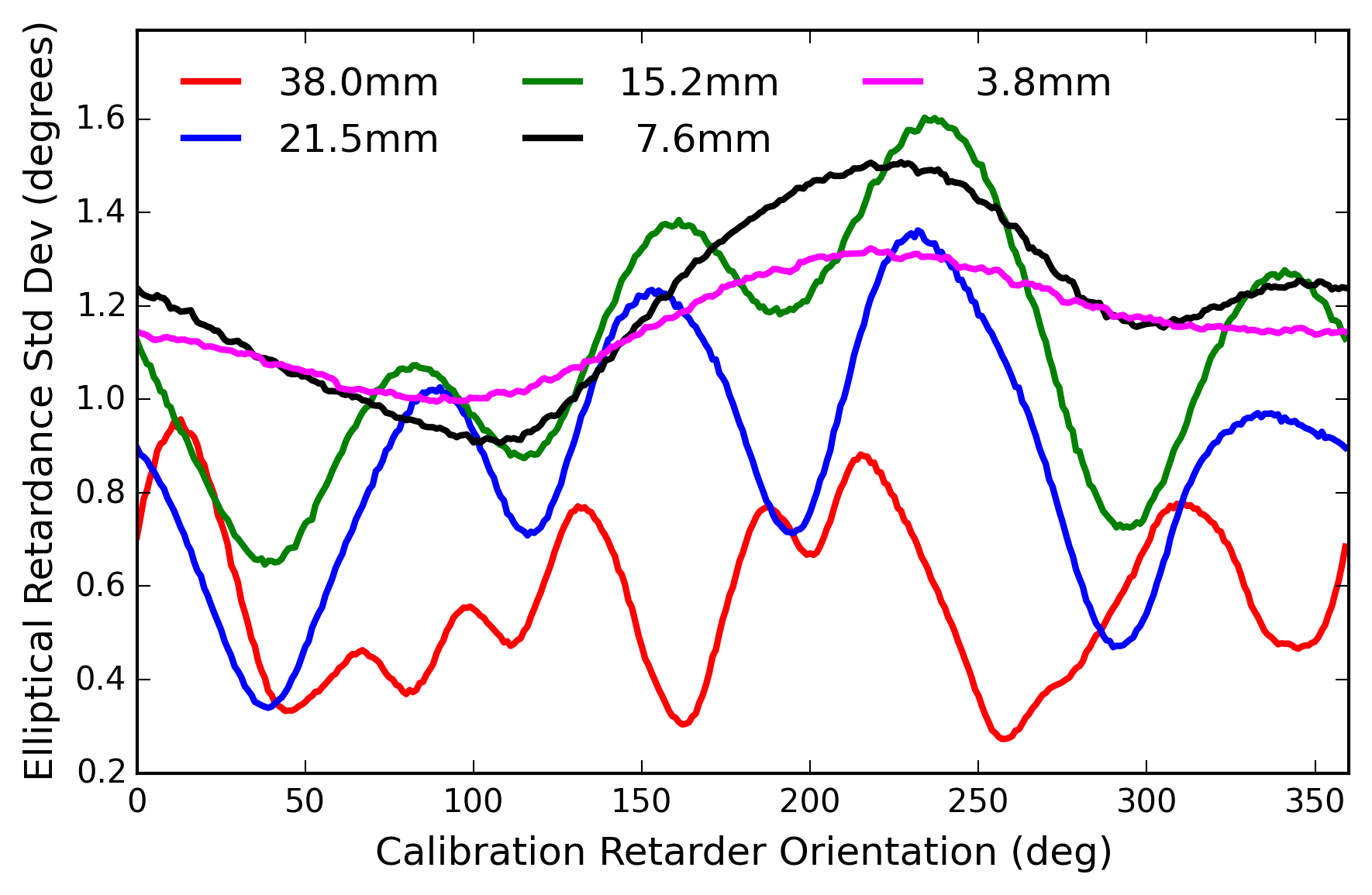}
}
\end{tabular}
\caption[] 
{ \label{fig:spatial_retardance_footprint_stddevv_ViSPsar_AsBuilt_420nm} 
The standard deviation of retardance spatial variation causing depolarization over a footprint as the ViSP calibration retarder spins during a calibration sequence at 420 nm wavelength.}
\vspace{-6mm}
\end{wrapfigure}

In Figure \ref{fig:spatial_retardance_footprint_stddevv_ViSPsar_AsBuilt_420nm} we compute the standard deviation of retardance across a footprint from a specific field angle at over a range of field angles and beam de-centration.  Even though the on-axis beam sees no change in footprint location as the optic rotates, the 26.6 mm diameter beam will see (constant) spatial variation.  As the field angle increases and the beam de-centers, the spatial variation changes and varies more with retarder orientation.

Figure \ref{fig:spatial_retardance_footprint_stddevv_ViSPsar_AsBuilt_420nm} also shows that we do not necessarily see the greatest spatial variation in individual footprints for the greatest field angles. The black and magenta curves correspond to the nearly stationary beam footprints.  However, the spatial variation is a nearly constant 1.2$^\circ$ to 1.5$^\circ$ retardance. For the largest field angle with the beam de-centered by 38 mm, the spatial variation oscillates between 0.3$^\circ$ and 0.9$^\circ$, significantly less than the on-axis beam.  For the as-built ViSP SAR at 420 nm wavelength, the outer field angles happen to see significantly less spatial variation than the on-axis footprint. 

Most calibration applications can ignore the spectral dependence of spatial variation over a narrow bandpass as these retardance maps are effectively monochromatic.  Polishing error will also introduce spectral variation to all retardance parameters similar to the above simulations.

\clearpage
\section{Impact to Instruments: Field Dependent Demodulation}
\label{sec:field_dependend_demod}

There is also significant impact to the DKIST modulation process. Depending on the instrument, different methods for determining modulation matrices use different styles of spatial and spectral fitting. High signal-to-noise ratios can be achieved by averaging data to lower spatial sampling.

\begin{wrapfigure}{r}{0.61\textwidth}
\centering
\vspace{-4mm}
\begin{tabular}{c} 
\hbox{
\hspace{-1.1em}
\includegraphics[height=8.0cm, angle=0]{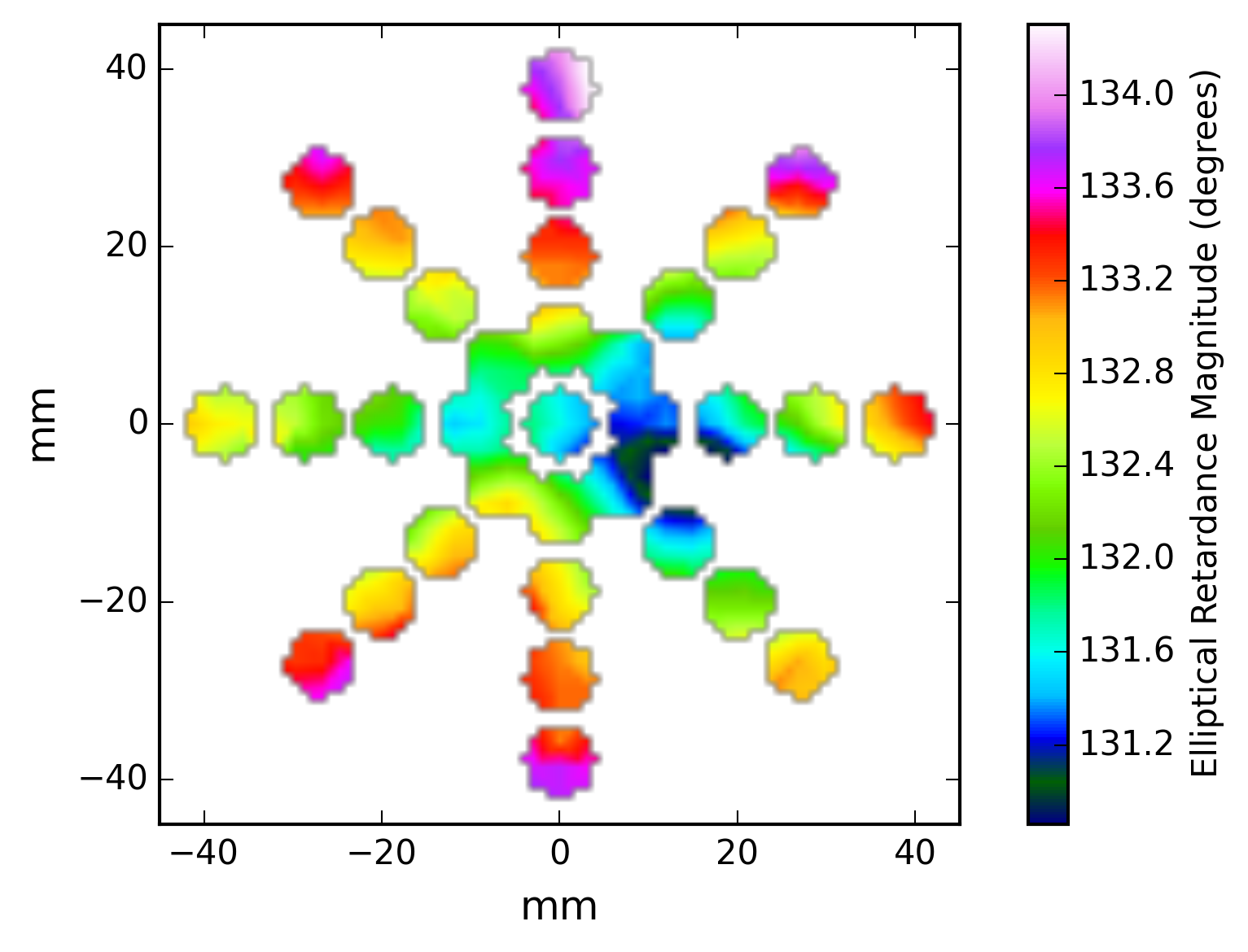}
}
\end{tabular}
\caption[] 
{ \label{fig:footprint_ViSP_PCM_420nm_field} 
The measured spatial retardance variation within footprints of the beam on the ViSP PCM during an eight-orientation modulation sequence at 420 nm wavelength. }
\vspace{-6mm}
\end{wrapfigure}

For a retarder placed closer to a focal plane, the footprints are smaller which increases the rotational variability while decreasing variance within a footprint.  For the ViSP instrument on DKIST, the modulator is in an F/ 32 beam and has 8.1 mm diameter footprints. This spectrograph nominally has a 2.0 arc minute tall entrance slit. The beam through illuminates a rectangle of roughly 80 mm by 8.1 mm on the modulating retarder after 218.5 mm of propagation at F/ 32.  

Figure \ref{fig:footprint_ViSP_PCM_420nm_field} shows the footprints and corresponding spatial pattern for modulation with the ViSP.  In this example, the 8.1 mm footprints are very small relative to field angle beam de-centeres.  We choose to display field of view footprints for 2.0, 1.5, 1.0, 0.5 and 0.25 arc minutes. Beam decentering ranges from 4.8 mm to 38.3 mm computed using the 131.8 meter effective focal length of the entire system.

\begin{figure}[htbp]
\begin{center}
\vspace{-1mm}
\hbox{
\includegraphics[height=5.5cm, angle=0]{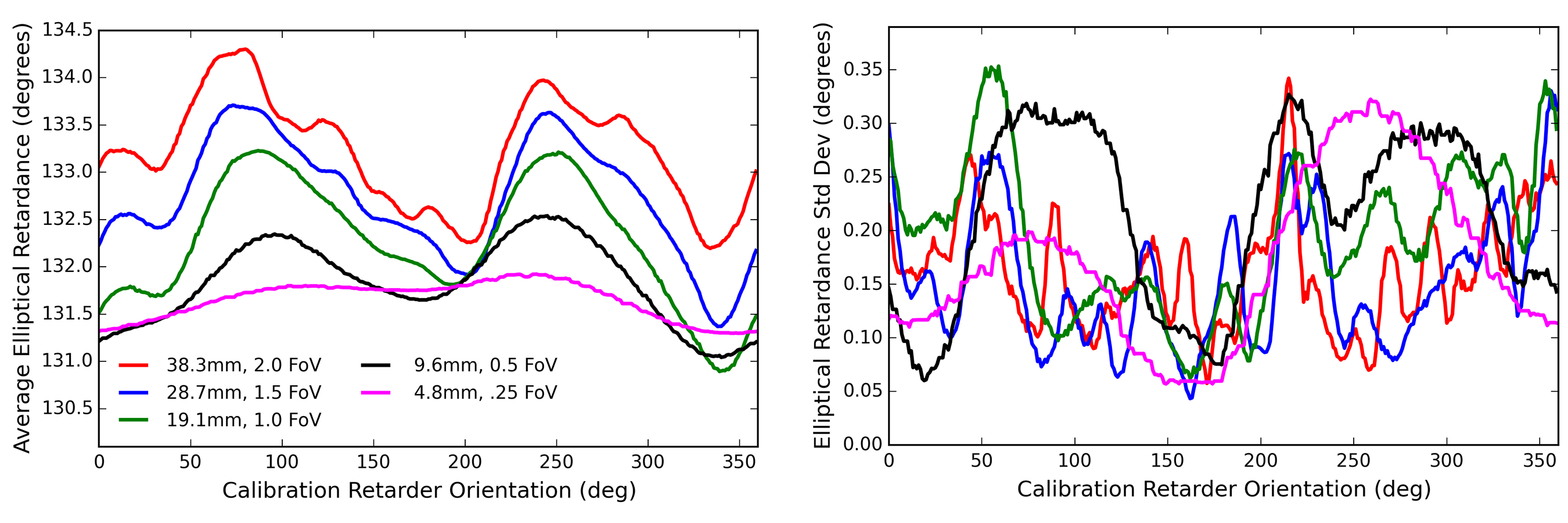}
}
\caption[] 
{\label{fig:footprint_ViSP_PCM_420nm_field_average_and_variance} ViSP modulator average retardance in a footprint at left and the standard deviation of the spatial variation across the footprint at right.}
\vspace{-7mm}
\end{center}
\end{figure}

The small fooprint also changes the relative amplitude of the variation with rotation.  Figure \ref{fig:footprint_ViSP_PCM_420nm_field_average_and_variance} shows the average retardance over a single footprint in the left plot.   The right plot of Figure \ref{fig:footprint_ViSP_PCM_420nm_field_average_and_variance} shows the standard deviation of the spatial variation across the footprint as the optic rotates.

As shown in the Appendix with our FFT analysis and data smoothing analysis of Figure \ref{fig:cryo_as_built_599nm_spatial_variation_difference}, the bulk of the spatial variation is at spatial scales greater than 8 mm. Thus the standard deviation of retardance variation within any individual footprint is now less than 0.9$^\circ$ peak with typical values in the range of 0.1$^\circ$ to 0.5$^\circ$.  As the footprint is only 8 mm, the average over the footprint does not reduce the variability much with rotation of the optic.  The peak-to-peak retardance variation is over 5$^\circ$.  We show in the next section how these spatial variations are small compared to the rotation of the fast axis caused by continuous rotation of the optic, and the main consequence is only a slight reduction of modulation efficiency.

\subsection{Comparison of Uniformity Impacts With Continuously Rotating Modulators}
\label{sec:compare_continuous_rotation}

When deciding on polishing retardance uniformity specifications, the discrete and continuously rotating retarder cases have very different sensitivities and calibration strategies. We need to include the depolarization caused by the rotation within a single exposure to correctly compare depolarization and field dependent demodulation techniques as they impact the optical calibration. The continuously changing modulation matrix must be sampled within the exposure time. The flux detected on the camera represents an integral of the modulated intensity propagated through the analyzer to the sensor through the exposure time. Depending on the accuracy required, the integral of flux on the sensor while the retarder spins represents an impact on efficiency and computational complexity. We compute the equation for integration of flux by integrating the transmission term of the system Mueller matrix. This method would represent how flux propagating through the linearly polarizing analyzer is recorded on an ideal sensor.

\begin{wrapfigure}{r}{0.45\textwidth}
\vspace{-6mm}
\begin{equation}
{\bf X}_{d} = {\bf D} {\bf O}  = {\bf D}  \left[ \Omega {\bf M}_{a} {\bf M}_{LR}(\theta,\phi) \right] 
\end{equation}
\begin{equation}
{\bf X}_{c} = {\bf D} {\bf O}  = {\bf D}  \int_{\theta_0} ^ {\theta_i} \Omega {\bf M}_{a} {\bf M}_{LR}(\theta,\phi) d\theta
\label{eqn:Flux_Integral_Rotate_Retarder}
\end{equation}
\vspace{-6mm}
\end{wrapfigure}

The polarization response matrix ({\bf $X_{ij}$}) is nominally the identity matrix for theoretically demodulated data where the exact input modulation matrix is known and used to demodulate without error.  Input vectors are transferred to output vectors exactly without change in magnitude or direction. We adopt a notation to match del Toro Iniesta and other works \cite{delToroIniesta:2000cg}. Computing the response as demodulation ({\bf $D_{ij}$}) multiplied by modulation ({\bf $O_{ij}$}) allows us to assess perturbations. The matrix $\Omega$ is [1,0,0,0] and represents the sensor only recording total flux without sensitivity to transmitted polarization state when propagated through an ideal analyzer (polarizer).

\begin{wrapfigure}{l}{0.62\textwidth}
\centering
\vspace{-3mm}
\begin{tabular}{c} 
\hbox{
\hspace{-1.0em}
\includegraphics[height=7.5cm, angle=0]{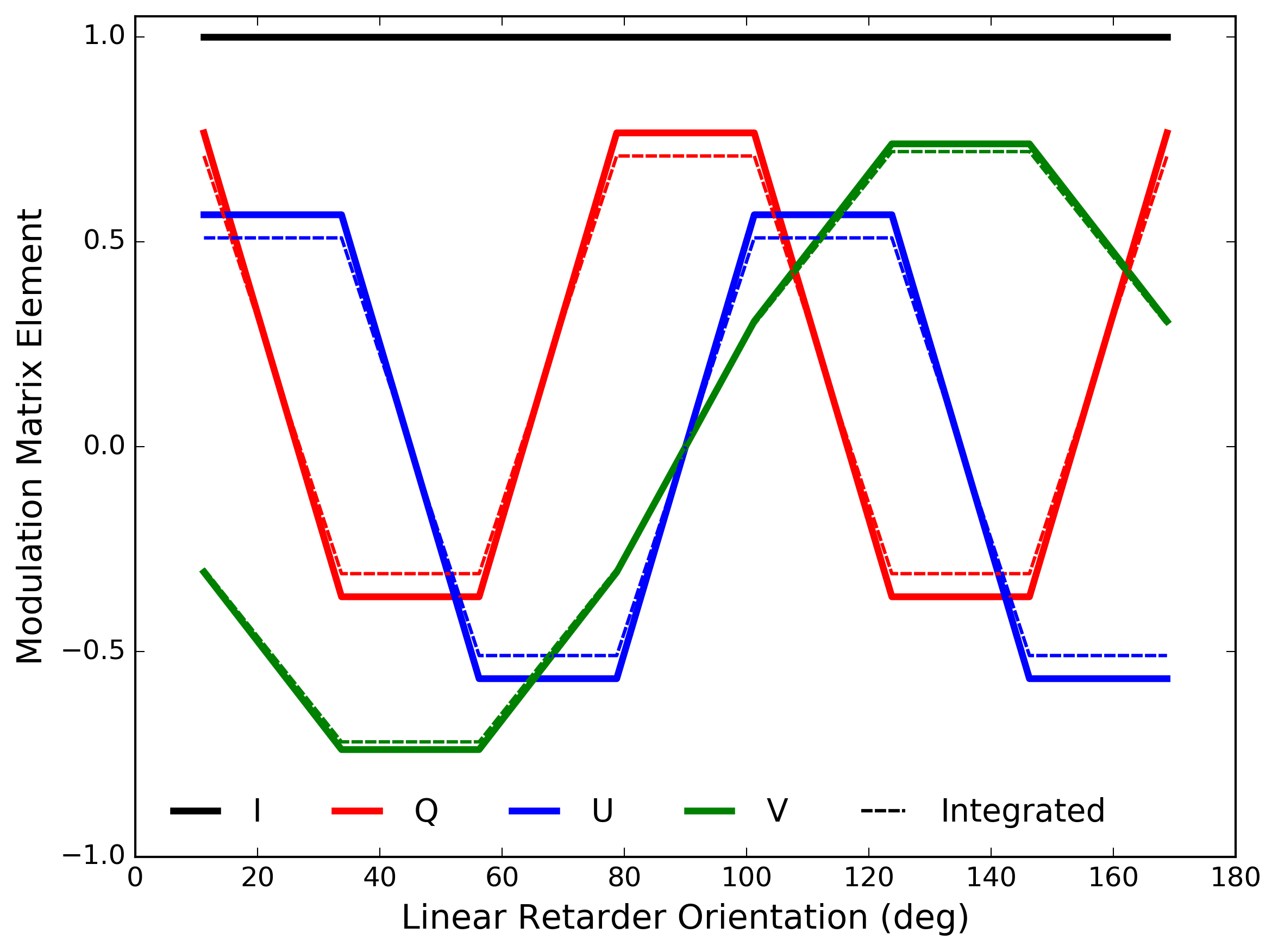}
}
\end{tabular}
\caption[] 
{ \label{fig:modulation_matrix_8state} The modulation matrix elements for the 8-state modulation cycle with integration beginning with the linear retarder fast axis at 0$^\circ$ and temporal sampling centered about the middle of each modulation state (e.g. 11.5$^\circ$ for the first state). Red shows $Q$.  Blue shows $U$. Green shows $V$. The solid black line at 1.0 shows $I$.  Solid lines show the instantaneous sampling.  Dashed lines show the impact of flux averaging within a subframe.}
\vspace{-2mm}
 \end{wrapfigure}

We can apply perturbations to simulate either imperfect knowledge or errors in determining the demodulation matrix.  Alternatively we can assess errors in modulation within individual or co-added modulation cycles with various types of instrumental and statistical errors. In Equation \ref{eqn:Flux_Integral_Rotate_Retarder} we show the discrete modulation response matrix as {\bf X}$_{d}$ and the continuous modulation case as {\bf X}$_{d}$ with an integral from the starting and ending orientations of the modulator.  These orientations are represented as $\theta_0$ and $\theta_i$.  In the most ideal case, the Mueller matrix of the modulator can be assumed to be of an ideal linear retarder ($M_{LR}$) and the Mueller matrix of the polarizer is a perfect analyzer ($MM_a$).

We can approximate the impact of continuous rotation as the flux integral from $\theta_0$ to $\theta_i$ through each subframe. If we use a standard 8-state modulation, the retarder is sampled in 22.5$^\circ$ steps through a 180$^\circ$ rotation. We propagate flux through the retarder and analyzer and then average flux as seen by the sensor. If we break the sub-frame into 50 individual samples, we can approximate the flux as the sum of the system transmission in 0.45$^\circ$ steps.  Figure \ref{fig:modulation_matrix_8state} shows the modulation matrix elements for an 8-state modulation cycle when using a 127$^\circ$ linear retarder and an angular offset of 11.25$^\circ$ to put half the rotation within a single modulation state centered on the 0$^\circ$ fast axis orientation. Solid lines show the modulation matrix elements when assuming discrete sampling and no depolarization. Dashed lines show the impact of flux averaging within a subframe as the optic rotates. The net effect is a slight depolarization and a reduction of efficiency. We note that this reduction of the modulation matrix magnitudes slightly decreases the efficiency and also occurs identically for all field angles.

We also note that this kind of simulation is also useful for establishing specifications on timing and synchronization. As a reference, the DKIST cameras should be able to synchronize the start and end of exposures with errors less than 10 micro seconds. Additionally, the rotation stage has following errors that can introduce rotational misalignments between the expected position of the retarder and the actual position. For the DKIST rotation stages we typically maintain the retarder location within a few arc seconds of the expected orientation throughout the modulation cycle.  Both the camera timing jitter and the rotational misalignments due to rotation stage imperfections are tiny compared to other polarization artifacts, introducing errors in the response matrix of order 10$^{-5}$ or less over typical modulation cycles at DKIST frame rates of 80 Hz or less.

\subsection{Errors in Modulation from Cal. Retarder Variation Across the Field of View}
\label{sec:sim_fov_errors_demod}

We next make an estimate of the impact of spatially variable retardance in the calibration optic to the modulation matrix errors. We include these errors by modeling a simple procedure for calibrating an instrument.  We create a model for the calibration optic including spatial variation of retardance magnitude and fast axis orientation variation. We then propagate this retardance variation through a specific sequence of calibration optic orientations.  An ideal instrument then modulates and analyzes this beam to record the perturbed intensities as a function of field.  We then take these modeled intensities and assess the impact to DKIST by modeling a simple calibration fitting procedure. We first average the detected flux over some specified field angle across the slit of this simulated instrument.  We then do a simultaneous fit for the three elliptical calibration retarder parameters as well as a field-averaged modulation matrix.  This is a simple but demonstrative example of instrument calibration, and of the application of this metrology to simulating future instrument performance.


\begin{wrapfigure}{l}{0.35\textwidth}
\vspace{-6mm}
\begin{equation}
{\bf M}_{Stokes} =
 \left ( \begin{array}{rrrr}
I_1	&Q_1	&U_1	& V_1	\\
I_2	&Q_2	&U_2	& V_2	\\
....	& ...		& ...		& ...		\\
I_n	&Q_n	&U_n	& V_n	\\
 \end{array} \right ) 
\label{eqn:input_stokes_vectors}
\end{equation}
\vspace{-8mm}
\end{wrapfigure}

As a first step, we must choose a specific sequence of calibration optic orientations to generate a diverse set of Stokes vectors.  For operational efficiency, DKIST must calibrate as many instruments as possible in a simultaneous configuration. There is a similar analogy between modulation efficiency and calibration efficiency.  In Selbing 2005 on the polarization calibration of the Swedish Solar Telescope, some simple optimization procedures are summarized \cite{Selbing:2005wc}.


\begin{wraptable}{r}{0.14\textwidth}
\vspace{-5mm}
\caption{Seq.}
\label{table:optimized_sequence}
\centering
\begin{tabular}{r r }
\hline \hline
Pol	& Ret		\\
\hline \hline	
0	& Out	 		 \\
45	& Out	 		\\
90	& Out	 		\\
135	& Out	 		\\
135	& 0				\\
135	& 45		 		\\
135	& 90		 		\\
0	& 180	 		\\
45	& 135	 		\\
180	& 180	 		\\
180	& 135	 		\\
180	& 45		 		\\
\hline
{\bf 128}	& {\bf 180} 	\\
{\bf 98}	& {\bf 128}	 	\\
\hline \hline		
\end{tabular}
\vspace{-4mm}
\end{wraptable}

In section 2.5.2 of Selbing \cite{Selbing:2005wc}, the orthogonality of the Stokes vectors created by the calibration unit is assessed in matrix form with one row per input state. Equation \ref{eqn:input_stokes_vectors} shows the Stokes vector matrix (${\bf M}_{Stokes}$) used to derive the condition number and the relative calibration efficiency of the exposure sequence. The pseudo-inverse of ${\bf M}_{Stokes}$ is created as $E_{i,j} $ = ($ {\bf M}^T_{Stokes}$${\bf M}_{Stokes}$)$^{-1}$${\bf M}^T_{Stokes}$. The efficiencies are computed from the pseudo-inverse as the usual sum of squared elements $e  = (n \sum_{1}^{n} E^2)^{-0.5}$ where n is the number of input states. This pseudo-inverse $E$ can also assessed by it's condition number. This is the same approach as for finding optimum demodulation matrices \cite{delToroIniesta:2000cg, Tomczyk:2010wta, 2010SPIE.7735E..4AD, Snik:2012jw}

The condition number of a matrix and the associated linear equation Ax = b gives an estimate on how errors propagate from individual values (photometric measurements) to the solution (fitted modulation matrix elements). This is commonly defined as the ratio of the largest to smallest singular value in the singular value decomposition of a matrix. The condition number of a function shows how much the output of the function can change with a change in the input arguments.  A matrix with a low condition number is said to be well-conditioned, while a problem with a high condition number is said to be ill-conditioned and ill-conditioned fits have very different noise properties in the resulting fit values. For DKIST calibration, we want to ensure the signal to noise ratios of the calibrated Stokes vectors are equal within a factor of a few, and roughly proportional to the SNR of the instrument photometric measurements, hence needing a condition number below a few. 

We make a simple example by optimizing an existing calibration sequence using this framework and the as-built ViSP SAR optical properties.  Optimizing an entire sequence of input calibration states would require significant development and computation time, and is beyond the scope of this paper.  We simply adapt a nominal 12-state sequence adapted from the Advanced Stokes Polarimeter instrument \cite{1992SPIE.1746...22E,Skumanich:1997eh}. We first consider this sequence noting that we can use 12 nominal orientations that would each create four individual $\pm$QUV inputs when using a quarter-wave linear retarder as is common in traditional calibration. We show in Table \ref{table:optimized_sequence} the orientations of the calibration polarizer (CP) and calibration retarder (CR) in the first two columns.  We note that in practice, no optic is perfectly achromatic or aligned and we do not need to assume that perfectly pure individual states are created for calibration.

\begin{figure}[htbp]
\begin{center}
\vspace{-1mm}
\hbox{
\includegraphics[height=6.2cm, angle=0]{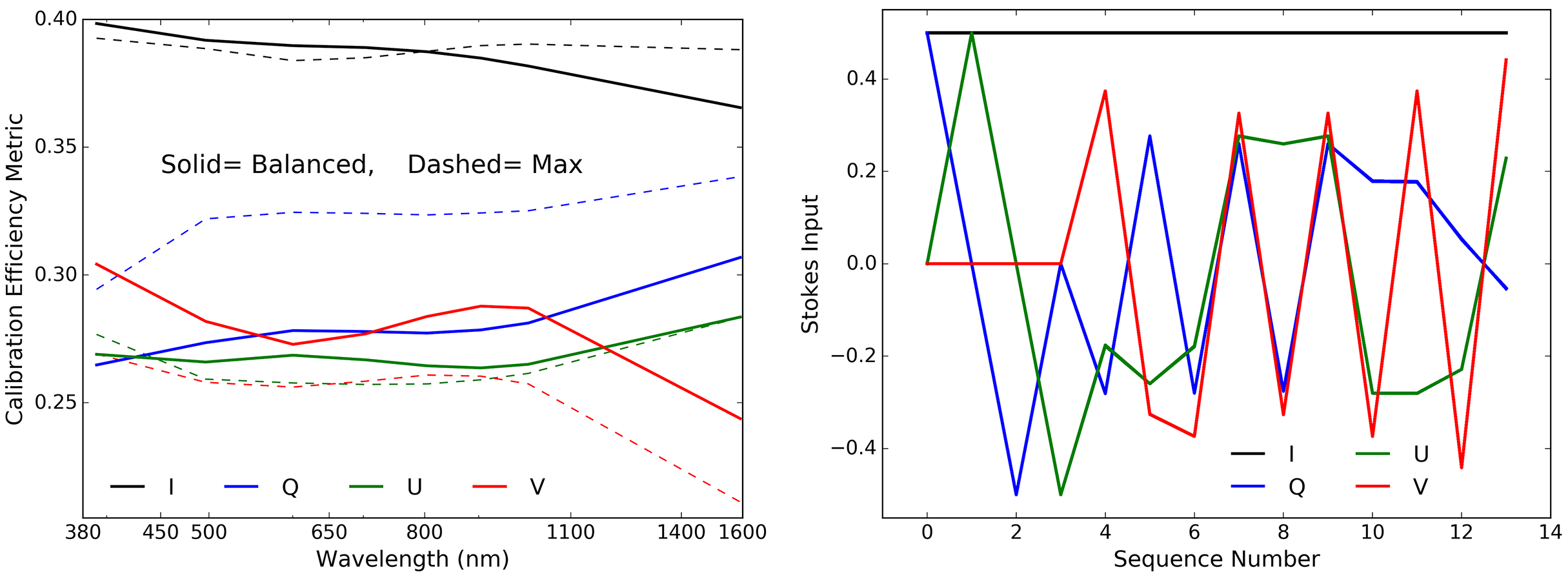}
}
\caption[] 
{\label{fig:vispSAR_cal_efficiency_14state} Fitting for optimized QUV calibration efficiency across all DKST wavelengths 380 nm to 1600 nm. The left hand graphic shows the sequence efficiency for best and worst choice of optic orientations for the two final exposures. The solid lines show the efficiency is maximized when the last two exposures are chosen as in Table \ref{table:optimized_sequence}. These two exposures have significant impact as the minimum efficiency shown in dashed lines is up to 40\% lower when a poor choice of optic orientations is made.  The right hand graphic shows the Stokes vectors created at 420 nm wavelength as the optics are oriented in sequence. Note that the default 12 exposures never created a state near -V.  The optimization added two more states with large negative V magnitudes providing more diverse inputs. }
\vspace{-5mm}
\end{center}
\end{figure}

We first search the calibration efficiency space using the as-built ViSP SAR optics and find good efficiency across all DKIST wavelengths 380 nm to 1600 nm when the retarder is used with a starting orientation of 129$^\circ$ for the fast axis using the NLSP as-built data. We then consider adding two additional exposures as free variables where the CP and the CR can be at any orientation between 0$^\circ$ and 180$^\circ$. We searched this space with 7.5$^\circ$ steps so we have 25 possible orientations for each optic in each step.  As we're searching two optic orientations for two exposures, we search 25$^4$ models (390,625) using brute force. The last two bold entries show the optimized sequence in Table \ref{table:optimized_sequence}.  The left hand graphic of Figure \ref{fig:vispSAR_cal_efficiency_14state} shows this maximized QUV efficiency as solid lines.  The dashed lines show the minimum QUV efficiency to demonstrate sensitivity to making the correct choice of optic orientation.  A 40\% difference in calibration of V is seen with the variation in choosing these two exposures.  The right hand graphic of Figure \ref{fig:vispSAR_cal_efficiency_14state} shows the input Stokes parameters at 420 nm wavelength for this two-exposure optimized sequence. The inputs are diverse and irregular, and span the space.

\vspace{-3mm}

\begin{figure}[htbp]
\begin{center}
\vspace{-0mm}
\hbox{
\includegraphics[height=6.2cm, angle=0]{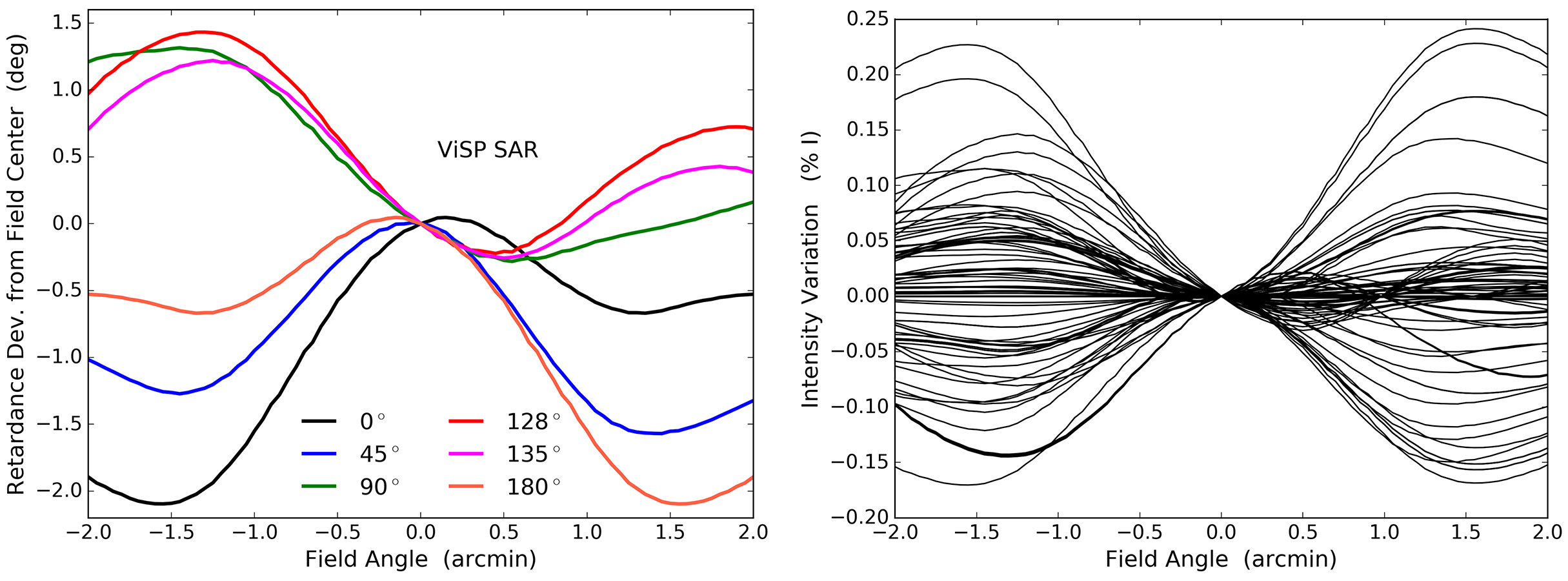}
}
\caption[] 
{\label{fig:visp_sar_ret_mod_deviate_vs_field_orientation} Left shows the ViSP SAR retardance deviation from average as a function of field angle for the six unique retarder orientations during this calibration sequence at 420 nm wavelength. Errors of the same orientation but at different magnitude are simulated at longer wavelengths. The right hand graphic show the detected intensity variation from the field center value across the field of view for all 8 modulation states and all 14 calibration inputs. The inner $\pm$0.5 arc minute field generally shows constant intensities to better than $\pm$0.05\%}
\vspace{-5mm}
\end{center}
\end{figure}

We take the Meadowlark Optics metrology for the ViSP SAR and derive the footprint-averaged retardance at the appropriate orientation and beam decenter, similar to that shown above in Figure \ref{fig:footprint_ViSPsar_AsBuilt_420nm}. With this spatial retardance error, we can create a perturbation of the as-built elliptical retardance parameters at any wavelength by adding a scaled amount of retardance and fast axis variation.  Though we showed above that the scaling is not strictly inversely proportional to the wavelength, the errors do scale roughly as $\lambda^{-1}$ as shown above within the demonstrated design sensitivity of factors of 2 to 3. We create a model here where we take the perturbation measured at Meadowlark optics and apply a scaled version to the as-built elliptical retardance values at the appropriate wavelength. The left hand graphic of Figure \ref{fig:visp_sar_ret_mod_deviate_vs_field_orientation} shows the variation of the footprint-averaged retarder magnitude as a function of field angle for the six unique retarder orientations at 420 nm wavelength.  The six unique retarder orientations are noted in the legend.  The fast axis variation is roughly similar in morphology with magnitudes generally less than 0.1$^\circ$.

We next simulate the simultaneous fit of a modulation matrix and the three elliptical retardance properties for the calibration retarder during the calibration process.  We use a theoretical modulation matrix produced by a perfect linear retarder and discrete modulation (no depolarization from smearing during rotation). We used a magnitude of 120$^\circ$ and 8 modulation states, evenly spaced by 22.5$^\circ$ as in the above section. For each calibration state as shown in Figure \ref{fig:vispSAR_cal_efficiency_14state}, there are 8 intensities recorded during modulation, giving 112 intensity points for fitting. The first four settings of the calibration sequence only use the calibration polarizer alone and do not show spatial variation across the field of view.  Once the calibration retarder is inserted for input states number 4 to 13, the as-built retardance magnitude, the 129$^\circ$ optic orientation offset and the elliptical retardance create uneven but efficiently diverse input Stokes vectors. The right hand graphic of Figure \ref{fig:visp_sar_ret_mod_deviate_vs_field_orientation} shows the variation in intensity propagated through this ideal modulation process as a function of field angle. The intensities recorded account for the spatially varied retardance in the left hand graphic at the nominal optic orientation and magnitude.  

Intensity variations of magnitudes below 0.15\% are seen across the ViSP slit field of view of $\pm$1 arc minute.  The field variation is not strictly symmetric, and we anticipate significant averaging across the field.  However, significant gains from averaging spatially along the slit are seen. The change in intensity detected at the instrument starts at zero and only increases to magnitudes of 0.02\% as we increase the field angle over which we sum intensities to over $\pm$1 arc minute.  The ideal CalPol, modulator and analyzer remain unchanged by field angle.  The calibration retarder has retardance varying as a function of field, which produces different modulated intensities across the slit, which are then spatially averaged. The right hand graphic of Figure \ref{fig:visp_sar_ret_mod_deviate_vs_field_orientation} shows substantial symmetry in some cases as well as a wide range in magnitude of field dependence for different modulation states.  These curves average down from magnitudes of 0.15\% at any individual field angle to values closer to 0.02\% when computing the field average uniformly weighted across the slit (linear 1-d weighting).

We take the intensity-perturbations to the calibration sequence and then perform a simultaneous fit to the calibration retarder and modulation. The calibration retarder has 3 variables for elliptical retardance at each wavelength. The modulation matrix has 32 variables from the 8 modulation states multiplied by the 4 Stokes vector components at each wavelength. We have 112 measured intensities as 14 input states multiplied by 8 modulation states. With 112 intensities and 35 variables at each wavelength, we have a highly constrained fit.  

\begin{wrapfigure}{r}{0.31\textwidth}
\centering
\vspace{-7mm}
\begin{equation}
\label{eqn:rss_error_metric}
\epsilon  = \sum_{i=1}^{m}  \sum_{\lambda=1}^{l} ( I_{model} - I_{synth} )^2
\end{equation}
\vspace{-8mm}
\end{wrapfigure}

We define an error metric as the RSS of the difference in intensities as shown in Equation \ref{eqn:rss_error_metric} for each wavelength. The intensity model ($I_{model}$) is created using each iterations guessed elliptical retardance parameters and modulation matrix elements. The synthetic data set ($I_{synth}$) is the reference created using the field-dependent, perturbed calibration retarder and ideal 120$^\circ$ modulator. This synthetic data includes the intensity average along the ViSP slit at the appropriate field angle for each of the $m$ detected flux values as modulated for each of the input states.  The elliptical retardance parameters are expected to change slightly from the as-built field-center values due to the added spatial variation.  However, the modulation matrix derived will not exactly match that of the perfect 120$^\circ$ ideal retarder.

\begin{figure}[htbp]
\begin{center}
\vspace{-3mm}
\hbox{
\includegraphics[height=6.1cm, angle=0]{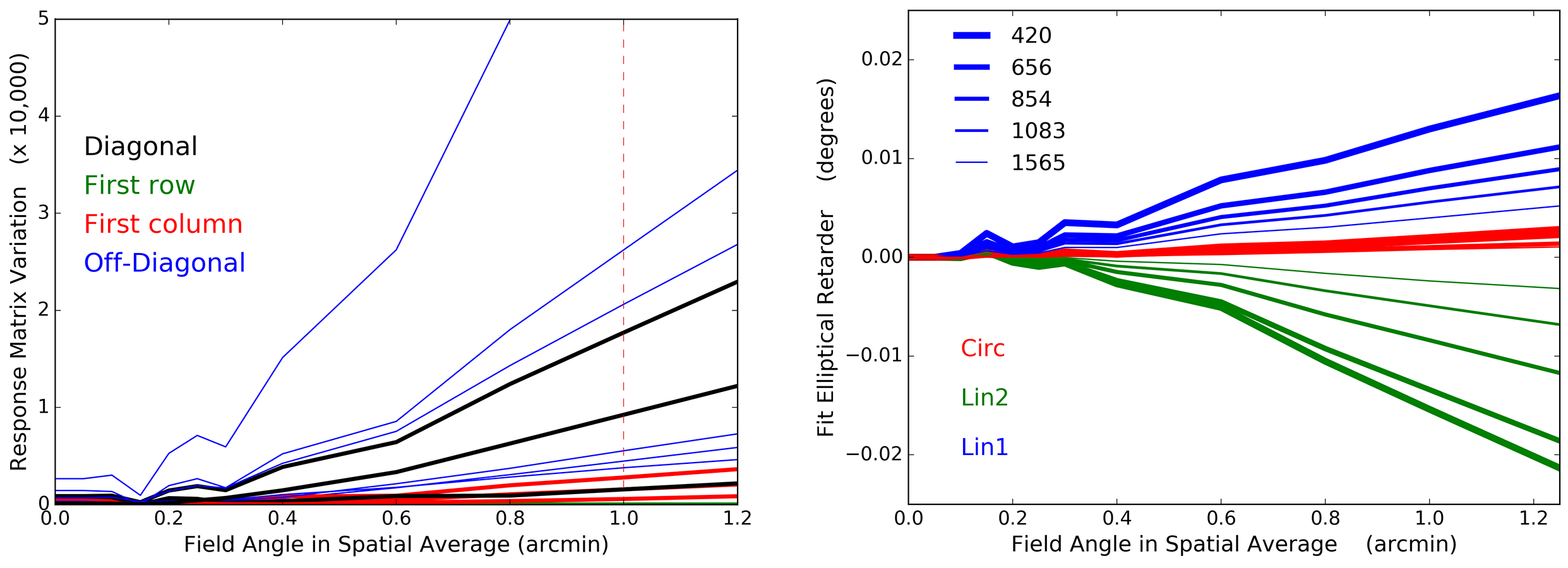}
}
\caption[] 
{\label{fig:demod_n_er_fitting_with_field_and_scaleLam} ViSP SAR:  Left shows the error matrix computed from the polarization response matrix ({\bf DO}) in fit demodulation matrix compared to the theoretical ideal modulation matrix from a 120$^\circ$ linear retarder. The errors are of order 0.01\% until field angles of 0.5 arc minutes are exceeded.  The right hand graphic shows the change in best-fit elliptical retardance parameters for the calibration retarder.  All fits are done with the spatially perturbed calibration retarder with the magnitude computes as the ratio of the wavelength to 420 nm. }
\vspace{-5mm}
\end{center}
\end{figure}

In the left hand graphic of Figure \ref{fig:demod_n_er_fitting_with_field_and_scaleLam} we show the error computed from the response matrix ({\bf DO}) when the best fit modulation matrix is inverted and used to demodulate at each field angle and wavelength in the simulation. We only show the 420 nm simulation for clarity. We used the as-built retarder properties at each wavelength so the chromatic variation of the ViSP SAR is included in this simulation. The blue lines correspond to the off-diagonal modulation matrix elements varying at 420 nm wavelength. The black lines show Stokes I with very low sensitivity.  

\begin{wrapfigure}{l}{0.42\textwidth}
\vspace{-6mm}
\begin{equation}
{\bf DO}_{420nm, 1'} =
 \left ( \begin{array}{rrrr}
 0   	& 0		&    0		&   0 		\\
 0 	& 0.4 	&  1.6 	&-4.6	\\
 -0.4 	&  2.0	&  -0.2	& -1.1	\\
 -0.1	&  0.1	& -1.1	&  0		\\
\end{array} \right ) 
\label{eqn:error_matrix_420nm_1arcmin}
\end{equation}
\vspace{-8mm}
\end{wrapfigure}

As an example of the error matrix, we show tabulated values in Equation \ref{eqn:error_matrix_420nm_1arcmin} multiplied by 10,000 and rounded to 1 decimal place (10$^{-5}$ magnitudes) at the 420 nm simulation wavelength and using 1 arc minute field averaging.  The first row varies less than 10$^{-5}$ while the first column errors are a few parts in 10$^{-5}$.  The off-diagonal elements are significant at magnitudes of a few 10$^{-4}$.  The diagonal elements show errors of a few 10$^{-5}$. 

The resulting variation of the derived elliptical retardance parameters for the calibration retarder is shown in the right hand graphic of Figure \ref{fig:demod_n_er_fitting_with_field_and_scaleLam}.  Blue shows the first component of linear retardance.  Green shows the second linear component.  Third shows circular retardance. The dominant trend is the field variation scales with the magnitude of the spatial retardance errors as seen by the thickness of the lines.  The error of 0.04$^\circ$ seen at 420 nm wavelength is roughy four times larger than the thin blue curve at 0.01$^\circ$ magnitude representing 1565 nm wavelength, scaling appropriately with the ratio of wavelengths.

This simple simulation shows how spatial polishing errors can be propagated through a calibration sequence as modulated by a polarimeter. The general trends are to have modulation matrix errors of order a few 10$^{-4}$ when averaging over an arcminute field at shorter wavelengths. Longer wavelengths generally decrease in linear proportion to the spatial retardance error, which itself scales roughly inversely with wavelength as shown above. Some expected changes in the best-fit elliptical retardance parameters are found at magnitudes below 0.1$^\circ$ given that we sample the optic non-uniformly and in spatial regions that change with the field angle used in the average.  There are several compounding factors we have ignored for this simple demonstration.  A real modulator is spatially variable and will impart it's own issues with calibration across a non-zero field of view. Each exposure will have highly variable signal to noise, which has been ignored presently.  The DKIST mirrors will impart their own field dependence we outlined previously \cite{Harrington:2017dj} as would any real system. This does not include the impact of thermal drifts as we have assessed previously \cite{Harrington:2018cx}. However, with the new metrology tools presented here and this demonstration of a simple system model, we can easily extend system level models to include multiple artifacts. This then will be used to choose an appropriate FoV for each calibration optic and sequence to achieve high SNR while keeping modulation artifacts below our required calibration thresholds.

\section{Summary: Calibration \& An Error Comparison}
\label{sec:summary}

Spectropolarimetric instrumentation has been pushing to to higher accuracy levels, wider wavelength ranges, larger fields of view and complex use cases. The design and fabrication of the calibration optics can be severely constrained. A calibrator must be known, stable under all operations to tight thresholds and completely modeled to ensure accuracy. We showed in this paper how the field of view available for calibration at high accuracy for DKIST is impacted by of spatial non-uniformity for our six crystal super achromatic calibration retarders. Calibration techniques relying on a rotating retarder are often chosen for their stability, durability of the crystals and ease of accurate characterization. However, wide wavelength requirements and high heat loads for can drive designs to large apertures and many-crystal polychromatic solutions mounted near focal planes that creates elliptical retardance and couples in spatial variation to calibration accuracy. Spectral metrology tools showed that net circular retardance over 5$^\circ$ and spectral oscillations of circular retardance of over $\pm$2$^\circ$ were detected in the DKIST calibration retarders. A fully elliptical calibration retarder model must be used to achieve accuracy below 1$^\circ$. With this additional degree of freedom, we showed that we can fully reproduce the as built optic Mueller matrix in a collimated beam. We showed in Section \ref{sec:spatial_polish} outlined beam footprint sizes and the impact of retarder optical locations on spatial errors.  We also include in Appendix \ref{sec:appendix_uniformity_polycarb_FLC} some additional information about footprints, materials and thermal calculations for four solar telescopes DKIST, GREGOR, GST and the DST. Simulations of polishing errors were outlined here to demonstrate super achromat design tolerances to typical manufacturing errors. Different designs can be more than twice as sensitive to polishing errors depending on which crystal pairs have which polishing errors. In addition, errors are exacerbated at short wavelengths, more than double what simple scaling relations predict. This analysis shows that future instruments placing retarders near focal planes must consider the impact of optic size, beam footprint size and the various scanning techniques required by instruments to cover the field of view.

We showed spatial and wavelength dependence using the as-built DKIST six crystal optics. Individual crystal materials can be softer or harder.  Different polishing techniques make individual crystals have specific spatial patterns, often tied to E- and O- beam directions. When stacking up six crystals oriented in specific ways, there are significant changes in magnitude, spatial morphology and wavelength dependence. In Appendix \ref{sec:appendix_spatial_maps} we show metrology and models for the rest of the DKIST calibration retarders not presented in the main text. The spatial retardance maps and spectral Mueller matrix measurements are shown along with clocking and polishing tolerance models for all optics. These errors were assessed for their impact to operations using optical footprints during calibration and modulation. The magnitudes of retardance variation as function of DKIST field of view show polishing errors in the six crystal designs translates to larger non-uniformity at shorter wavelengths with amplitudes easily over one degree elliptical retardance. These errors can be comparable other expected issues from thermal drifts, crystal clocking errors, mirror-induced field dependent polarization, that must also be considered in a system-level error budget. The DKIST optics impart their own field dependence we outlined previously \cite{Harrington:2017dj} and the impact of thermal drifts must similarly be considered on a per-optic and per-use-case basis \cite{Harrington:2018cx}. 

We compare these spatial non-uniformity errors simulated here to other known sources of error measured for our retarders. The longer-wavelength optimized Cryo-NIRSP SAR optic showed over 10$^\circ$ spatial variation at visible wavelength metrology while retarders with less net retardance like the ViSP SAR and DL-NIRSP SAR showed 3$^\circ$ retardance variation at the same metrology wavelength. The polishing techniques used on the SiO$_2$ crystals was different than for the MgF$_2$ crystals, resulting in double to an order of magnitude worse error in individual two crystal subtraction pairs. These manufacturing errors need to be combined with additional errors from the converging beam and non-zero field angles introducing depolarization at levels approaching 1\% and field dependent elliptical retardance \cite{Sueoka:2016vo,Sueoka:2014cm}. To achieve the DKIST required accuracy while instruments use their sub-fields scan the delivered field of view, we require field dependent calibration and demodulation. A significant challenge is how to correctly calibrate the system when we cannot make the assumption of a constant calibration retarder. Retarder properties vary with time through temperature, with rotation of the optic through polishing errors and optical stations near focal planes. 

We find that there is minimal impact for spatial non-uniformity when modulating near focal planes. Many common demodulation techniques already compensate for spatially variable modulation provided the spatial averaging is limited to scales within which the optic does not change significantly. We showed a simple simulation above in Section \ref{sec:sim_fov_errors_demod} that we see modulation matrix errors of order a few parts in 10$^{-4}$ when averaging over an arc minute field at blue wavelengths. This kind of metrology and simulation is crucial for DKIST as the MgF$_2$ calibration retarder has over four times the spatial retardance error but with much more aperture symmetry and also strongly variable wavelength dependence. We have shown here the general tools to assess a polarimeter for impact of these types of spatial variation.  The depolarization introduced by averaging over a non-uniform footprint is typically very small, often decreasing efficiency only by a few percent. We showed in Section \ref{sec:compare_continuous_rotation} that the modulation efficiency is impacted for continuously rotating optics analogous to the aperture-average induced depolarization. For modulators, spatial non-uniformity of the optic faces far less stringent requirements. 

These polishing errors can be compared to temporal drifts caused by temperature changes of the optics \cite{Harrington:2018cx}. The retardance variation is roughly a few degrees retardance error for $\pm$20$^\circ$C of temperature change. Models we presented \cite{Harrington:2018cx} showed operation from 0$^\circ$C to 40$^\circ$C from the baseline 20$^\circ$C along with depth gradients in the range of 0$^\circ$C to 4$^\circ$C. The variation in retardance from calibration on a cold summit morning to a warm afternoon will be roughly the same magnitude as the polishing errors. The clocking errors are known but are not substantially temperature dependent. Both the clocking and the polishing errors are stable and constant hence possibly removable through calibration. In addition, the optical properties of beam deflection can introduce image motion during modulation and also destabilize the beam on downstream optics. When considering calibration strategies, several other optical performance measures must be optimized.

Ambiguous solutions and complications with retarder model sign conventions are discussed in Appendix \ref{sec:appendix_eliret}. We show the {\it axis-angle} formalism for elliptical retarders as rotation matrices. We also show how multiple Euler angle formalism conventions are equally applicable to fitting retarder Mueller matrices and achieves the same fits to NLSP measurements of the ViSP SAR. We utilize the axis-angle formalism as is common in many other references as it seems more straightforward to interpret the various components of the resulting QUV rotation.

In Appendix \ref{sec:appendix_uniformity_polycarb_FLC} we show uniformity measurements for polycarbonate zero order and super-achromatic retarders as well as ferro-electric liquid crystal type retarders. The DKIST instrument VTF will use two polycarbonate zero-order retarders and two FLC retarders to make an achromatic modulator. The GST at Big Bear Solar Observatory uses the same stretched polycarbonate from the same company for their modulators and calibration retarders.  Our laboratory spectropolarimeter uses five-layer super achromatic retarders made of the polycarbonate. A larger version if this optic will most likely also be used as a replacement modulator for the DL-NIRSP instrument on DKIST. 

Solar telescopes will continue to reduce calibration errors, explore wider wavelength ranges and explore diverse optical solutions. The spatial variation of retardance for the six crystal DKIST retarders is potentially significant calibration error when using traditional algorithms at larger field angles. We showed basic simulations of calibration when averaging over field angle. Calibrations derived for an ideal polarimeter with field angles less than half an arc minute with the as-built ViSP SAR properties are not impacted at magnitudes above 10$^{-4}$ as the beam does not sample a spatially varying region of the retarder aperture. At longer wavelengths, these impacts decrease inversely proportional to the wavelength as modified by the design sensitivities outlined in Section \ref{sec:spatial_sensitivity_ordering}. DKIST and other instruments will aim to calibrate the full field of view. The spatial non-uniformities presented here can be dominant error source if spatial variation is large and the optic is placed at a location that strongly couples spatial non-uniformities into the calibration. Using a fully elliptical retardance model for the calibration optic adds an extra degree of freedom but accounts for both net circular retardance as well as spectral oscillations that may be present away from metrology wavelengths. With accurate metrology and knowledge of the various competing error sources, we can now improve our system-level modeling tools and compare alternate calibration strategies.  These spatial metrology tools also allow us to assess alternate optics to improve the calibration accuracy and ultimately the science output of the observatory.

\section{Acknowledgements}

This work was supported by the DKIST project. The DKIST is managed by the National Solar Observatory (NSO), which is operated by the Association of Universities for Research in Astronomy, Inc. (AURA) under a cooperative agreement with the National Science Foundation (NSF). We thank Tom Schad for his initial simulations of calibration efficiency of several common polarization sequences. This inspired our creation of an efficient but short calibration sequence for DKIST with a single optic using as-built elliptical properties.  We thank David Elmore for his assistance, guidance and insight into the long history of work on the DKIST project.  We thank Doug Gilliam and the dedicated staff at the Dunn Solar Telescope for their knowledge of the systems. We thank Manolo Collados for his participation in DKIST design reviews and discussions about the GREGOR systems. We thank Kwangsu Ahn at Big Bear Solar Observatory for assistance understanding their polarimeter optical performance. We thank Christian Beck for his frequent discussions on polarization calibration. This research made use of Astropy, a community-developed core Python package for Astronomy (Astropy Collaboration, 2013). This research made use of the Physical Optics Propagation in PYthon (POPPY) Python package developed and maintained by Marshall Perrin, Joseph Long, and collaborators.

\appendix
\clearpage

\section{Elliptical Retarders \& Rotation Matrices}
\label{sec:appendix_eliret}

Here we describe elliptical retarder models and rotation matrix formalisms in common use for describing retarders. In the {\it axis-angle} representation of a rotation, two quantities are typically given. The first is a unit vector $e$ indicating the direction of an axis for the rotation. The second is an angle $\theta$ describing the magnitude of the rotation about the axis. Only two numbers are needed to define the direction of a unit vector because the magnitude of $e$ is a specified constraint. The equation for the rotation in matrix notation is thus a magnitude times the basis vector {\bf r} = $\theta e$. Alternatively, the three components of the vector can be specified and the magnitude computed from the vector components. 

In Sueoka \cite{Sueoka:2016vo}, fits for elliptical retardance are applied to the DKIST six crystal retarder optics. We rewrite this rotation matrix equation in the published form while adopting a slightly different notation and fixing a typographical error in Equation \ref{eqn:sueoka_ellipse}. We use a notation where $\cos$($\theta$) is denoted $C_\theta$ and $\sin$($\theta$) is denoted $S_\theta$.  We adopted the notation substitution $r_H$ = $r_x$, $r_{45}$ = $r_y$ and $r_z$ = $r_r$ to explicitly denote an xyz coordinate frame for the rotation matrix (${\bf \mathbb{R}}_{ij}$). This substitution makes the notation similar to other references on rotation matrices as the $(H,45,R)$ notation corresponds to naming conventions of {\it horizontal} as x or preservation of Stokes Q, the {\it 45} as y or preservation of Stokes U and {\it R} as z or preservation of Stokes V. The rotations are about $(x,y,z)$ axes respectively when the Poincar\'{e} sphere is represented in $(x,y,z)$ coordinates. 

\vspace{-6mm}
\begin{Large}
\begin{equation}
 {\bf \mathbb{R}}_{ij} = \left ( \begin{array}{ccc}
 \frac{r_x^2}{d^2} + \left ( \frac{r_y^2 + r_z^2}{d^2}  \right )C_\theta
     & \frac{r_x r_y}{d^2} \left ( 1-C_\theta \right ) + \frac{r_z}{d} S_\theta  
     & \frac{r_x r_z}{d^2} \left ( 1-C_\theta \right ) - \frac{r_y}{d} S_\theta  
      \\
 \frac{r_x r_y}{d^2} \left ( 1-C_\theta \right ) - \frac{r_z}{d} S_\theta  
     &  \frac{r_y^2}{d^2} + \left ( \frac{r_x^2 + r_z^2}{d^2}  \right )C_\theta   
     &  \frac{r_y r_z}{d^2} \left ( 1-C_\theta \right ) + \frac{r_x}{d} S_\theta
     \\ 
\frac{r_x r_z}{d^2} \left ( 1-C_\theta \right ) + \frac{r_y}{d} S_\theta  
     &  \frac{r_y r_z}{d^2} \left ( 1-C_\theta \right ) - \frac{r_x}{d} S_\theta
     &  \frac{r_z^2}{d^2} + \left ( \frac{r_x^2 + r_y^2}{d^2}  \right )C_\theta   \\ 
\end{array} \right ) 
\label{eqn:sueoka_ellipse}
\end{equation}
\end{Large}
\vspace{-2mm}

This equation is an {\it axis-angle} version of a rotation matrix. In the following paragraphs we show some properties of this rotation matrix formalism. The matrix equation for an {\it axis-angle} rotation using a unit vector $\bf{u}$ = $(u_x, u_y, u_z)$ and the angle (magnitude of rotation) $\theta$.

\begin{wrapfigure}{r}{0.40\textwidth}
\centering
\vspace{-7mm}
\begin{equation}
{\bf \mathbb{R}}_{ij} = \mathbb{II}  C_\theta  + \bf{u_\otimes} S_\theta  + \left ( \bf{u \otimes u} \right ) \left ( 1-C_\theta \right )
\end{equation}
\vspace{-9mm}
\end{wrapfigure}

The rotation matrix ${\bf \mathbb{R}}_{ij}$ is generated by using the {\it axis} specified by the unit vector $\bf{u}$ and the angle $\theta$. We can generate this rotation matrix using the identity matrix $\mathbb{II}$, the cross-product matrix of ${\bf u}$ as ${\bf u_\otimes}$ and the tensor product matrix of ${\bf u}$ as ${\bf u \otimes u}$.

\begin{wrapfigure}{r}{0.37\textwidth}
\centering
\vspace{-7mm}
\begin{equation}
\label{eqn:cross_product_matrix}
{\bf u_\otimes} =  \left [ \begin{array}{ccc}
0  	& -u_z 	& u_y  \\
u_z 	& 0		& -u_x \\
-u_y 	& u_x 	& 0 	    \\
\end{array} \right ]
\end{equation}
\begin{equation}
\label{eqn:tensor_product_matrix}
{\bf u \otimes u} =  \left [ \begin{array}{ccc}
u_x^2  	& u_x u_y 	& u_x u_z  \\
u_x u_y 	& u_y^2		& u_y u_z  \\
u_x u_z 	& u_y u_z 	& u_z^2	    \\
\end{array} \right ]
\end{equation}
\vspace{-5mm}
\end{wrapfigure}

The cross-product matrix ${\bf u_\otimes}$ and the tensor-product matrix ${\bf u \otimes u}$ can written as in Equations \ref{eqn:cross_product_matrix} and \ref{eqn:tensor_product_matrix} respectively. By expanding out the rotation matrix, we recover a standard axis angle notation for ${\bf \mathbb{R}}_{ij}$ as in Equation \ref{eqn:wikipedia_axis_angle_matrix}.  We can use a few simple substitutions to show Equation \ref{eqn:sueoka_ellipse} of Sueoka \cite{Sueoka:2016vo} is a version of this {\it axis-angle} equation for a rotation matrix. 

Since $\bf{u}$ is a unit vector, we can always substitute $1-u_x^2$ as $u_y^2 + u_z^2$. A similar substitution applies to $1-u_y^2$ and $1-u_z^2$. We can rewrite the diagonal elements of Equation \ref{eqn:wikipedia_axis_angle_matrix} to create a rotation matrix form more similar to Sueoka \cite{Sueoka:2016vo}. We note that the unit vector ${\bf u}$ must obey a normalization relation of $ 1 = u_x^2 + u_y^2 + u_z^2 $.   In the Sueoka 2016 equation, we note that the {\it retarder vector} is denoted as ( $r_H, r_{45}, r_R$) and is not itself a unit vector. It does however obey a similar normalization process.

\vspace{-5mm}
\begin{Large}
\begin{equation}
\label{eqn:wikipedia_axis_angle_matrix}
 {\bf \mathbb{R}}_{ij} = \left ( \begin{array}{ccc}
C_\theta + u_x \left ( 1-C_\theta \right ) 
     & u_x u_y \left ( 1-C_\theta \right ) - u_z S_\theta
     & u_x u_z \left ( 1-C_\theta \right ) + u_y S_\theta
      \\
 u_x u_y \left ( 1-C_\theta \right ) + u_z S_\theta
     &  C_\theta + u_y \left ( 1-C_\theta \right )  
     & u_y u_z \left ( 1-C_\theta \right ) - u_x S_\theta
     \\ 
u_x u_z \left ( 1-C_\theta \right ) - u_y S_\theta
     & u_y u_z \left ( 1-C_\theta \right ) + u_x S_\theta
     &  C_\theta + u_z \left ( 1-C_\theta \right )   \\
\end{array} \right ) 
\end{equation}
\end{Large}
\vspace{-1mm}

\begin{wrapfigure}{r}{0.42\textwidth}
\centering
\vspace{-2mm}
\begin{equation}
d = \sqrt{ r_H^2 + r_{45}^2 + r_R^2} = \sqrt{u_x^2 + u_y^2 + u_z^2 }
\label{eqn:normalize_retmag}
\end{equation}
\vspace{-8mm}
\end{wrapfigure}

If we change notation from ($r_H, r_{45}, r_R$) to ( $u_x, u_y, u_z$) as common in geometrical texts and apply a normalization, we can explicitly show the normalization similar to Sueoka \cite{Sueoka:2016vo}. We allow the vector ${\bf u}$ to not have unit length but to be normalized explicitly in the matrix equation. We then explicitly normalize every component of ${\bf u}$ or ${\bf r}$ by $d$ as in Equation \ref{eqn:normalize_retmag}. This equation is identical to Equation \ref{eqn:sueoka_ellipse} after collecting terms and normalizing by $d$. 

\vspace{-4mm}
\begin{large}
\begin{equation}
{\bf \mathbb{R}}_{ij} =  \left ( \begin{array}{ccc}
u_x^2 +  \left ( u_y^2 + u_z^2 \right ) C_\theta 
     & u_x u_y \left ( 1-C_\theta \right ) - u_z S_\theta
     & u_x u_z \left ( 1-C_\theta \right ) + u_y S_\theta
      \\
 u_x u_y \left ( 1-C_\theta \right ) + u_z S_\theta
     &  u_y^2 + \left ( u_x^2 + u_z^2 \right ) C_\theta
     & u_y u_z \left ( 1-C_\theta \right ) - u_x S_\theta
     \\ 
u_x u_z \left ( 1-C_\theta \right ) - u_y S_\theta
     & u_y u_z \left ( 1-C_\theta \right ) + u_x S_\theta
     &  u_z^2 + \left ( u_x^2 + u_y^2 \right ) C_\theta  \\
\end{array} \right ) 
\label{eqn:wikipedia_axis_angle_matrix_diag}
\end{equation}
\end{large}

In Sueoka \cite{Sueoka:2016vo}, the magnitude of the rotation {\it angle} was specified as $d$, the overall magnitude of the rotation. Equation \ref{eqn:sueoka_ellipse} contains many terms as cosine or sine of the {\it angle} $d$, and as such, d can run from 0 to 2$\pi$ without ambiguity. Typically, rotation matrices are described by 4 numbers but with constraints.  There are only 3 degrees of freedom in the {\it axis-angle} expression. In one typically chosen expression, the input {\it angle} is specified along an {\it axis} vector but the axis required to be a unit vector which removes one degree of freedom through the equation $1 = u_x^2 + u_y^2 + u_z^2 $.  In another convention (followed in Sueoka 2016), you specify the magnitude of the {\it axis} components ($u_x^2, u_y^2, u_z^2$).  The {\it angle} is then derived as the magnitude of the {\it axis}.  This convention is also convenient as it projects the retardance components onto an easily interpreted basis.  The $x$ axis points along $+Q$ on the Poincar\'{e} sphere.  If $u_x$=90$^\circ$ and other components are 0$^\circ$ (${\bf u}=[90,0,0]$), we recover a quarter-wave linear retarder that would rotate a pure $+U$ input vector in to $+V$.  The $y$ axis points along $+U$ on the Poincar\'{e} sphere.  If $u_y$=90$^\circ$ and other components are 0$^\circ$ (${\bf u}=[0,90,0]$), we recover a quarter-wave linear retarder with the fast axis at 45$^\circ$ orientation that would rotate a pure $+Q$ input vector in to $+V$.  If $u_z$=90$^\circ$ and other components are 0$^\circ$ (${\bf u}=[0,0,90]$), we recover a quarter wave circular retarder that would rotate a pure $+Q$ input vector in to $+U$ (a linear polarization rotator).

\begin{wrapfigure}{r}{0.60\textwidth}
\centering
\vspace{-5mm}
\begin{tabular}{c} 
\hbox{
\hspace{-0.2em}
\includegraphics[height=7.3cm, angle=0]{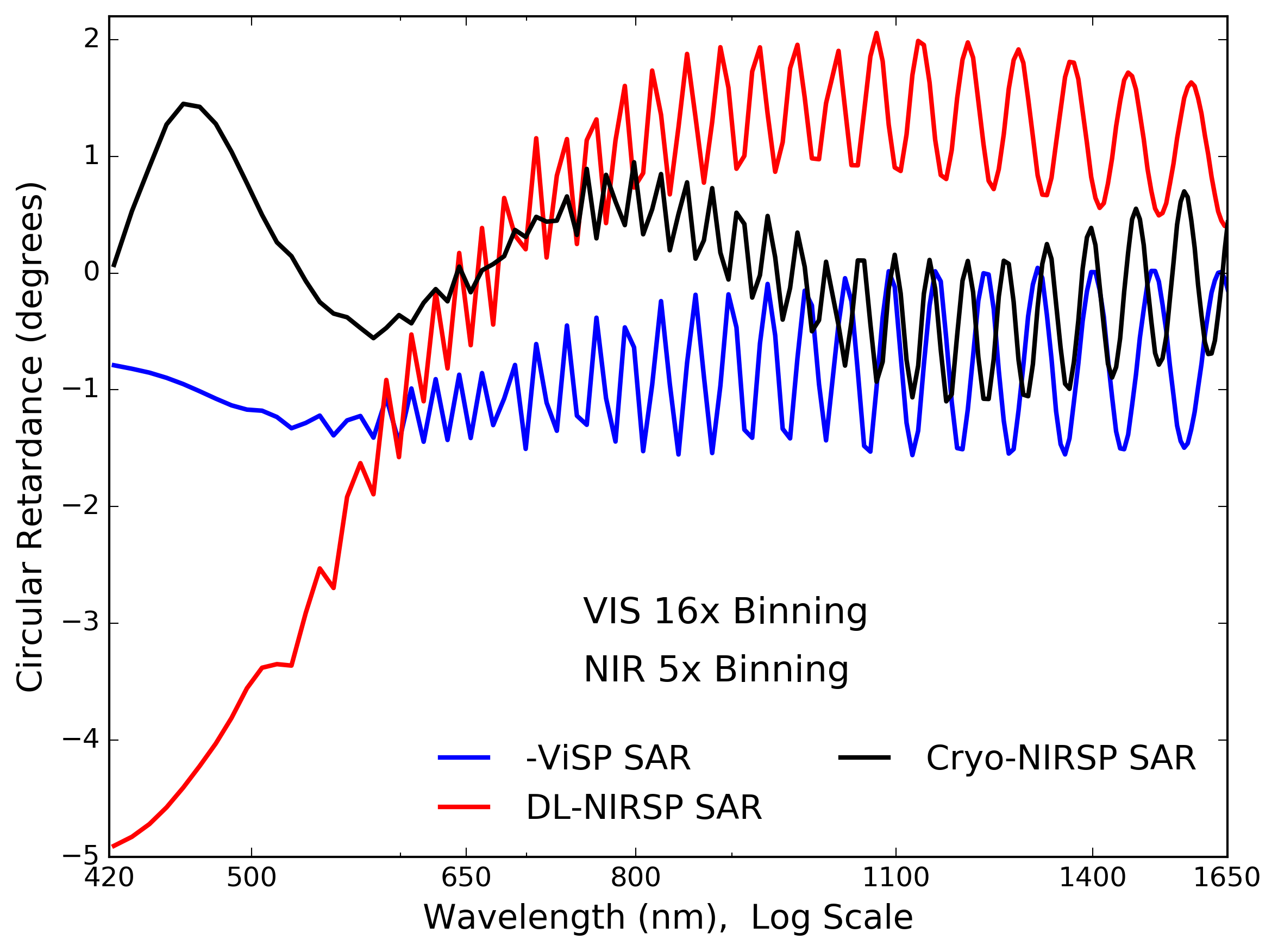}
}
\end{tabular}
\caption[Retarders] 
{ \label{fig:nlsp_circular_retardance_SARs} 
The circular retardance component of the elliptical retarder model fit to the NLSP measured Mueller matrices. Each color shows a different DKIST calibration retarder. Blue shows the ViSP SiO$_2$ retarder covering visible wavelengths multiplied by -1 for clarity. Red shows the DL-NIRSP SiO$_2$ retarder covering near infrared wavelengths.  Black shows the Cryo-NIRSP MgF$_2$ retarder designed for 2500 nm to 5000 nm wavelength. Reduced resolving power smooths clocking ripples leaving design errors visible at short wavelengths.}
\vspace{-5mm}
 \end{wrapfigure}

We have done preliminary acceptance testing on all six DKIST retarders and have data at a range of incidence angles for a single footprint near the center of the optic. Circular retardance is not present in the calibration retarder (SAR) designs so the presence of circular retardance is a good proxy for manufacturing errors. 

Figure \ref{fig:nlsp_circular_retardance_SARs} shows the best-fit circular retardance components to the measured Mueller matrix for the three DKIST calibration retarders. Measurements used a 4 mm diameter beam footprint near the aperture center of the DKIST calibration retarders after solving for normal incidence angle. No circular retardance is present in the designs but polishing and clocking errors combine to create significant circular retardance in all optics, even though the optics meet manufacturing specifications. For these measurements, the spectral resolving power of the visible spectrograph was limited to 9 nm which corresponds to a single sample in the curves of Figure \ref{fig:nlsp_circular_retardance_SARs}. The nominal dispersion is 0.57 nm per pixels but we spectrally average sixteen pixels to sample roughly 9 nm per bin, equivalent to 1 measurement per instrument profile full width half maximum (FWHM).  The near infra-red channel delivers a 12 nm FWHM instrument profile with sampling between 1.1 nm per pixel at 950 nm wavelength to 1.6 nm per pixel at 1650 nm wavelength. Binning of five spectral pixels preserves the spectral resolving power of the system with two points per instrument profile FWHM. The infrared wavelengths in Figure \ref{fig:nlsp_circular_retardance_SARs} shows that we clearly detect these oscillations at amplitudes of roughly 1$^\circ$.  The oscillations are stable and can be calibrated.

In Figure \ref{fig:nlsp_circular_retardance_SARs}, we see that there are non-zero trends in circular retardance as well as spectral oscillations in all SAR optics clearly detected at wavelengths longer than 800 nm. The smooth variation with wavelength can be attributed to polishing errors in addition to mis-alignments of the A-B-A pairs.  But the oscillations are caused by misalignments between individual crystals. The Cryo-NIRSP SAR circular retardance is shown in black for both Figures \ref{fig:fit_nominal_ER_nlsp_cryo_sar_CIRCULAR} and \ref{fig:nlsp_circular_retardance_SARs}. The lower resolving power of the measurements in Figure \ref{fig:nlsp_circular_retardance_SARs} leads to a removal of the ripples from the measurements at shorter wavelengths. The net circular retardance magnitude is up to 5$^\circ$ with oscillations potentially doubling the magnitude.

\subsection{Limiting Cases: Linear \& Circular Retarders}
\label{sec:linear_circular_retarders}

\begin{wrapfigure}{r}{0.55\textwidth}
\centering
\vspace{-6mm}
\begin{equation}
 \left ( \begin{array}{ccc}
\frac{u_x^2}{d^2} 
     & 0
     & 0
      \\
0
     & \left ( \frac{u_x^2}{d^2} \right ) C_\theta
     & - \frac{u_x}{d} S_\theta
     \\ 
0
     &  \frac{u_x}{d} S_\theta
     &  \left ( \frac{u_x^2 }{d^2} \right ) C_\theta  \\
\end{array} \right )  \sim 
 \left ( \begin{array}{ccc}
 1
     & 0
     & 0
      \\
0
     &  C_\theta
     & - S_\theta
     \\ 
0
     & S_\theta
     &  C_\theta  \\
\end{array} \right ) 
\label{eqn:linear_rotator}
\end{equation}
\vspace{-8mm}
\end{wrapfigure}

Limiting cases illuminate the three degrees of freedom. If we set the y and z components of an elliptical retarder to zero ($u_y = u_z = 0$), we put the {\it axis} of rotation in the $+Q$ direction ($x$) recover the equation for a simple linear retarder of Equation \ref{eqn:linear_rotator} with the fast axis along $Q$ such that the optic rotates $U$ into $V$. With only the $x$ component in the retarder magnitude vector, $u_x$ = $d$.  We leave in the explicit normalization to demonstrate the notation.

\begin{wrapfigure}{r}{0.55\textwidth}
\centering
\vspace{-6mm}
\begin{equation}
 \left ( \begin{array}{ccc}
 \left ( \frac{u_z^2}{d^2} \right ) C_\theta
     & \frac{u_z}{d} S_\theta
     & 0
     \\ 
-\frac{u_z}{d} S_\theta
     &  \left ( \frac{u_z^2 }{d^2} \right ) C_\theta  
     & 0   \\
0 & 0 & \frac{u_z^2}{d^2}  \\
\end{array} \right )  \sim
 \left ( \begin{array}{ccc}
  C_\theta
     &  S_\theta
     & 0
     \\ 
 -S_\theta
     &  C_\theta 
    & 0  \\
0
     & 0
     & 1  \\
\end{array} \right ) 
\label{eqn:circular_rotator}
\end{equation}
\vspace{-8mm}
\end{wrapfigure}

Similarly, if we set the x and y components of the retardance to zero, we recover a pure circular retarder with the {\it axis} pointing along $+$z. This optic rotates Stokes $Q$ into Stokes $U$ following Equation \ref{eqn:circular_rotator}. This optic is called a circular retarder or sometimes confusingly a {\it polarization rotator} as it rotates the plane of linear polarization $Q$ into $U$ while leaving $V$ unperturbed.

\begin{wrapfigure}{r}{0.45\textwidth}
\centering
\vspace{-6mm}
\begin{equation}
 \left ( \begin{array}{ccc}
 \frac{1}{2} \left (1+C_d \right )
     & -\frac{1}{\sqrt{2}} S_d  
     &  \frac{1}{2} \left (1-C_d \right )
      \\
\frac{1}{\sqrt{2}} S_d
     & C_d
     & -\frac{1}{\sqrt{2}} S_d
     \\ 
 \frac{1}{2} \left (1-C_d \right )
     & \frac{1}{\sqrt{2}} S_d  
     &  \frac{1}{2} \left (1+C_d \right )
      \\
\end{array} \right ) 
\label{eqn:linear_circular_rotator}
\end{equation}
\vspace{-8mm}
 \end{wrapfigure}

There are other interesting limiting cases.  Suppose we have an equal mix of linear and circular retardance magnitude.  Imagine the linear retardance is oriented about the $x$ direction.  In this case, we have the retarder vector as $u_x$ = $u_z$ and $u_y = 0$. The normalization constraint simplifies many terms. The ratios $\frac{u_x u_y}{d^2}$  = $\frac{u_x^2}{d^2}$ = $\frac{u_z^2}{d^2}$ = $\frac{1}{2}$. The other terms simplify as $\frac{u_x}{d}$ = $\frac{u_z}{d}$ = $\frac{1}{\sqrt{2}}$. With the explicit normalization included and the magnitude of the {\it angle} as $d$, we obtain Equation \ref{eqn:linear_circular_rotator} for an elliptical retarder with the rotation {\it axis} pointing equally in between $+Q$ and $+V$ ($u_x$=$u_z$).

\begin{figure}[htbp]
\begin{center}
\vspace{-0mm}
\hbox{
\includegraphics[height=16.5cm, angle=0]{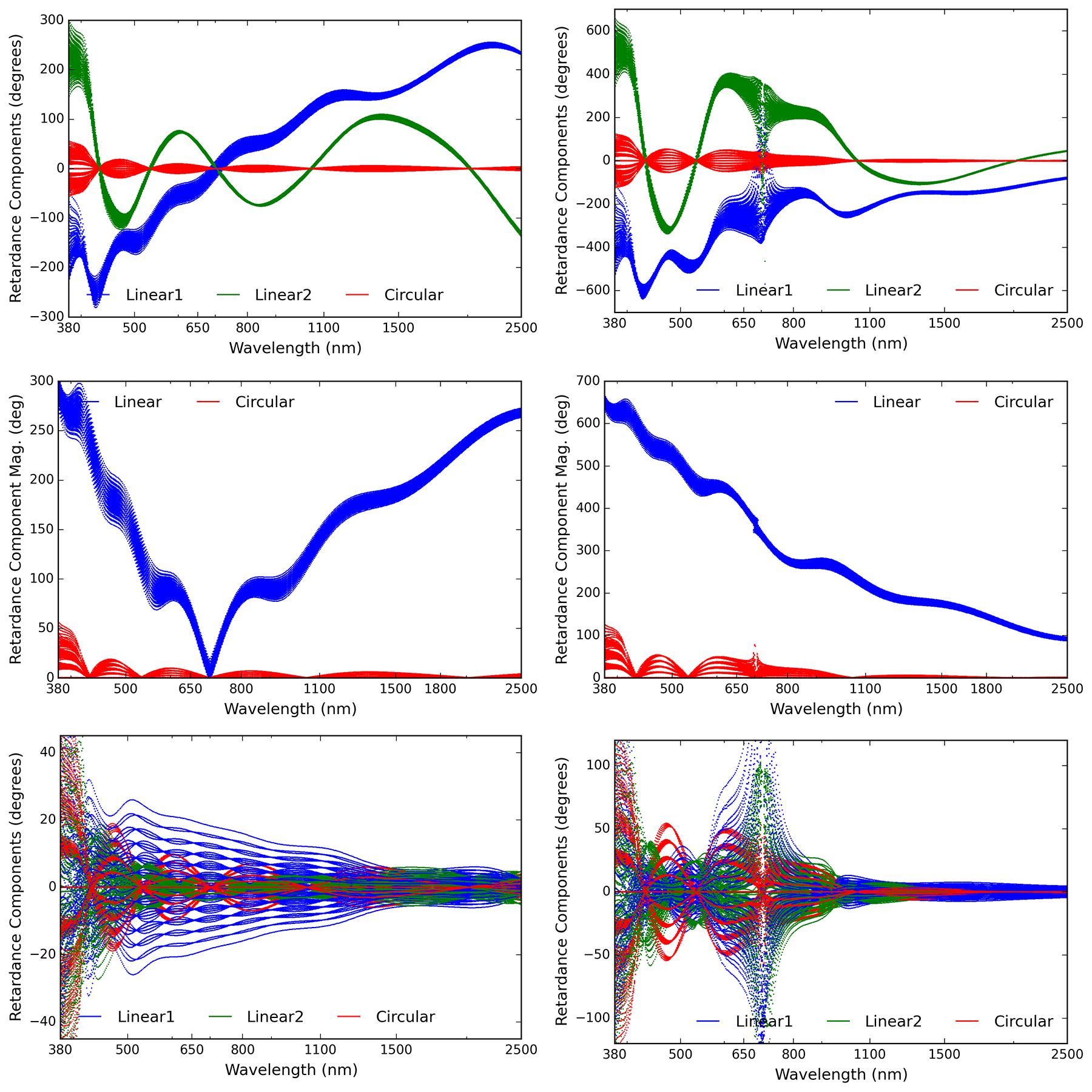}
}
\caption[] 
{\label{fig:compare_solutions_cryosar} A comparison of errors when fitting the Cryo-NIRSP SAR Mueller matrix to different elliptical retardance models. The left side shows a model with retardance components restricted to be less than one wave. The right side shows a model where the retardance components were guided to be over 1.5 waves following the analytical Pancharatnam solution. The right hand graphics have over twice the scale as the left. The top graphics show the three elliptical retarder components ($u_x, u_y, u_x$). Blue shows the first linear component of retardance.  Green shows the second linear component of retardance. Red shows circular retardance. The middle graphics show the magnitude of linear retardance ($\sqrt{u_x^2 + u_y^2}$) in blue and the magnitude of circular retardance in red ($u_z$). The left hand design starts with a magnitude around 270$^\circ$ linear retardance with values increasing decreasing through zero around 700 nm then increasing toward 270$^\circ$ wave at 2500 nm wavelength. The right hand design starts near 90$^\circ$ linear retardance with the magnitude rising steadily with decreasing wavelength. The bottom panel shows the elliptical retarder component errors derived from the same polishing errors as a function of wavelength. In both cases left and right, the optic has the same Mueller matrix. The assumed elliptical retardance model does impact the derived sensitivities as seen in the lower two graphics.}
\vspace{-6mm}
\end{center}
\end{figure}

\subsection{Degenerate Solutions: Sign conventions and Rotations Of Multiple Waves}
\label{sec:degenerate_solutions}

Rotation formalisms have ambiguities. As for any kind of rotation, there are often several substitutions that create identical mappings of unrotated to rotated coordinates. For each of these ambiguous solutions, the magnitude of the computed errors in our various simulations can be impacted deriving different sensitivities to polishing errors. Note that the ambiguous retardance components in this axis-angle formalism are not simply related by a sign change. Simply enforcing the spectral smoothness of a solution does not guarantee the correct solution.

For the DKIST six crystal retarders, our task is also complicated by having infrared-optimized science instruments while laboratory acceptance testing is only performed at shorter wavelengths. We must compute in advance the retardance magnitude and design solutions appropriately. We show two equally accurate elliptical retarder fits to the Cryo-NIRSP calibration retarder (SAR) Mueller matrix in Figure \ref{fig:compare_solutions_cryosar}. In both cases, the identical Mueller matrices were fit including 0.001 waves polishing error at 633 nm wavelength. The 3$^6$ models had this polishing error applied to each crystal against every other crystal.  In the left hand set of graphics, the elliptical retardance parameters were restricted to the domain $\pm\pi$ and the retardance magnitude was forced to be less than one wave.

\begin{wrapfigure}{r}{0.59\textwidth}
\centering
\vspace{-4mm}
\begin{tabular}{c} 
\hbox{
\hspace{-1.0em}
\includegraphics[height=7.2cm, angle=0]{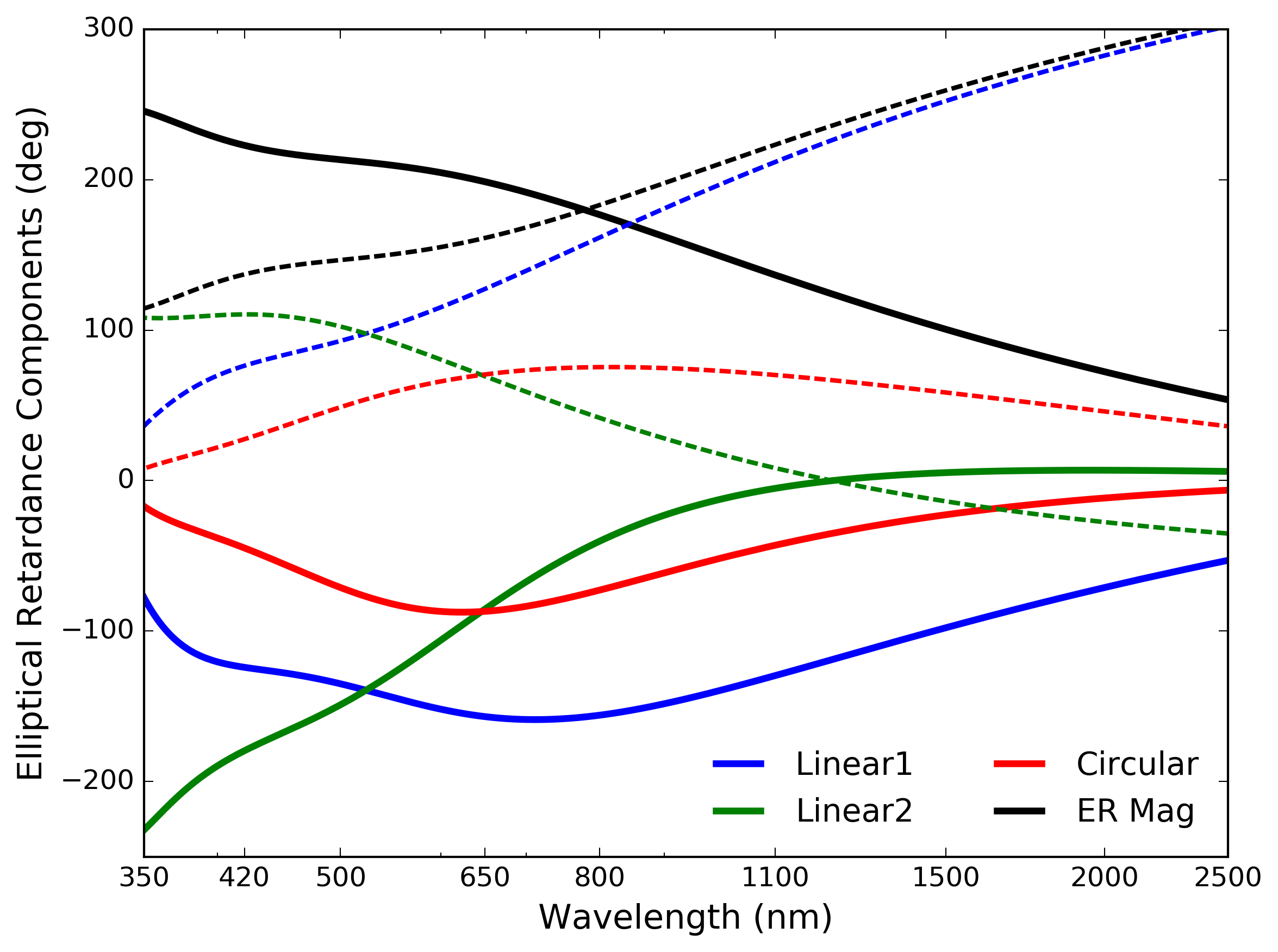}
}
\end{tabular}
\caption[] 
{ \label{fig:visp_pcm_multiple_ER_solutions}  Multiple ER solutions for the ViSP PCM design. Solid lines show one family of solutions with mostly negative signs. Dashed lines show an alternative sign solution that is an equally good fit to the Mueller matrix. The two solutions have a different wavelength dependence and total magnitude trend.  Using physically correct parameters is required when fitting a Mueller matrix and subsequently deriving tolerance sensitivities.}
\vspace{-4mm}
 \end{wrapfigure}

The solution begins near one wave magnitude at short wavelengths, falls to zero around 700 nm wavelength and then rises toward 3/4 wave net retardance at long wavelengths.  In the right hand set of plots, we enforce a solution that has a net retardance magnitude that rises from 1/4 wave at 5000 nm wavelength, through 1.0 waves magnitude at 700 nm reaching nearly 2.0 waves retardance magnitude at 380 nm.  Notice that the right hand plots have typically double the y-axis range.  The derived sensitivity to polishing errors is more than double, and has significantly different wavelength dependence.  Given that we are fitting the {\it identical} Mueller matrix, we consider this numerical fitting issue to be critical to assessing retarder design sensitivities.

An example of the ViSP PCM is shown in Figure \ref{fig:visp_pcm_multiple_ER_solutions}. The solid lines show one solution where all three components of elliptical retardance are negative at visible wavelengths. The solid black line shows the elliptical retardance magnitude falling from roughly three quarter wave at 350 nm wavelength to about quarter wave at 2500 nm wavelength. This decreasing magnitude is physically realistic as the net retardance of the six quartz crystals in the ViSP PCM decrease significantly over this large wavelength range. The dashed lines Figure \ref{fig:visp_pcm_multiple_ER_solutions} show an identically good fit to the optic Mueller matrix, but with all components having positive sign at visible wavelengths. The elliptical retardance magnitude in this case starts near 110$^\circ$ at 350 nm wavelength and rises towards almost a full wave at 2500 nm wavelength. The Mueller matrices are identical for these two solutions, and the individual components are not related by a simple exchange of sign. If one simply enforces spectral smoothness and / or quality of fit, identifying the physically realistic model is not guaranteed. We developed scripts to ensure that the as-built metrology informed the physical model of the optic and that the appropriate elliptical retardance parameters were fit. For the longer wavelength DKIST instruments of Cryo-NIRSP and DL-NIRSP, this care was required as the net retarder magnitude can be quite high.

To ensure correct solutions, additional knowledge of the optic as well as physical constraints must be included.  For instance, the net retardance of the quartz crystals falls with wavelength, thus the solid line solution from Figure \ref{fig:visp_pcm_multiple_ER_solutions} is a more representative model as the net retardance falls with wavelength and approaches zero retardance in the near infrared. We note though that the Mueller matrices of both models is entirely equivalent and that describing retarders as rotations has inherent ambiguity.

\subsection{Euler Angles as an equivalent rotation formalism}

In earlier works, we have used an Euler formalism to describe rotation matrices \cite{Harrington:2017eja,Harrington:2017dj,Harrington:2015cq,Harrington:2015dl,Harrington:2011fz,Harrington:2010km,2008PASP..120...89H,Harrington:2006hu}. In the Euler angle rotation formalism, there are three successive rotations about specific coordinate axes. We denote the three Euler angles as ($\alpha, \beta, \gamma$) and use a short-hand notation where $cos(\gamma)$ is shortened to $c_\gamma$. We had specified the rotation matrix (${\bf \mathbb{R}}_{ij}$) using the ZXZ convention where rotations are applied in sequence, first about the Z axis, second about the X axis then third about the new Z axis. Equation \ref{eqn_definerot} shows this rotation matrix. Note that this rotation matrix is identical to a circular retarder followed by a linear retarder with the fast axis at 0$^\circ$ followed by another circular retarder. There are 12 independent conventions for applying a sequence of rotations about a coordinate axis by the Euler angles such as XZX, YZY, etc. We note that the retarder matrix is a rotation formalism for $QUV$ rotating into $QUV$ and several models for rotation matrices are available.

\begin{equation}
\begin{footnotesize}
{\bf \mathbb{R}}_{ij} =
 \left ( \begin{array}{rrr}
 c_\gamma	& s_\gamma	& 0		\\
 -s_\gamma	& c_\gamma	& 0		\\
 0			& 0			&1		\\ 
 \end{array} \right ) 
 \left ( \begin{array}{rrr}
 1			& 0			& 0		\\
 0			& c_\beta		& s_\beta	\\
 0			& -s_\beta		& c_\beta	\\ 
 \end{array} \right ) 
 \left ( \begin{array}{rrr}
 c_\alpha		& s_\alpha	& 0		\\
 -s_\alpha		& c_\alpha	& 0		\\
 0			& 0			&1		\\ 
 \end{array} \right )  =
 \left ( \begin{array}{ccc}
 c_\alpha c_\gamma - s_\alpha c_\beta s_\gamma		& s_\alpha c_\gamma + c_\alpha c_\beta s_\gamma		& s_\beta s_\gamma		\\
-c_\alpha s_\gamma - s_\alpha c_\beta c_\gamma 		&-s_\alpha s_\gamma + c_\alpha c_\beta c_\gamma 	& s_\beta c_\gamma 	\\
 s_\alpha s_\beta								& -c_\alpha s_\beta								& c_\beta				\\ 
 \end{array} \right ) 
\label{eqn_definerot}
\end{footnotesize}
\end{equation}

\begin{equation}
\begin{footnotesize}
{\bf \mathbb{R}}_{ij} =
 \left ( \begin{array}{rrr}
 c_\phi		& s_\phi	& 0		\\
 -s_\phi		& c_\phi	& 0		\\
 0			& 0			&1		\\ 
 \end{array} \right ) 
 \left ( \begin{array}{rrr}
 c_\theta		& 0			&  s_\theta		\\
 0			& 1			& 0	\\
 -s_\theta		& 0			& c_\theta	\\ 
 \end{array} \right ) 
 \left ( \begin{array}{rrr}
 1			& 0		& 0			\\
 0			& c_\psi	& s_\psi		\\
 0			& -s_\psi	&c_\psi		\\ 
 \end{array} \right )  =
 \left ( \begin{array}{ccc}
c_\theta c_\psi 		& -c_\phi s_\psi + s_\phi s_\theta c_\psi	& s_\phi s_\phi + c_\phi s_\theta c_\psi	\\
c_\theta s_\psi		& c_\phi c_\psi + s_\phi s_\theta s_\psi	& -s_\phi c_\psi + c_\phi s_\theta s_\psi	\\
-s_\theta			& s_\phi c_\theta					& c_\phi c_\theta					\\
 \end{array} \right ) 
\label{eqn_define_rotationmatrix_intrinsic_euler}
\end{footnotesize}
\end{equation}

In this Euler angle formalism, the rotation matrix is identical under exchange of $\alpha$ with -180$^\circ$+$\alpha$, $\beta$ with -$\beta$ and $\gamma$ with -180$^\circ$+$\gamma$. Solutions are also identical when adding multiples of 2$\pi$ to any individual component. With this Euler angle formalism, there is no normalization and no natural relation to components of linear or circular retardance as with the {\it axis-angle} formalism.

\begin{figure}[htbp]
\begin{center}
\vspace{-2mm}
\hbox{
\hspace{-1.0em}
\includegraphics[height=6.2cm, angle=0]{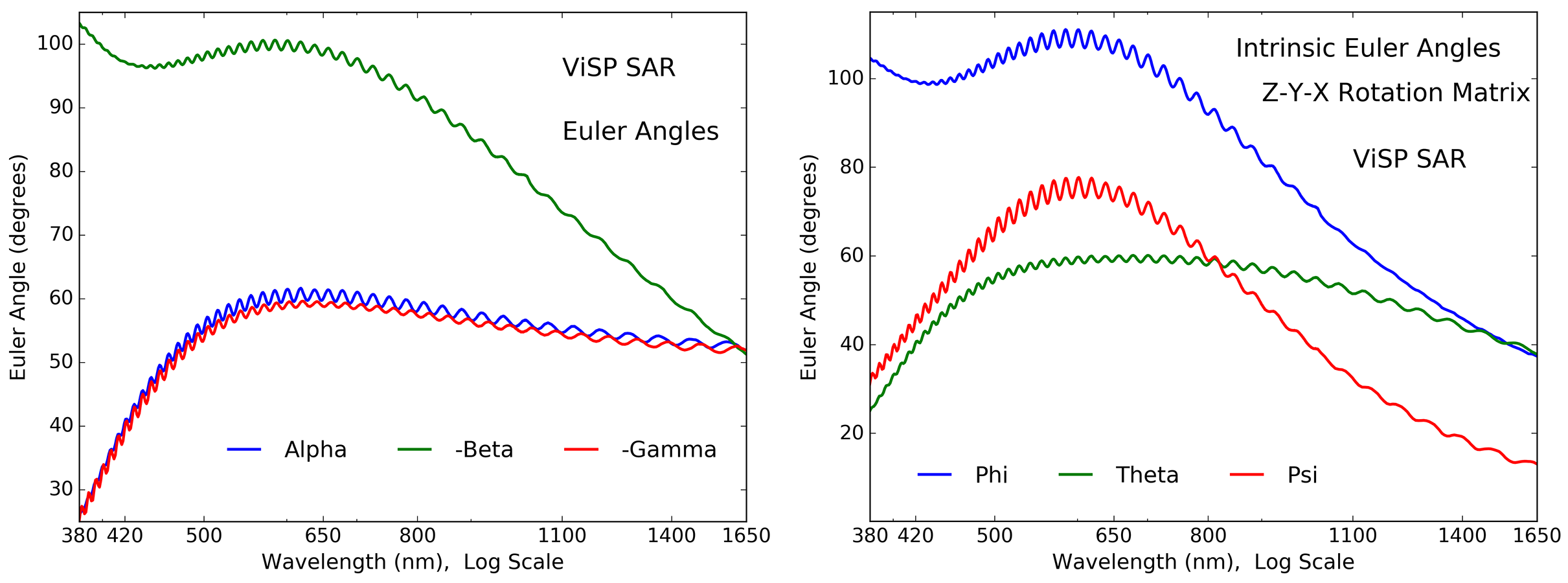}
}
\caption[] 
{\label{fig:visp_sar_euler_angle_solution} Left shows Euler angles derived in a Z-X-Z type rotation matrix fit to the NLSP measurements of the ViSP SAR. Blue shows $\alpha$, green shows -$\beta$ and red shows -$\gamma$. Right shows intrinsic Euler Angles using the Z-Y-X convention.  Blue shows $\psi$, green shows $\theta$ and red shows $\psi$ when using the Z-Y'-X'' convention.  }
\vspace{-6mm} 
\end{center}
\end{figure}

In the left hand graphic of Figure \ref{fig:visp_sar_euler_angle_solution}, we show an Euler Angle rotation matrix model fit to the NLSP measurements of the ViSP SAR data set. The blue, green and red curves show $\alpha$, -$\beta$ and $\gamma$ respectively. The clocking error introduced spectral oscillations (ripples) are immediately visible as is the wavelength dependent magnitude of the three rotation angles. 

\begin{wrapfigure}{r}{0.40\textwidth}
\centering
\vspace{-5mm}
\begin{equation}
\theta = \cos^{-1} (  [\mathbb{R}_{00} + \mathbb{R}_{11} + \mathbb{R}_{22} - 1 ] / 2 )
\label{eqn:convert_rotmat_to_axisangle}
\end{equation}
\begin{equation}
u_x = (\mathbb{R}_{32} - \mathbb{R}_{23}) / 2\sin\theta
\end{equation}
\begin{equation}
u_y = (\mathbb{R}_{13} - \mathbb{R}_{31}) / 2\sin\theta
\end{equation}
\begin{equation}
u_z = (\mathbb{R}_{21} - \mathbb{R}_{12}) / 2\sin\theta
\label{eqn:rot_vector}
\end{equation}
\vspace{-8mm}
 \end{wrapfigure}

A common rotation matrix formalism often just called the {\bf Rotation Matrix} uses an intrinsic Euler angle arrangement where rotation components are specified in Z-Y'-X'' ordering.  The angles are denoted in the (z,y,x) ordering of ($\phi,\theta,\psi$) and the {\bf Rotation Matrix} is defined as in Equation \ref{eqn_define_rotationmatrix_intrinsic_euler}. This Equation is just as easily described as a linear retarder with fast axis orientation at 0$^\circ$ followed by a linear retarder with a fast axis at 45$^\circ$ followed by a circular retarder.

With this particular intrinsic Z-Y'-X'' Euler angle based Rotation Matrix, we can relate the individual matrix elements back to the {\it axis-angle} rotation matrix formalism common in retarder Mueller matrix fitting routines.  Equations \ref{eqn:convert_rotmat_to_axisangle} through \ref{eqn:rot_vector} show the analytic solution for the magnitude of the rotation $\theta$ and the three retarder components of the {\it axis} vector ($u_x,u_y,u_z$) denoted either normalized or un-normalized as ($r_H,r_{45},r_R$) or simply (H,45,R) in Sueoka \cite{Sueoka:2016vo,Sueoka:2014cm}.

In the right hand graphic of Figure \ref{fig:visp_sar_euler_angle_solution}, we fit the ViSP SAR Mueller matrix measurements and show the intrinsic Euler angles ($\psi,\theta,\psi$) when using the {\it Rotation Matrix} convention of Z-Y'-X''. These Euler angles also oscillate but with very different wavelength dependence than the Z-X-Z convention of Figure \ref{fig:visp_sar_euler_angle_solution}. When using Equations \ref{eqn:convert_rotmat_to_axisangle} through \ref{eqn:rot_vector} to compare the {\it axis-angle} formalism to the Z-Y'-X'' version of the Euler angle rotation matrix, we completely reproduce the correct results.  The NLSP Mueller matrix measurements are equally well fit within numerical settings of the {\it axis-angle} formalism. Both rotation matrix representations of Figure \ref{fig:visp_sar_euler_angle_solution} are equally good reproductions of the Mueller matrix, as are the axis-angle rotation terms presented above.

\clearpage

\section{Spatial \& Spectral Retardance of DKIST Retarders}
\label{sec:appendix_spatial_maps}

In this appendix, we analyze the spatial maps of the retardance across the rest of the DKIST calibration and modulation optics not described in the main text.   There is significant spatial variation between retarders of the same materials, and between retarders different materials.  Spatial distributions of retardance across each optic are unique and are presented here for completeness. We also show that 90\% of the retardance magnitude spatial variation occurs at spatial scales larger than 10 mm for the optics.

\subsection{Spatial Retardance Variation for a MgF$_2$ Retarder: The Cryo-NIRSP SAR}
\label{sec:sub_cryoSAR_spatial}

\begin{figure}[htbp]
\begin{center}
\vspace{-0mm}
\hbox{
\includegraphics[height=6.5cm, angle=0]{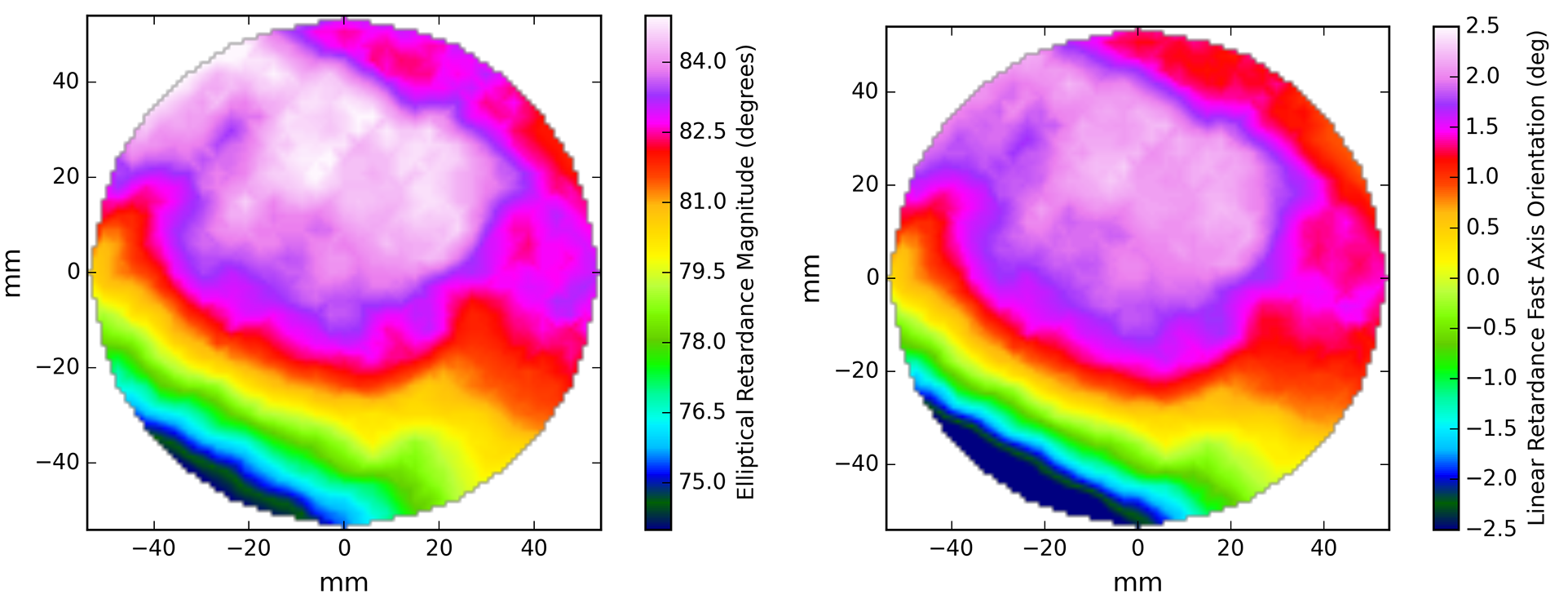}
}
\caption[] 
{\label{fig:mlo_map_cryo_sar_599nm}  The spatial measurements of retardance properties for the Cryo-NIRSP SAR as-built.  The left graphics show maps measured at 598.9 nm using a $\sim$10 nm bandpass narrow-band filter and spatial sampling of $\sim$3 mm. The part was nominally designed as a $\sim$90$^\circ$ linear retarder. The left panel shows the spatial map of elliptical retardance magnitude with a color scale varying from 74$^\circ$ to 85$^\circ$. The right panel shows the fast axis of linear retardance with all points were within a range of $\pm$2.5$^\circ$.}
\vspace{-6mm}
\end{center}
\end{figure}

The Cryo-NIRSP superchromatic calibration retarder (SAR) was designed to be a quarter-wave linear retarder for wavelengths longer than 2500 nm. We also plan to calibrate the visible and near infrared instrumentation with this optic given our recent thermal modeling, fringe amplitude estimates and plans for calibrating the DKIST primary and secondary mirror \cite{Harrington:2017jh, Harrington:2018cx}. 

\begin{wrapfigure}{l}{0.52\textwidth}
\centering
\vspace{-5mm}
\begin{tabular}{c} 
\hbox{
\hspace{-1.2em}
\includegraphics[height=5.8cm, angle=0]{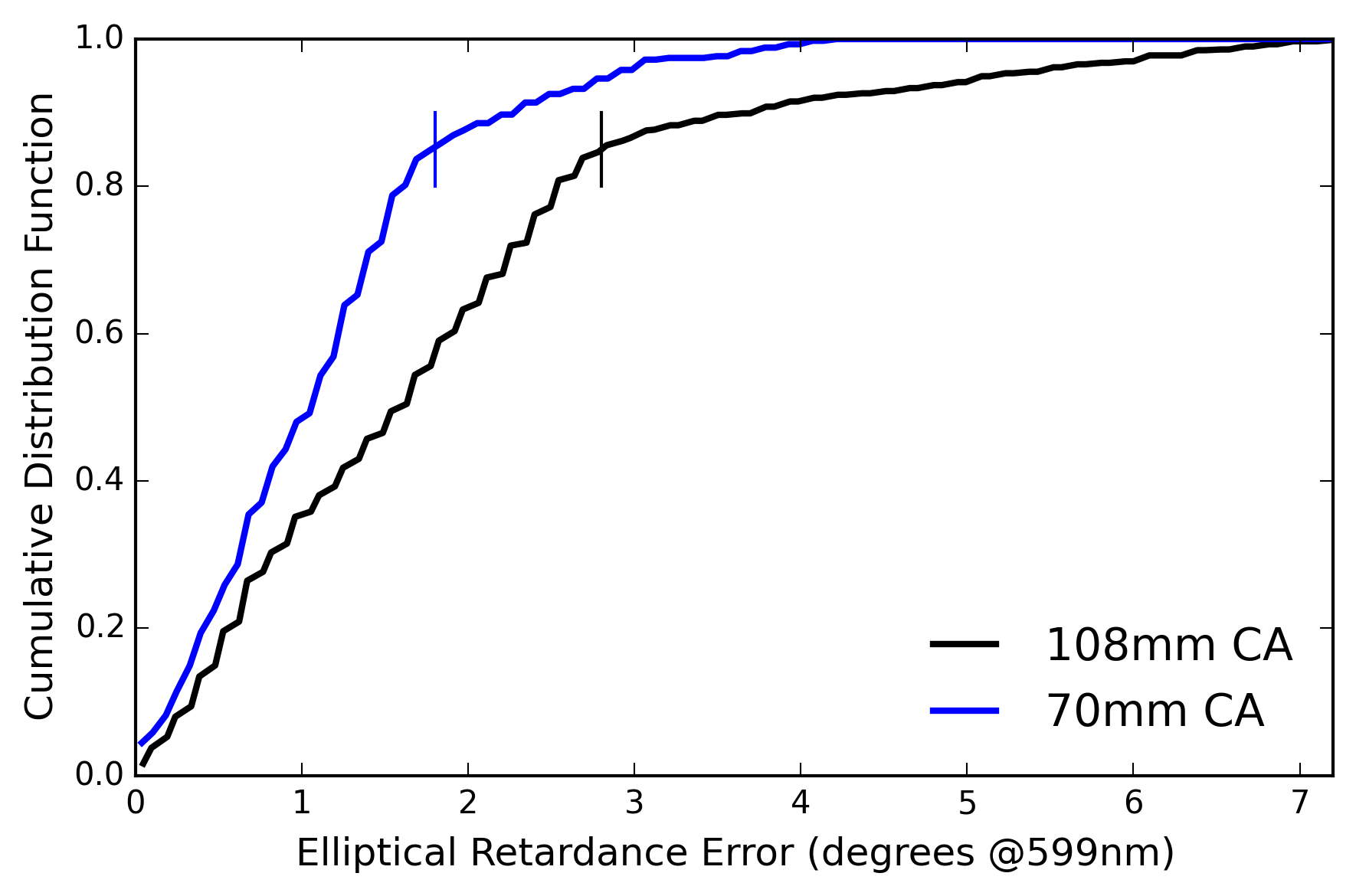}
}
\end{tabular}
\caption[] 
{\label{fig:cryo_as_built_cdf_599nm}  The cumulative distribution of the retardance magnitude spatial variation from average for the as-built Cryo-NIRSP SAR. See text for details.}
\vspace{-6mm}
 \end{wrapfigure}

This optic can be used in the full 300 Watt DKIST beam, without any polarizers or heat rejection filters with minimal influence of absorption on the temperature of the optic. With the benefits of this optics greatly reduced heat loads (NIR transparancy), thermal stability and reduced fringes we examine in this subsection how spatial retardance errors will restrict the field of view for calibrations with this optic much more strongly for these visible wavelength use cases.   

This retarder uses MgF$_2$ crystals with thickness of (2273.14$\mu$m and 2153.11$\mu$m) for A pairs and (2333.21$\mu$m and 2153.11$\mu$m) for B pairs. These correspond to net retardance of 40 waves per plate. The A-B-A pairs are polished to net retardances of 2.230, 3.346, 2.230 waves retardance at 633.443 nm wavelength. The orientations for the crystal pairs are (0$^\circ$, 107.75$^\circ$, 0$^\circ$). Our as-built measurements used 3 mm spatial sampling with a 3 mm diameter probe beam footprint. We covered a 108 mm diameter aperture with 1009 separate measurements in a grid. The wavelength was controlled using a 598.9 nm filter with a 10 nm FWHM band pass.  

Figure \ref{fig:mlo_map_cryo_sar_599nm} shows the elliptical retardance magnitude and linear retardance fast axis orientation spatial maps. The elliptical retardance magnitude varies from under 74$^\circ$ to over 85$^\circ$ giving 11$^\circ$ peak-to-peak spatial variation. The linear retardance fast axis also rotates by over 5$^\circ$ peak-to-peak as seen by the right graphic of Figure \ref{fig:mlo_map_cryo_sar_599nm}. The spatial variation is reasonably smooth over small spatial scales. No abrupt variations between individual spatial measurements is seen.  The edges of the clear aperture have stronger deviations as expected for polishing errors. 

The corresponding cumulative distribution of errors is seen in Figure \ref{fig:cryo_as_built_cdf_599nm}.  We show two clear apertures to demonstrate how the inner part of the optic is more uniform.  The blue curve shows the CDF for a 70 mm clear aperture with 80\% of the measurements within 1.2$^\circ$ retardance from average and all spatial points within 4.2$^\circ$.  The black curve shows a 108 mm clear aperture. All spatial measurements are within 11$^\circ$ of the average but this distribution has a very significant tail corresponding to the outer edge of the aperture seen in Figure \ref{fig:mlo_map_cryo_sar_599nm}.  For the 108 mm clear aperture, 80\% of those points are less than 2.8$^\circ$ retardance error, but the lower left part of the aperture shows a very strong deviation from average. 

We performed a Fourier analysis of the retardance spatial variation to assess the spatial distribution of errors.  Both 1 and 2 dimensional FFT power spectra show that 90\% of the spatial variation is contained with spatial frequencies larger than 10 mm.  For our laboratory acceptance testing as well as system modeling, this shows that we can effectively sample at significantly more coarse spatial resolution and still capture the retardance spatial variation behavior. 

\begin{wrapfigure}{r}{0.55\textwidth}
\centering
\vspace{-0mm}
\begin{tabular}{c} 
\hbox{
\hspace{-1.0em}
\includegraphics[height=7.2cm, angle=0]{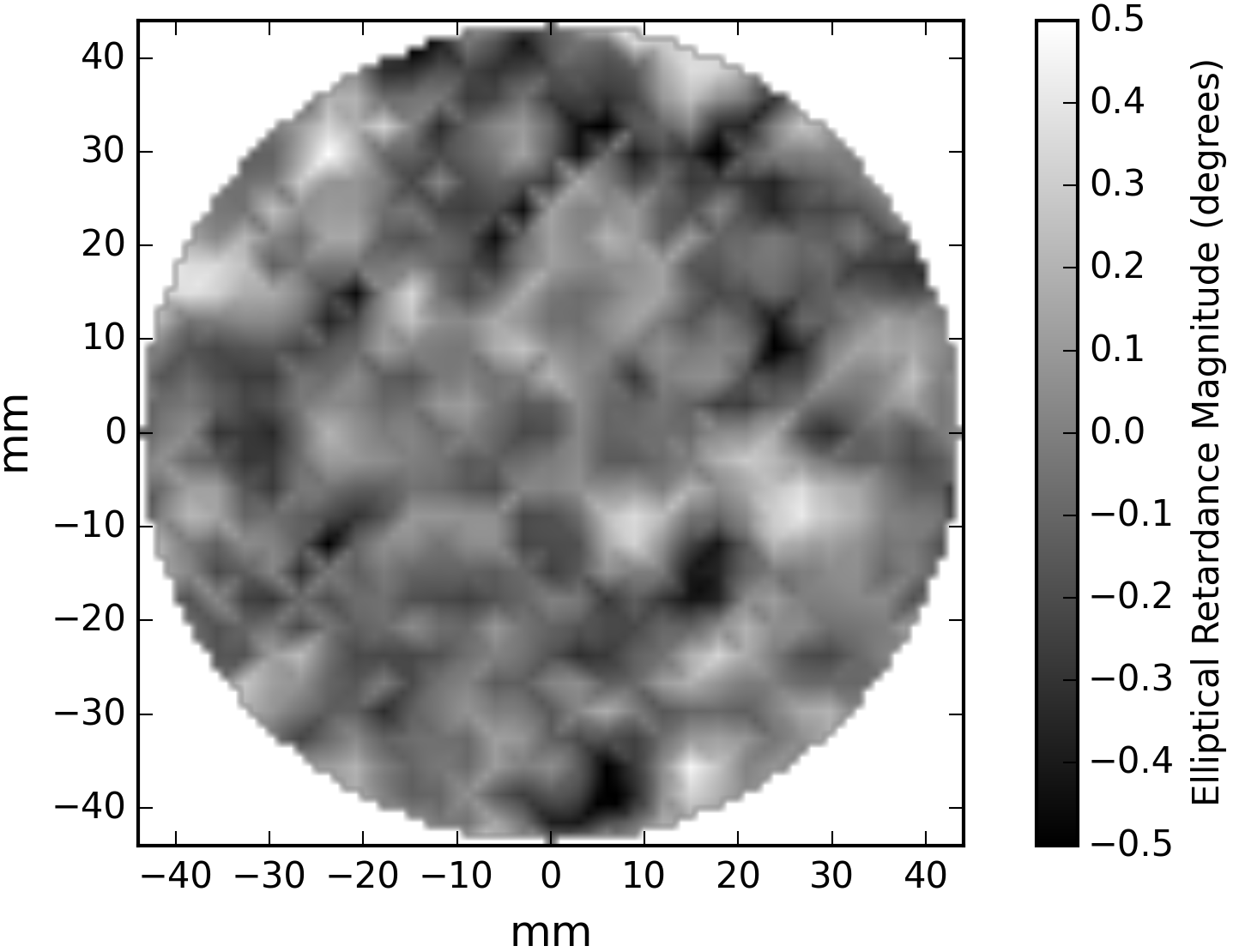}
}
\end{tabular}
\caption[] 
{\label{fig:cryo_as_built_599nm_spatial_variation_difference}  The spatial variation difference between retardance maps constructed at 3 mm spatial resolution and 8 mm spatial resolution. The linear grey scale covers $\pm$0.5$^\circ$ while the magnitude measured was $\pm$7.5$^\circ$ in Figure \ref{fig:cryo_as_built_cdf_599nm}. }
\vspace{-5mm}
 \end{wrapfigure}

Figure \ref{fig:cryo_as_built_599nm_spatial_variation_difference} shows the difference between 3 mm spatial sampling and 8 mm spatial sampling on a linear grey scale. The difference between maps is seen as higher spatial frequency noise at magnitudes of less than $\pm$0.5$^\circ$ degrees peak to peak variation though the retardance varied across the optic by 11$^\circ$ peak to peak as seen in the high spatial resolution map of Figure \ref{fig:mlo_map_cryo_sar_599nm}.  The RMS variation across the aperture of Figure \ref{fig:cryo_as_built_599nm_spatial_variation_difference} is 0.14$^\circ$.  Thus we capture 11$^\circ$ spatial variation with an RMS variation 100x smaller when using 8 mm sampling.

\begin{figure}[htbp]
\begin{center}
\vspace{-0mm}
\hbox{
\includegraphics[height=12.2cm, angle=0]{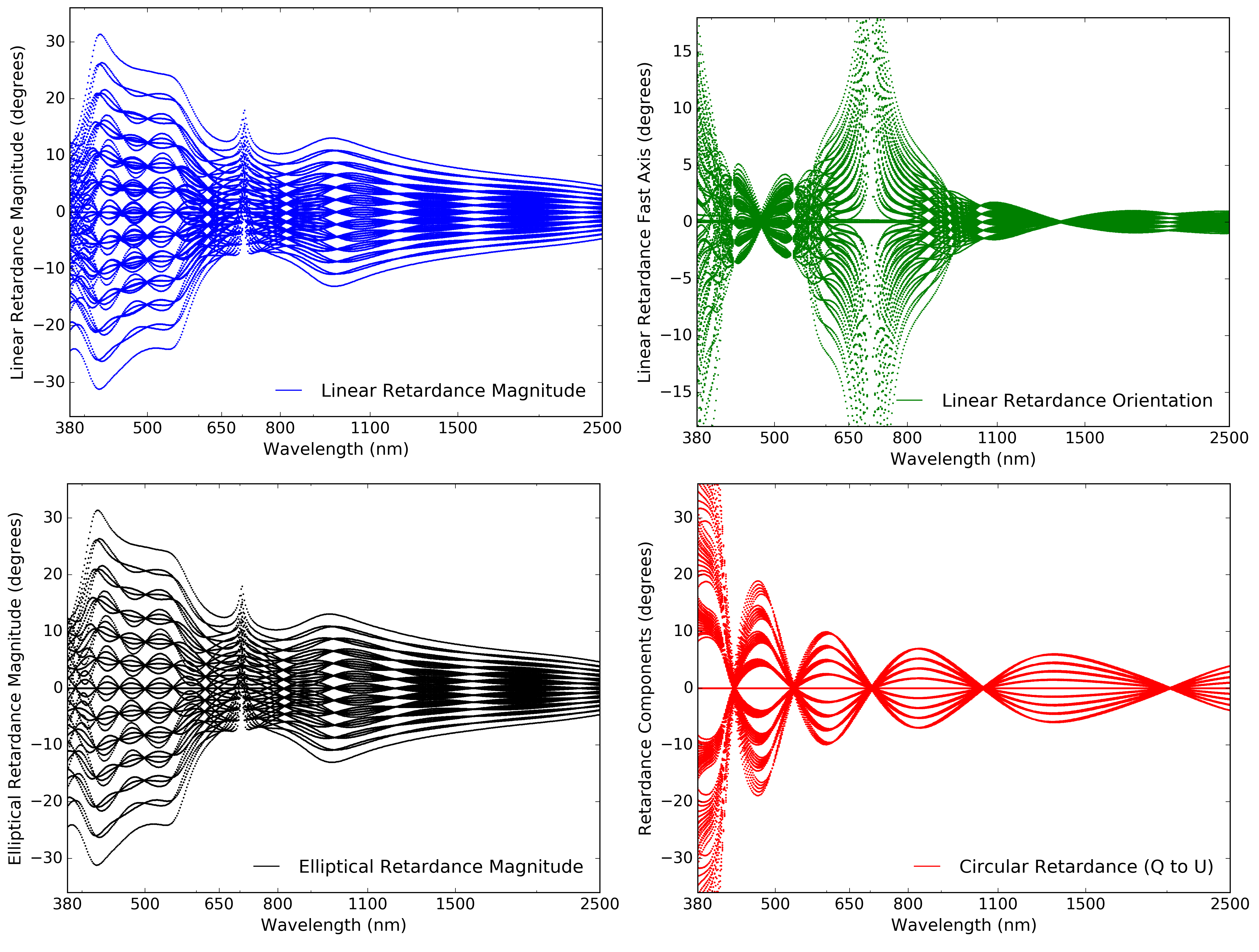}
}
\caption[] 
{\label{fig:polish_residual_ER_components_cryoSAR} The spectral variation of elliptical retardance errors in the polishing simulations for the Cryo-NIRSP SAR. The polishing error of 0.01 waves retardance was optionally placed on each of the six crystals and an elliptical retarder model fit to the resulting optic Mueller matrix. The blue curves at top left show the linear retardance magnitude error.  The green curves in the top right show the fast axis of linear retardance with a discontinuity near 700 nm wavelength where the linear retardance goes to zero and the inverse tangent of small errors is greatly amplified. The circular retardance error is shown at bottom right with several nulls. The elliptical magnitude error is shown in black at bottom left with close resemblance to the linear retardance magnitude except at the shortest wavelengths.}
\vspace{-8mm}
\end{center}
\end{figure}

We simulated the impact of polishing errors by fitting elliptical retardance to a large grid of toleranced models. We put errors of either none or $\pm$0.01 waves of retardance error at 633 nm wavelength on all six of the crystals. This gave us 729 individual models (3 choices for each of the 6 crystals gives 3$^6$ models). To highlight the residual errors, we simply difference the fit elliptical retardance parameters from the nominal design. The residual errors can be over 30$^\circ$ retardance for some components for a worst-case stack up of errors. The polishing simulations shown in Figure \ref{fig:polish_residual_ER_components_cryoSAR} illustrate how this design has varying sensitivity to circular retardance from polishing errors. The top left graphic shows the linear retardance magnitude computed as the root square sum of the two linear components. The green curves in the top right shows the fast axis orientation of linear retardance as the inverse tangent of the two linear components. The red curves shown at lower right are circular retardance error with generally increasing sensitivity at shorter wavelengths combined with several sensitivity nulls.  The elliptical retardance at lower left closely tracks the linear retardance magnitude errors.  The issue in the degeneracy of the axis-angle rotation formalism is easily seen where the part reaches one full wave net retardance around 700 nm wavelength. A full wave of rotation gives identical results regardless of the axis of rotation.

\begin{figure}[htbp]
\begin{center}
\vspace{-2mm}
\hbox{
\hspace{-0.0em}
\includegraphics[height=10.5cm, angle=0]{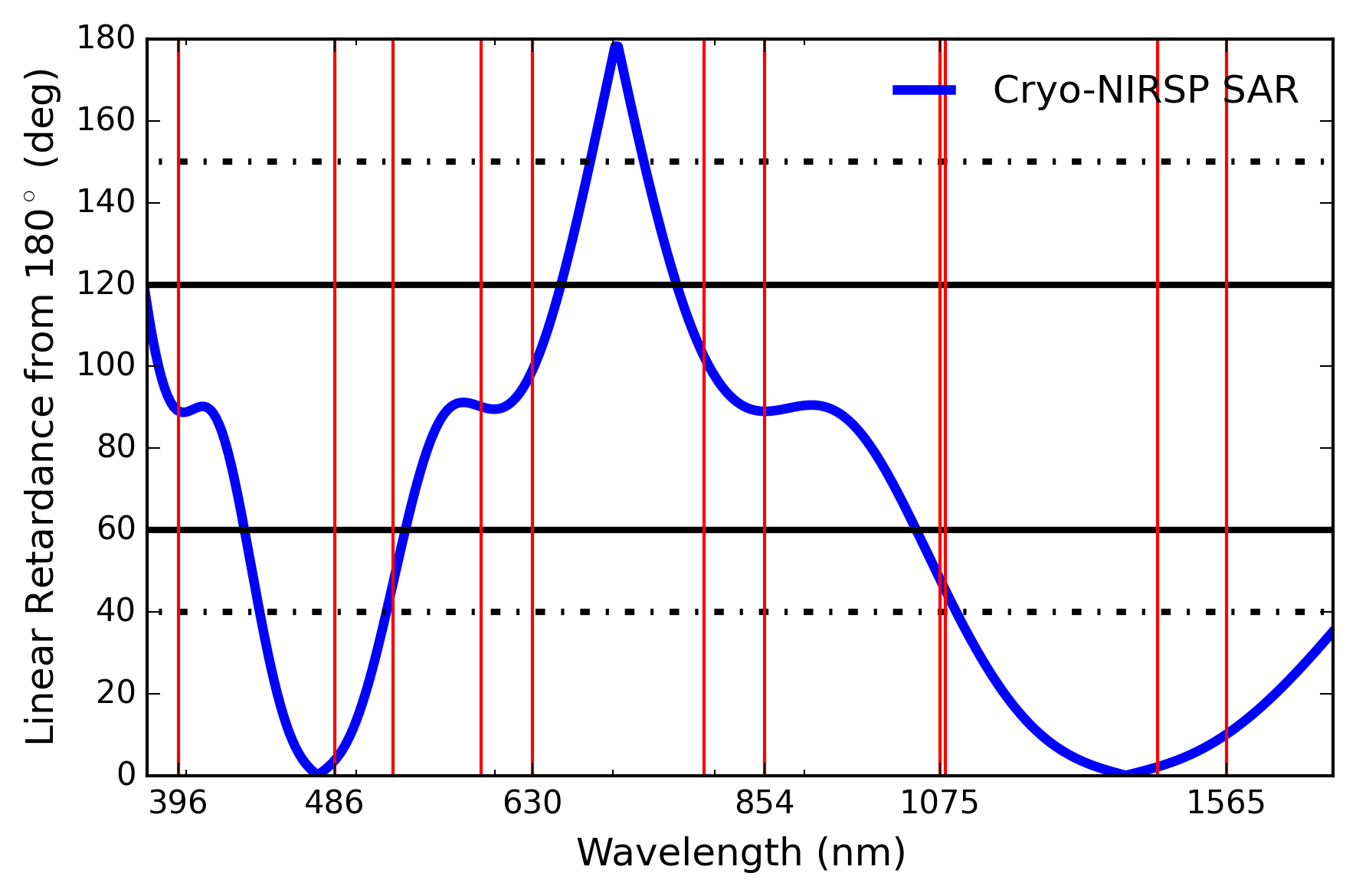}
}
\caption[Cryo SAR Usefulness] 
{ \label{fig:cryo_sar_linear_retardance_limits} 
The linear retardance difference of the Cryo-NIRSP MgF$_2$ calibration optic from 180$^\circ$. Thick black lines show the traditional 90$\pm$30$^\circ$ calibration retardance. Dashed horizontal black lines show the range 40$^\circ$ to 150$^\circ$ where calibration is reasonably feasible for DKIST. Red vertical lines show expected wavelengths for DKIST instrumentation.}
\vspace{-8mm}
\end{center}
\end{figure}

Many authors have shown more efficient calibration and modulation schemes based on retarders that are either elliptical or values that optimize the condition number of the demodulation matrix given an observing set \cite{1990JOSAA...7..693G,Kupinski:2014ek,2012OptL...37.1097L, LaCasse:2011kv, Twietmeyer:2005bz, delToroIniesta:2000cg, Tomczyk:2010wta,Wijn:2011wt,deWijn:2010fh,2010SPIE.7735E..4AD, Compain:1999do, Harrington:2010km}. We can easily implement similar calibration schemes with our retarders when they do not happen to be the traditional quarter-wave linear retarder. As an example of the utility for this MgF$_2$ retarder for on-axis calibration with strong thermal behavior benefits, we show the linear retardance magnitude away from half-wave in Figure \ref{fig:cryo_sar_linear_retardance_limits}. The retardance magnitude has been restricted to be within the interval 0$^\circ$ to 180$^\circ$ and has had half a wave subtracted from the theoretical curve. 

Typical calibration and modulation schemes, when using linear retardance, often find reasonable efficiency when the retardance is more than 30$^\circ$ away from half-wave integer multiples. This particular MgF$_2$ retarder has strongly varying retardance in the 380 nm to 2500 nm wavelength range, as seen in Figure \ref{fig:cryo_sar_linear_retardance_limits}. However, the retardance is near some multiple of 1/4 to 1/3 wave at many wavelengths throughout the bandpass. DKIST instruments work at specific narrow bandpasses configured on a daily to monthly basis.

As examples, Figure \ref{fig:cryo_sar_linear_retardance_limits} shows vertical red lines corresponding to typical solar observation wavelengths of 396 nm, 486 nm, 525 nm, 589 nm, 630 nm, 789 nm, 854 nm, 1079 nm, 1083 nm, 1430 nm and 1565 nm.  These lines correspond to atomic and molecular transitions of various constituents of the solar atmosphere useful for observation. The DKIST instruments often would be configured with the dichroic beam splitters to observe multiple simultaneous wavelengths. An example would be to configure wavelengths of 396 nm, 615 nm and 630 nm with the three cameras in ViSP, 854 nm with VTF and the 1083 nm 1565 nm pair with two of the three cameras in DL-NIRSP. For this particular configuration, the retardance would likely be sufficient to calibrate all but the 1565 nm camera using a single optic simultaneously. The optic is near 0.5 waves retardance at 1400 nm wavelength, at 1.0 waves around 700 nm and near 1.5 waves near 470 nm wavelength. The thermal properties of this retarder are strongly advantageous to SiO$_2$ based optics due to the greatly reduced absorption \cite{Harrington:2018cx}. The oil layers between crystals also matches refractive indices much better significantly reducing spectral fringes. 

This optic can be very useful in calibration with a reduced field of view to avoid calibration issues caused by polishing error.  In particular, the DKIST Visible Tunable Filter (VTF) requires fringe stability for the duration of calibration. The VTF is a Fabry-Perot (FP) imaging instrument which steps through wavelengths centered around specific spectral lines of interest. The four main lines are 525 nm, 630 nm, 656 nm and 854 nm with the instrument having a $\sim$1 nm free spectral range about these lines.  The instrument can sample in wavelength steps of roughly 3 picometers and has a spectral resolving power of R=$\lambda$/$\delta\lambda$ = 100,000.  The Cryo-NIRSP calibration retarder magnitude shown in Figure \ref{fig:cryo_sar_linear_retardance_limits} shows VTF can very efficiently be calibrated with this optic at all four primary wavelengths, even though the optic is many multiples of quarter-wave retardance. This  instrument cannot as easily spectrally filter fringes as spectrograph instruments typically do given the different timescales for completing a spatial / spectral data set. Not only does VTF need reduced fringe amplitudes, but it also needs stable retardance fringes for effective removal.  This optic, at reduced FoV, may provide such a stable calibration.

\begin{wrapfigure}{r}{0.59\textwidth}
\centering
\vspace{-0mm}
\begin{tabular}{c} 
\hbox{
\hspace{-1.2em}
\includegraphics[height=8.0cm, angle=0]{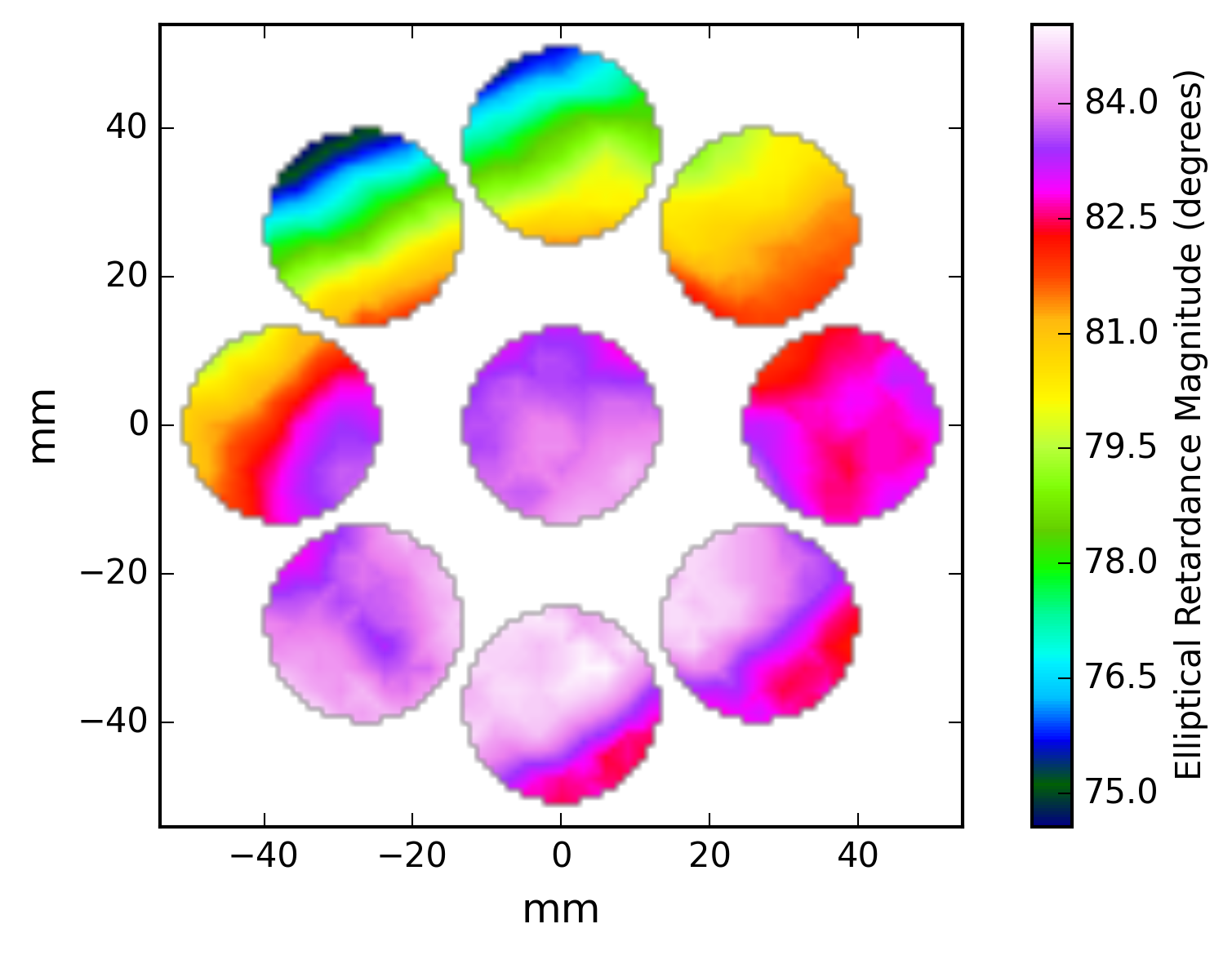}
}
\end{tabular}
\caption[] 
{ \label{fig:footprint_CRYOsar_AsBuilt_600nm} 
The spatial retardance variation within footprints of the beam on the Cryo-NIRSP SAR during an eight-orientation calibration sequence at 598.9 nm wavelength.}
\vspace{-4mm}
\end{wrapfigure}

Simulation of the impact to calibration can be done for this optic the same as for the ViSP SAR shown in the main text. This optic has significantly greater spatial variation with a spatial pattern that leads to significant differences in average retardance as a function of field angle. Figure \ref{fig:footprint_CRYOsar_AsBuilt_600nm} shows the spatial variation of retardance for the eight footprints on the Cryo-NIRSP calibration retarder at 598.9 nm wavelength as this optic rotates during a typical calibration sequence. The center footprint remains constant for the on-axis beam and would see the same average retardance for all retarder orientations. For the footprint corresponding to the 5 arc-minute field edge, clear spatial separation is seen between footprints along with significant retardance variation between steps. 

\begin{wrapfigure}{l}{0.56\textwidth}
\centering
\vspace{-3mm}
\begin{tabular}{c} 
\hbox{
\hspace{-1.1em}
\includegraphics[height=6.3cm, angle=0]{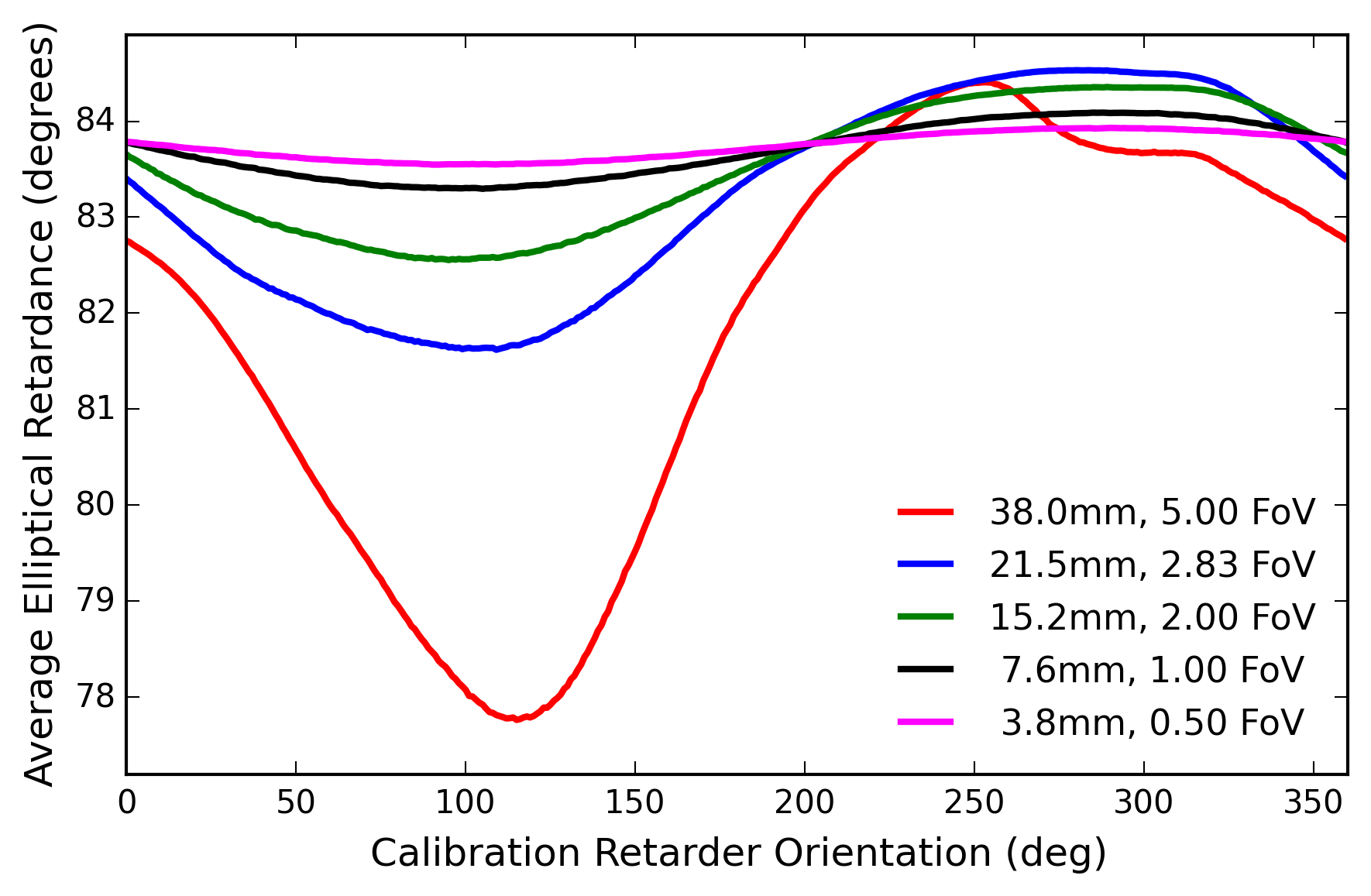}
}
\end{tabular}
\caption[] 
{ \label{fig:spatial_retardance_variation_CRYOsar_AsBuilt_600nm} 
The retardance spatial variation computed as an average over the footprint as the Cryo-NIRSP SAR spins during a calibration sequence at 598.9 nm wavelength.}
\vspace{-5mm}
\end{wrapfigure}

We quantify the retardance variability during calibration by taking spatial averages over the footprint for varying FoV.  Figure \ref{fig:spatial_retardance_variation_CRYOsar_AsBuilt_600nm} shows an example of the average retardance across a footprint as the calibration retarder rotates. The various curves correspond to select field angles using the Cryo-NIRSP calibration retarder at 600 nm wavelength. Each individual footprint is 26.6mm at this optical location in the converging F/ 13 beam. Magenta shows a very small field of view at 0.5 arc minutes diameter with a beam decenter of 3.8 mm and 0.5$^\circ$ peak to peak spatial variation.  Black shows a 1 arc minute field of view corresponding to a beam center 7.6 mm away from the retarder center and 1$^\circ$ peak to peak spatial variation.  Green shows a 2.0 arc minute field of view with 15.2 mm of decenter.  Blue shows the edge of the 2.83 arc minute field of view and a 21.5 mm beam decenter.   Red corresponds to the maximal DKIST field of 5.0 arc minutes and a 38.0 mm beam decenter. 

For this MgF$_2$ retarder, the variation of retardance as the optic rotates is roughly 2$^\circ$ peak-to-peak for a 2 arc minute field of view at 15.2 mm radius, the green curve in Figure \ref{fig:spatial_retardance_variation_CRYOsar_AsBuilt_600nm}.  For larger field angles however, the retardance variation with optical orientation rapidly increases to 7$^\circ$ peak to peak at the 5 arc minute field edge. In addition, the red curve is not symmetric about the average value. The net retardance of the optic would be considered different as a function of effective field angle, complicating any calibration technique that assumes constant retardance. With these curves we can conclude that field dependent calibrations are required for high accuracy.  We also find that the field angle for which we can ignore this spatial dependence is only for the inner half arc minute, at magnitudes below 0.5$^\circ$ retardance.

\clearpage
\subsection{Measuring a MgF$_2$ Elliptical Retarder: The Cryo-NIRSP Modulator}
\label{sec:appendix_spatial_maps_cryo_pcm}

\begin{wraptable}{l}{0.60\textwidth}
\vspace{-2mm}
\caption{Measured Cryo-NIRSP PCM Crystal Properties}
\label{table:CNpcm}
\centering
\begin{tabular}{l | l l | l l | l l | l}
\hline \hline
N		& Des.	& Meas	& Ret		& Meas	& Des		& Meas			& Ori				\\
		& mm	& mm	& Bias		& Waves	& Pair		&$\delta\pm$0.01	& $\pm0.3^\circ$	\\
\hline \hline
2E$_s$	& 2.26	& 2.302	& 42$\pm$1	& 42.776	& 			& 				& 0$^\circ$		\\
2E$_b$	& 2.15	& 2.200	& 40$\pm$1	& 40.885	& 1.893		& 1.8873			& 90$^\circ$		\\
\hline
3F$_s$	& 2.22	& 2.234	& 41$\pm$1	& 41.514	& 			& 				& 71.86$^\circ$	\\
3F$_b$	& 2.15	& 2.165	& 40$\pm$1	& 40.235	& 1.282		& 1.2921			& 161.86$^\circ$	\\
\hline
4E$_s$	& 2.26	& 2.297	& 42$\pm$1	& 42.689	& 			& 				& 30.39$^\circ$		\\
4E$_b$	& 2.15	& 2.196	& 40$\pm$1	& 40.803	& 1.893		& 1.8841			& 120.39$^\circ$		\\
\hline \hline		
\end{tabular}
\vspace{-2mm}
\end{wraptable}

The Cryo-NIRSP PCM is designed to be an efficient modulating retarder rotating in front of an ideal Stokes Q analyzer for wavelengths between 1000 nm and 5000 nm. The efficiency is high and balanced across QUV when sampled with 5 or more evenly spaced images in a 180$^\circ$ rotation. We measured the spatial variation of elliptical retardance magnitude for this optic at 600 nm wavelength. The design predicts a 0.3529 wave linear retarder (127$^\circ$) with no circular retardance component present in the design.  Table \ref{table:CNpcm} shows the design and Meadowlark metrology of the components used to assemble the optic. The E-F-E style retarder mirrors the Cryo-NIRSP SAR but without the requirement that E pairs be parallel. 

\begin{figure}[htbp]
\begin{center}
\vspace{-0mm}
\hbox{
\includegraphics[height=6.0cm, angle=0]{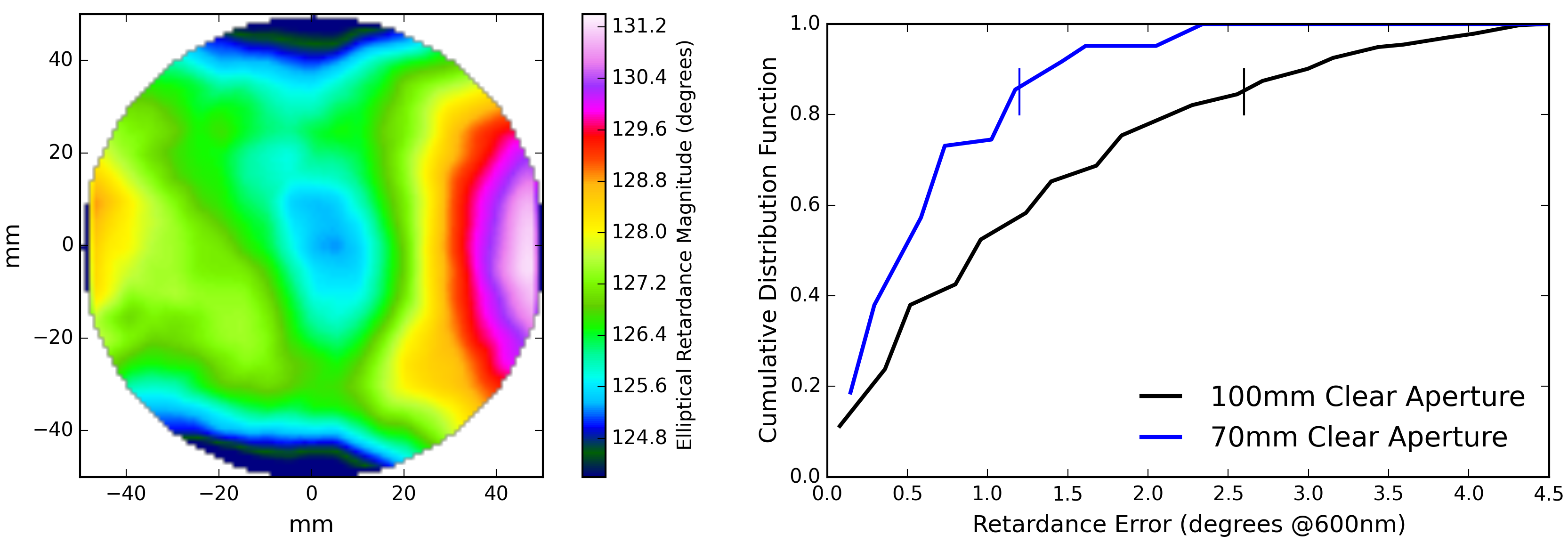}
}
\caption[] 
{\label{fig:mlo_map_cryo_pcm_600nm} The spatial measurements of retardance for the Cryo-NIRSP modulator after final mounting in the cell.  A map was measured at 598.9 nm wavelength and spatial sampling of 5 mm out to 100 mm diameter clear aperture. The left panel shows the spatial map of elliptical retardance magnitude. The right panel shows the cumulative distribution function of retardance errors from the spatial average.  A larger 100 mm clear aperture sees 80\% of the aperture within 2.8$^\circ$ retardance of nominal while a restricted 70 mm clear aperture sees 80\% of the aperture within 1.3$^\circ$ of nominal.}
\vspace{-6mm}
\end{center}
\end{figure}

Measurements at MLO used 5 mm spatial sampling and a 3 mm beam footprint covering roughly a 90 mm area with 374 samples. Wavelength was controlled using a 598.9 nm interference filter with a 10 nm FWHM bandpass.  Figure \ref{fig:mlo_map_cryo_pcm_600nm} shows the elliptical retardance magnitude map on the left with a scale running from 124.7$^\circ$ to 131.2$^\circ$. This map was one of our first before the expanded capability to also output linear retardance fast axis was available. The spatial patter is very roughly saddle like with low values running in a line from top to bottom of the clear aperture and high values on the left and right aperture edges. The spatial variation is reasonably smooth over small spatial scales. No abrupt variations between individual spatial measurements is seen. The edges of the clear aperture have stronger deviations than the middle.

The resulting spatial error cumulative distribution functions (CDFs) are shown on the right of Figure \ref{fig:mlo_map_cryo_pcm_600nm}. We show two clear apertures to demonstrate how the inner part of the optic is more uniform. The blue curve shows the CDF for a 70 mm clear aperture with 80\% of the measurements within 1.2$^\circ$ retardance from average and all spatial points within 2.5$^\circ$ error. The black curve shows a 90 mm clear aperture. All spatial measurements are within 4.5$^\circ$ of the average and 80\% of those points are less than 2.6$^\circ$ retardance error. 

\begin{wrapfigure}{l}{0.55\textwidth}
\centering
\vspace{-2mm}
\begin{tabular}{c} 
\hbox{
\hspace{-1.0em}
\includegraphics[height=6.8cm, angle=0]{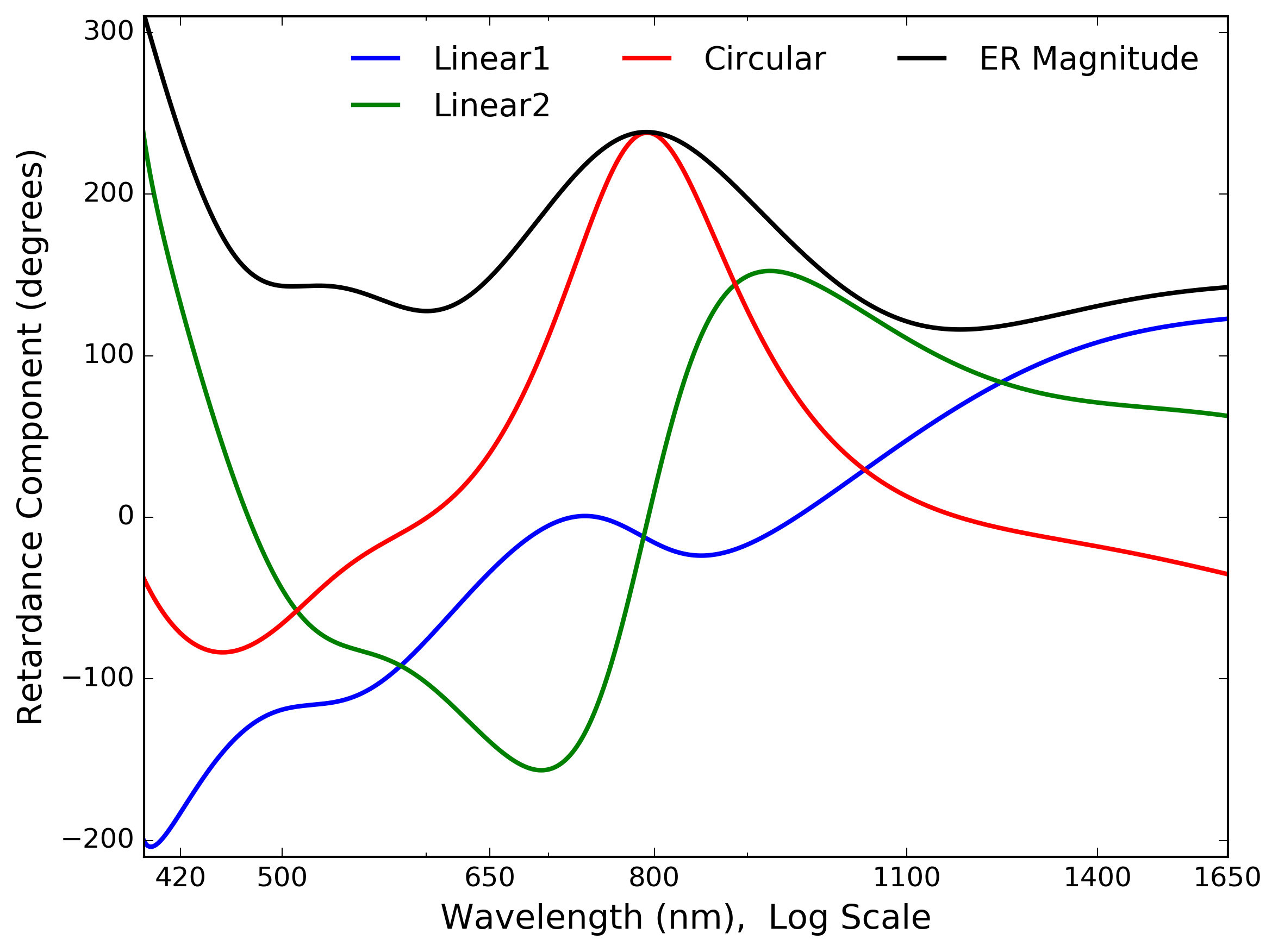}
}
\end{tabular}
\caption[] 
{ \label{fig:cryoPCM_elliptical_retarder_model_fits} 
The elliptical retarder model fit to the NLSP measurements of the Cryo-NIRSP modulator Mueller matrix at zero incidence angle.}
\vspace{-6mm}
 \end{wrapfigure}

This retarder has significantly smaller clocking errors than the ViSP SAR although this data set was recorded with significantly lower spectral resolving power. Figure \ref{fig:cryoPCM_elliptical_retarder_model_fits} shows the elliptical retarder model fits to our NLSP measurements of the optic Mueller matrix. The curves look incredibly smooth as the signal to noise ratio of NLSP data is $>$ 10,000 for all wavelengths.

We simulated errors with amplitudes of the contracted 0.01 waves retardance at 633 nm wavelength. This corresponds to a physical thickness error of 0.54 $\mu$m on each the E-F-E crystal pairs. Given that we apply $\pm$0.02 waves retardance error or zero error to each crystal against every other crystal, we get 3$^6$ models of varying behavior. We found generally increasing sensitivity at shorter wavelengths with notable nulls around 400 nm and 800 nm wavelength.   At these wavelengths, the retarder has zero linear retardance component and is entirely a circular retarder.

\begin{wrapfigure}{r}{0.61\textwidth}
\centering
\vspace{-3mm}
\begin{tabular}{c} 
\hbox{
\hspace{-1.2em}
\includegraphics[height=8.4cm, angle=0]{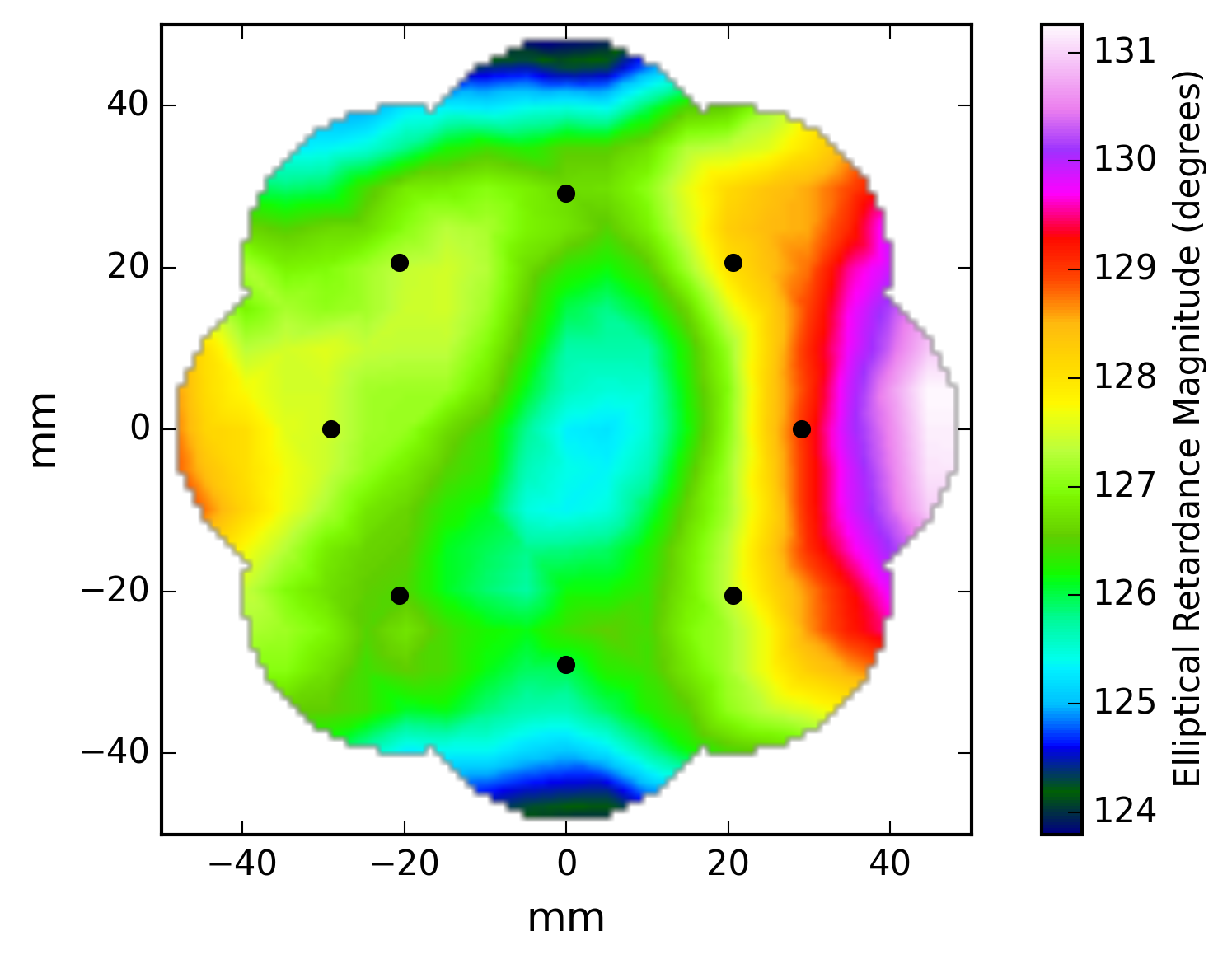}
}
\end{tabular}
\caption[] 
{ \label{fig:footprint_CryoPCM_600nm_field} The spatial retardance variation within footprints of a single 1.4 arc minute field angle beam on the Cryo-NIRSP modulator during an eight-orientation modulation sequence at 600 nm wavelength. Black dots show the footprint centers. A beam de-center of 29.1 mm is used corresponding to a 1.4 arc minute radius field angle. }
\vspace{-5mm}
\end{wrapfigure}

For the Cryo-NIRSP modulator, the footprints are substantially bigger and average over more of the polishing non-uniformity. The optic is located 631 mm upstream of the F/ 18 focal plane on the spectrograph entrance slit. An individual beam illuminates a 39.2 mm circular footprint on the retarder. Figure \ref{fig:footprint_CryoPCM_600nm_field} shows the footprints corresponding to the edge of the modulator clear aperture at a field angle of 1.4 arc minutes as well as the footprint at field center.

With an effective focal length of 72,000 mm, the beam de-center is 31.1 mm at the edge of the un-vignetted field at 1.5 arc minute radius. Within a $\pm$1 arc minute radius field, we illuminate $\pm$40.4 mm of the optic with 20.8 mm of decenter and a 19.6 mm radius footprint. Even for the largest un-vignetted field angles with Cryo-NIRSP, the beams overlap.  For this optic, spatial averaging is most significant for all DKIST retarders. This reduces the variation footprint-to-footprint but increases the variance across any individual footprint. For this inner arc minute we computed field of view variation from the mirror induced polarization in Zemax following the procedures in Harrington \& Sueoka 2017 \cite{Harrington:2017dj}. The mirrors introduce some field dependence to the telescope Mueller matrix model, but at levels significantly below one degree retardance.

\begin{figure}[htbp]
\begin{center}
\vspace{-0mm}
\hbox{
\includegraphics[height=5.3cm, angle=0]{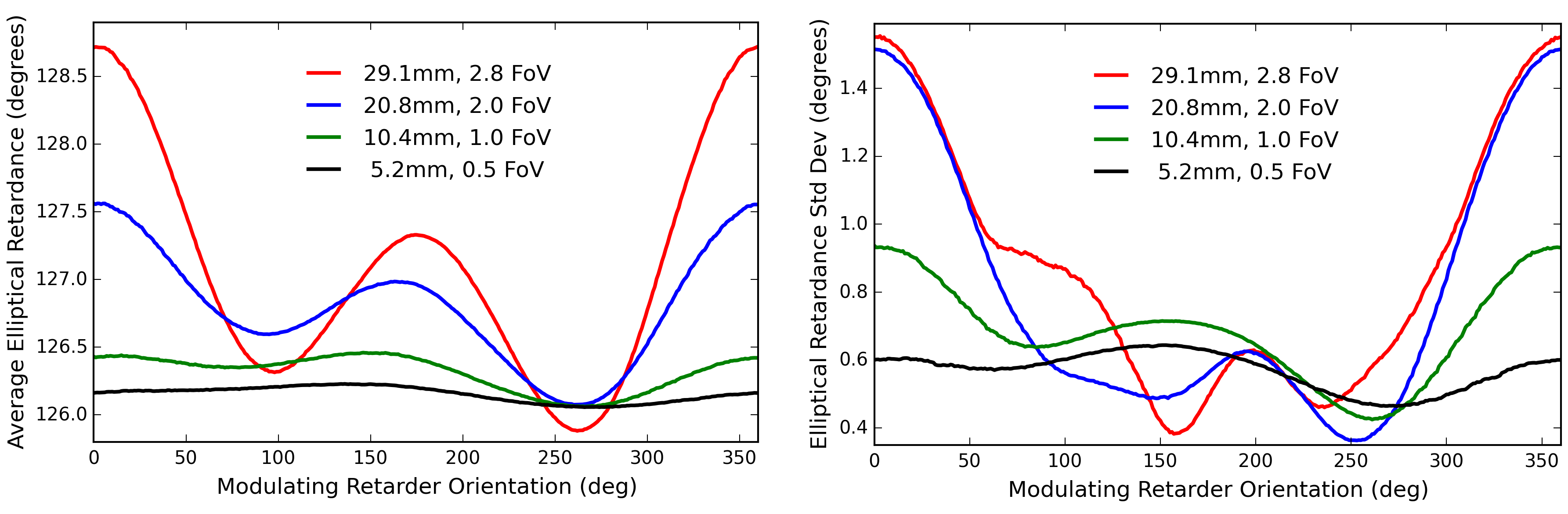}
}
\caption[] 
{\label{fig:footprint_CryoPCM_600nm_field_average_and_variance} Cryo-NIRSP Modulator: The average retardance in a footprint is shown at left for various field angles. The standard deviation of the spatial variation across the footprint is at right.}
\vspace{-6mm}
\end{center}
\end{figure}

Figure \ref{fig:footprint_CryoPCM_600nm_field_average_and_variance} shows the average retardance over a single footprint in the left plot. The right plot of Figure \ref{fig:footprint_CryoPCM_600nm_field_average_and_variance} shows the standard deviation of the spatial variation across the footprint as the optic rotates. With such a large footprint, the standard deviation of retardance variation across any individual footprint is up to 1.5$^\circ$ peak with typical values in the range of 0.5$^\circ$ to 1.0$^\circ$. As the footprint is 39 mm, the average over the footprint does reduce the variability somewhat and smooths the variation with rotation of the optic. However, the peak-to-peak retardance variation is over 2$^\circ$.  Though these values are significantly larger than desired for calibration use cases, these variations are easily removed provided a field dependent demodulation is accomplished.

\begin{figure}[htbp]
\begin{center}
\vspace{-0mm}
\hbox{
\includegraphics[height=6.5cm, angle=0]{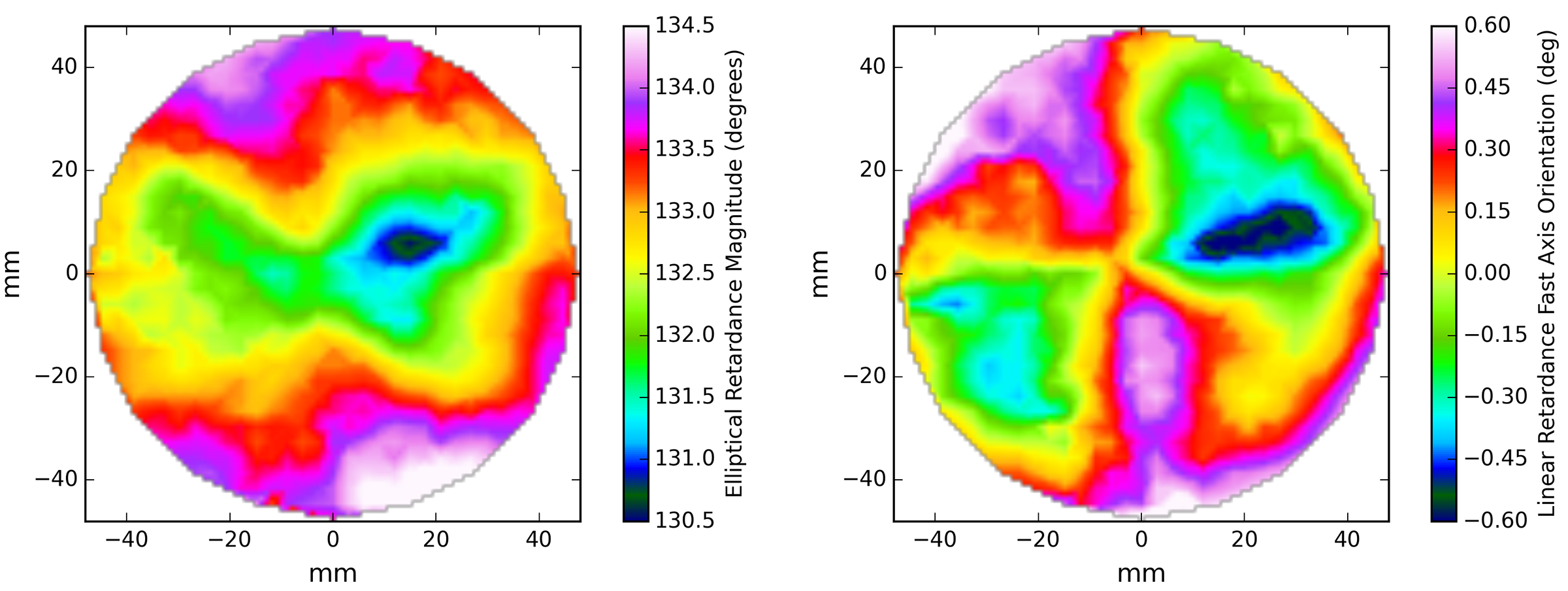}
}
\caption[] 
{\label{fig:mlo_map_visp_pcm_420nm} The spatial measurements of elliptical retardance magnitude and linear retardance fast axis for the ViSP PCM.  A retardance map was measured at 420.0 nm wavelength and spatial sampling of 3 mm to a 96 mm diameter clear aperture. The left panel shows the spatial map of elliptical retardance magnitude with spatial variations of over $\pm$2$^\circ$. The right panel shows the fast axis spatial variation for linear retardance with errors over $\pm$0.6$^\circ$. }
\vspace{-6mm}
\end{center}
\end{figure}

\subsection{Measuring a SiO$_2$ Elliptical Retarder: The ViSP Modulator}
\label{sec:appendix_spatial_maps_visp_pcm}

The ViSP PCM retarder was designed to be an efficient modulator for wavelengths between 380 nm and 1000 nm. Table \ref{table:ViSPpcm} shows the Meadowlark metrology and the design for the components in this optic. The three retarder A-B-A crystal compound retarders were designed to be mounted at 0$^\circ$, 41.2$^\circ$ and 148.2$^\circ$ orientation. The ViSP PCM is designed to be a 120$^\circ$ wave linear retarder with roughly 26$^\circ$ of circular retardance at 420 nm wavelength. The Meadowlark mapping technique measures elliptical retardance magnitude so all components are included. Spatial measurements used 3 mm spatial sampling and a 3 mm beam footprint covering 96 mm diameter aperture with 797 separate samples. Wavelength was set with a 420.0 nm interference filter with a 10 nm FWHM bandpass.

\begin{wraptable}{l}{0.60\textwidth}
\vspace{-2mm}
\caption{Measured ViSP PCM Crystal Properties}
\label{table:ViSPpcm}
\centering
\begin{tabular}{l | l l | l l | l l | l}
\hline \hline
N		& Des.	& Meas	& Ret		& Meas	& Des		& Meas			& Ori				\\
		& mm	& mm	& Bias		& Waves	& Pair		&$\delta\pm$0.01	& $\pm0.3^\circ$	\\
\hline \hline
2A$_s$	& 2.13	& 2.188	& 31$\pm$1	& 31.26	& 			& 				& 0$^\circ$		\\
2A$_b$	& 2.10	& 2.153	& 30$\pm$1	& 30.78	& 0.476		& 0.4732			& 90$^\circ$		\\
\hline
3B$_s$	& 2.12	& 2.153	& 31$\pm$1	& 30.78	& 			& 				& 41.18$^\circ$	\\
3B$_b$	& 2.10	& 2.135	& 30$\pm$1	& 30.46	& 0.328		& 0.3239			& 131.18$^\circ$	\\
\hline
4A$_s$	& 2.13	& 2.187	& 31$\pm$1	& 31.26	& 			& 				& 148.23$^\circ$		\\
4A$_b$	& 2.10	& 2.154	& 30$\pm$1	& 30.78	& 0.476		& 0.4758			& 58.23$^\circ$		\\
\hline \hline		
\end{tabular}
\vspace{-2mm}
\end{wraptable}

Figure \ref{fig:mlo_map_visp_pcm_420nm} shows the elliptical retardance magnitude map. The spatial variation is roughly 4$^\circ$ peak to peak elliptical retardance variation. The linear retardance fast axis rotates by $\pm$0.6$^\circ$ across the clear aperture. Even though the quartz crystals tend to polish more uniformly, for this optic, we see substantial spatial variation near the center of the clear aperture. We computed cumulative distribution functions for retardance magnitude error using a 96 mm a 60 mm clear aperture.

\begin{figure}[htbp]
\begin{center}
\vspace{-0mm}
\hbox{
\includegraphics[height=5.8cm, angle=0]{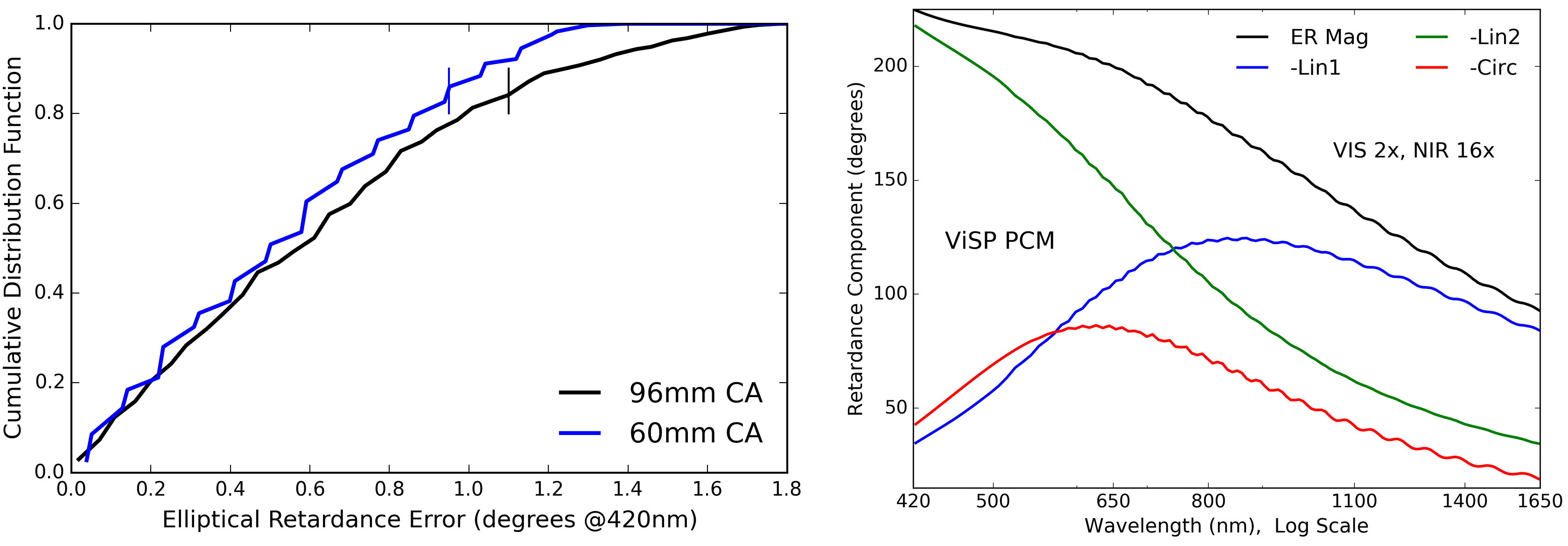}
}
\caption[] 
{\label{fig:visp_pcm_uniformity_polish_CDF} The left graphic shows the cumulative distribution function of elliptical retardance magnitude spatial non-uniformity in the ViSP PCM retarder as-built.  The black line shows a 96 mm clear aperture while the dashed lines show a reduced 60 mm clear aperture. As this optic has substantial variation near the center of the part, the CDFs are quite similar. The right graphic elliptical retarder model fit to the NLSP Mueller matrix measurements for the ViSP modulator. Black shows the total elliptical retardance magnitude.  Blue and green show the first and second components of linear retardance.  Red shows circular retardance. }
\vspace{-6mm}
\end{center}
\end{figure}

Figure \ref{fig:visp_pcm_uniformity_polish_CDF} shows the cumulative distribution function (CDF) for elliptical retardance magnitude spatial variation. For both of these apertures, 80\% of the area is within 1$^\circ$ retardance of the average.  The 96 mm clear aperture has all points within 1.8$^\circ$ retardance magnitude while the 60 mm clear aperture has all points within 1.3$^\circ$ of average. As the ViSP PCM contains significant spatial variation near the center of the part, reducing the clear aperture does not reduce the variation as significantly as with other parts above. The demodulation matrix will contain significant variation as shown above in Figures \ref{fig:footprint_ViSPsar_AsBuilt_420nm}  and \ref{fig:spatial_retardance_variation_ViSPsar_AsBuilt_420nm}.

We computed polishing error sensitivity using 0.005 waves of error per crystal and clocking errors of 0.3$^\circ$ per crystal. The retardance polish error corresponds to a physical thickness error of 0.35 $\mu$m on each the A-B-A crystal pairs. For this optic the peak to peak variation was measured to be 0.010 waves retardance error with an RMS of 0.0029 for the five spatial locations measured during acceptance metrology. The ViSP PCM design shows strong circular retardance sensitivity at short wavelengths even though the retardance magnitude is quite low. The design sensitivity to clocking errors is significantly smaller at short wavelengths for this particular design.

In the right hand side of Figure \ref{fig:visp_pcm_uniformity_polish_CDF} we show the elliptical retarder model fit to the measured Mueller matrix. We use the physically motivated solution from Appendix \ref{sec:appendix_eliret} and Figure \ref{fig:visp_pcm_multiple_ER_solutions} where the elliptical retardance magnitude generally decreases with increasing wavelength.  The ViSP PCM has greater than half-wave retardance magnitude for wavelengths shorter than about 700 nm and with significant circular retardance at this particular wavelength. For this data set the visible spectrograph was binned to a spectral sampling of roughly 1.1 nm per pixel but with an instrument profile of several times that at about 9 nm FWHM. Thus the visible data set does not resolve the ripples from the clocking oscillations. Clocking oscillations are clearly seen in the near infrared spectral region where the FWHM of the instrument profile was closer to 12 nm sampled at 1.6 nm per pixels.  We spectrally averaged (binned) this data set by a factor of 16 to increase the SNR.

\begin{figure}[htbp]
\begin{center}
\vspace{-0mm}
\hbox{
\includegraphics[height=6.5cm, angle=0]{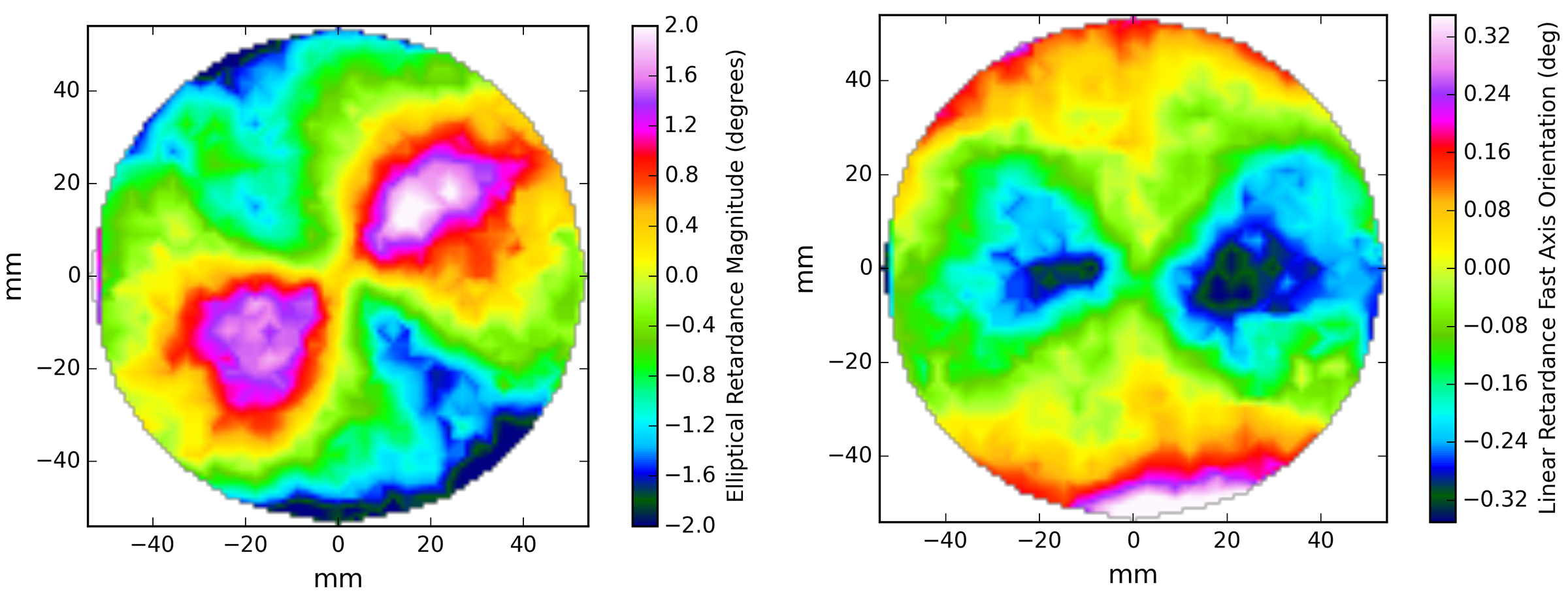}
}
\caption[] 
{\label{fig:mlo_map_dl_pcm_600nm} The spatial measurements of elliptical retardance magnitude and linear retardance fast axis orientation for the DL-NIRSP PCM. The wavelength was 600 nm wth spatial sampling of 3 mm to cover a 108 mm diameter aperture.The left panel shows the spatial map of retardance magnitude with spatial variations over $\pm$2$^\circ$. The right panel shows the linear retardance fast axis orientation varying by over $\pm$0.32$^\circ$.  }
\vspace{-6mm}
\end{center}
\end{figure}

\subsection{Measuring another SiO$_2$ Elliptical Retarder: The DL-NIRSP Modulator}
\label{sec:appendix_spatial_maps_dl_pcm}

\begin{wraptable}{l}{0.60\textwidth}
\vspace{-2mm}
\caption{Measured DL-NIRSP PCM Crystal Properties}
\label{table:DLpcm}
\centering
\begin{tabular}{l | l l | l l | l l | l}
\hline \hline
N		& Des.	& Meas	& Ret		& Meas	& Des		& Meas			& Ori				\\
		& mm	& mm	& Bias		& Waves	& Pair		&$\delta\pm$0.01	& $\pm0.3^\circ$	\\
\hline \hline
2C$_s$	& 2.17	& 2.230	& 31$\pm$1	& 31.78	& 			& 				& 0$^\circ$		\\
2C$_b$	& 2.10	& 2.164	& 30$\pm$1	& 30.78	& 1.000		& 0.9991			& 90$^\circ$		\\
\hline
3D$_s$	& 2.15	& 2.217	& 31$\pm$1	& 31.46	& 			& 				& 42.20$^\circ$	\\
3D$_b$	& 2.10	& 2.159	& 30$\pm$1	& 30.78	& 0.683		& 0.6836			& 132.20$^\circ$	\\
\hline
4C$_s$	& 2.17	& 2.235	& 31$\pm$1	& 31.79	& 			& 				& 152.51$^\circ$		\\
4C$_b$	& 2.10	& 2.162	& 30$\pm$1	& 30.79	& 1.00		& 1.0033			& 62.51$^\circ$		\\
\hline \hline		
\end{tabular}
\vspace{-2mm}
\end{wraptable}

The DL-NIRSP polychromatic modulator (PCM) is nominally designed to cover the 500 nm to 2500 nm bandpass with 8 or 10 samples in a 180$^\circ$ rotation of the optic and an ideal Stokes Q analyzer. The three retarder A-B-A crystal subtraction pairs were designed to be mounted at 0$^\circ$, 42.2$^\circ$ and 152.5$^\circ$ orientation.  Table \ref{table:DLpcm} shows the Meadowlark metrology and design properties of the components in this optic.


Measurements used 3 mm spatial sampling and a 3 mm beam footprint covering 108 mm diameter aperture with 1009 separate samples. Wavelength was set with a 598.9 nm interference filter with a 10 nm FWHM bandpass. At this wavelength, the retarder should nominally have 241$^\circ$ total retardance magnitude with 18.9$^\circ$ of circular retardance. Figure \ref{fig:mlo_map_dl_pcm_600nm} shows the elliptical retardance magnitude map. The spatial variation is roughly 2$^\circ$ peak to peak elliptical retardance variation. The linear retardance fast axis rotates by over $\pm$0.3$^\circ$ across the clear aperture.

\begin{figure}[htbp]
\begin{center}
\vspace{-0mm}
\hbox{
\includegraphics[height=5.8cm, angle=0]{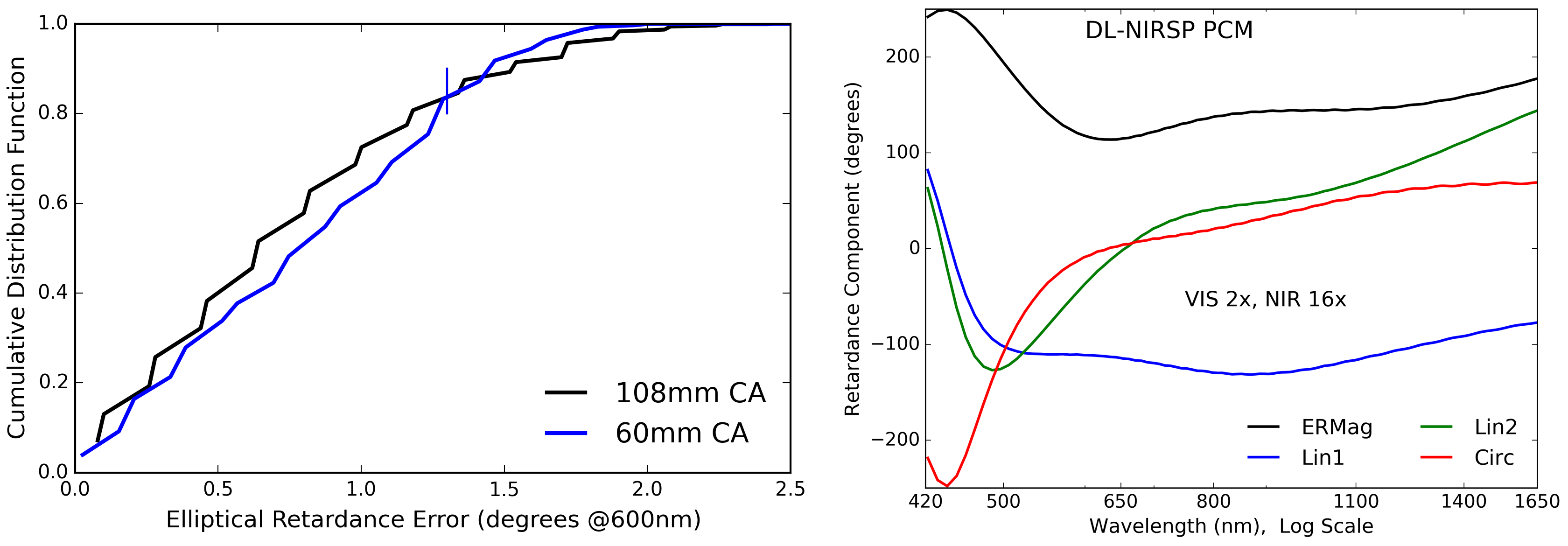}
}
\caption[] 
{\label{fig:DL_pcm_uniformity_polish_CDF} The left hand graphic shows the cumulative distribution function of elliptical retardance magnitude spatial non-uniformity in the DL-NIRSP PCM retarder.  The black line shows a 108 mm clear aperture while the dashed lines show a reduced 60 mm clear aperture. As this optic has substantial spatial variation near the center of the part, the CDFs are quite similar The right hand graphic shows the elliptical retarder model fit to the NLSP Mueller matrix measurements for the DL-NIRSP modulator. Black shows the total elliptical retardance magnitude.  Blue and green show the first and second components of linear retardance.  Red shows circular retardance. }
\vspace{-6mm}
\end{center}
\end{figure}

We computed cumulative distribution functions for retardance magnitude error using a 108 mm a 60 mm clear aperture. Figure \ref{fig:DL_pcm_uniformity_polish_CDF} shows the cumulative distribution function (CDF) for elliptical retardance magnitude spatial variation. For both of these apertures, 80\% of the area is within 1.4$^\circ$ retardance of the average.  Both the 108 mm and 60 mm clear aperture have nearly all points within 2.0$^\circ$ retardance magnitude. Similar to the ViSP PCM, the DL-NIRSP PCM contains significant spatial variation near the center of the part. Reducing the clear aperture does not reduce the variation as significantly as with other parts above.  

This optic will be sampled somewhat unusually in the nominal DL-NIRSP optical design. The DL-NIRSP was originally designed as a multi-slit instrument with optics sized for a large slit mask. Recently, the project pursued an integral field unit (IFU) made of polarization preserving optical fibers, replacing the slit mask. This IFU is significantly smaller, sampling less than 25 mm of the clear aperture of this retarder.  The various resolution optical beams work at F/ 24 to F/ 60 which gives rise to footprints on this retarder of only a few millimeters. We expect spatial variation on this part to be very small.

\begin{figure}[htbp]
\begin{center}
\vspace{-0mm}
\hbox{
\includegraphics[height=6.5cm, angle=0]{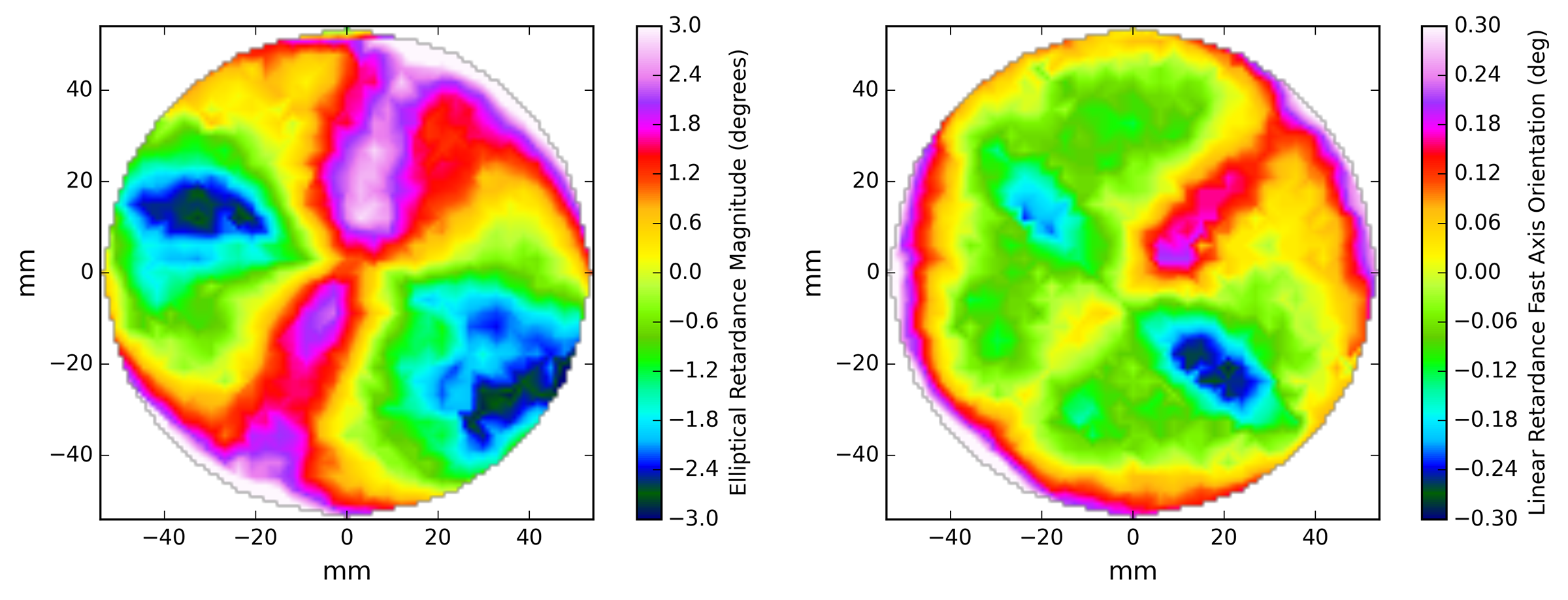}
}
\caption[] 
{\label{fig:mlo_map_dl_sar_660nm} The spatial measurements of retardance for the DL-NIRSP SAR.  A map was measured at 660 nm wavelength and spatial sampling of 3 mm out to 108 mm diameter clear aperture. The left panel shows the spatial map of elliptical retardance magnitude errors. The right panel shows the map of inear retardance fast axis errors.  }
\vspace{-8mm}
\end{center}
\end{figure}

We compared simulations of crystal polishing errors to crystal clocking errors for the DL-NIRSP PCM. The polishing errors were scaled to 0.005 waves per crystal while clocking errors are 0.3$^\circ$ per crystal. For this optic the peak to peak variation was measured to be 0.0066 waves retardance error with an RMS of 0.0023 for the five spatial locations measured during acceptance metrology. The circular retardance sensitivity to polishing error is high around 900 nm with greater linear retardance sensitivity at shorter wavelengths. The clocking error simulations show varying sensitivity across different wavelengths roughly equally spread between wavelengths.

The right hand graphic of Figure \ref{fig:DL_pcm_uniformity_polish_CDF} shows the elliptical retarder fits to the NLSP measurements of the optic Mueller matrix. This retarder is almost purely circular at wavelengths around 420 nm. Similar to the ViSP PCM, this data set had lower spectral resolving power of 9 nm FWHM for the visible data set and 12 nm for the near infrared data set.  Spectral pixels were averaged to achieve higher SNRs.  The clocking oscillations are significantly lower in this optic, compared to the ViSP PCM, even though this data set had lower resolving power and under estimates the spectral ripple from crystal mis-alignments.

\subsection{Measuring the DL-NIRSP SiO$_2$ Calibration Retarder (SAR)}
\label{sec:appendix_spatial_maps_dl_sar}

\begin{wraptable}{l}{0.60\textwidth}
\vspace{-2mm}
\caption{Measured DL-NIRSP SAR Crystal Properties}
\label{table:DLsar}
\centering
\begin{tabular}{l | l l | l l | l l | l}
\hline \hline
N		& Des.	& Meas	& Ret		& Meas	& Des		& Meas			& Ori				\\
		& mm	& mm	& Bias		& Waves	& Pair		&$\delta\pm$0.01	& $\pm0.3^\circ$	\\
\hline \hline
2D$_s$	& 2.15	& 2.202	& 31$\pm$1	& 31.47	& 			& 				& 0$^\circ$		\\
2D$_b$	& 2.10	& 2.162	& 30$\pm$1	& 30.78	& 0.683		& 0.6909			& 90$^\circ$		\\
\hline
3C$_s$	& 2.17	& 2.223	& 31$\pm$1	& 31.78	& 			& 				& 65$^\circ$	\\
3C$_b$	& 2.10	& 2.153	& 30$\pm$1	& 30.78	& 1.00		& 1.0008			& 155$^\circ$	\\
\hline
4D$_s$	& 2.15	& 2.203	& 31$\pm$1	& 31.47	& 			& 				& 0$^\circ$		\\
4D$_b$	& 2.10	& 2.156	& 30$\pm$1	& 30.79	& 0.683		& 0.6868			& 90$^\circ$		\\
\hline \hline		
\end{tabular}
\vspace{-2mm}
\end{wraptable}

The DL-NIRSP superachromatic calibration retarder (SAR) is required to cover the 900 nm to 2500 nm bandpass. The DKIST design achieves linear retardance magnitudes between 70$^\circ$ and 150$^\circ$ over the 600 nm to 2500 nm bandpass. The retarder is nominally a 120$^\circ$ wave linear retarder at 660 nm wavelength. No circular retardance is present in the design though our laboratory spectropolarimetry testing of this optic shows it is present at magnitudes up to 2$^\circ$.  Table \ref{table:DLsar} shows the Meadowlark metrology and design of the components in this optic.

\begin{wrapfigure}{r}{0.60\textwidth}
\centering
\vspace{-4mm}
\begin{tabular}{c} 
\hbox{
\hspace{-1.0em}
\includegraphics[height=7.3cm, angle=0]{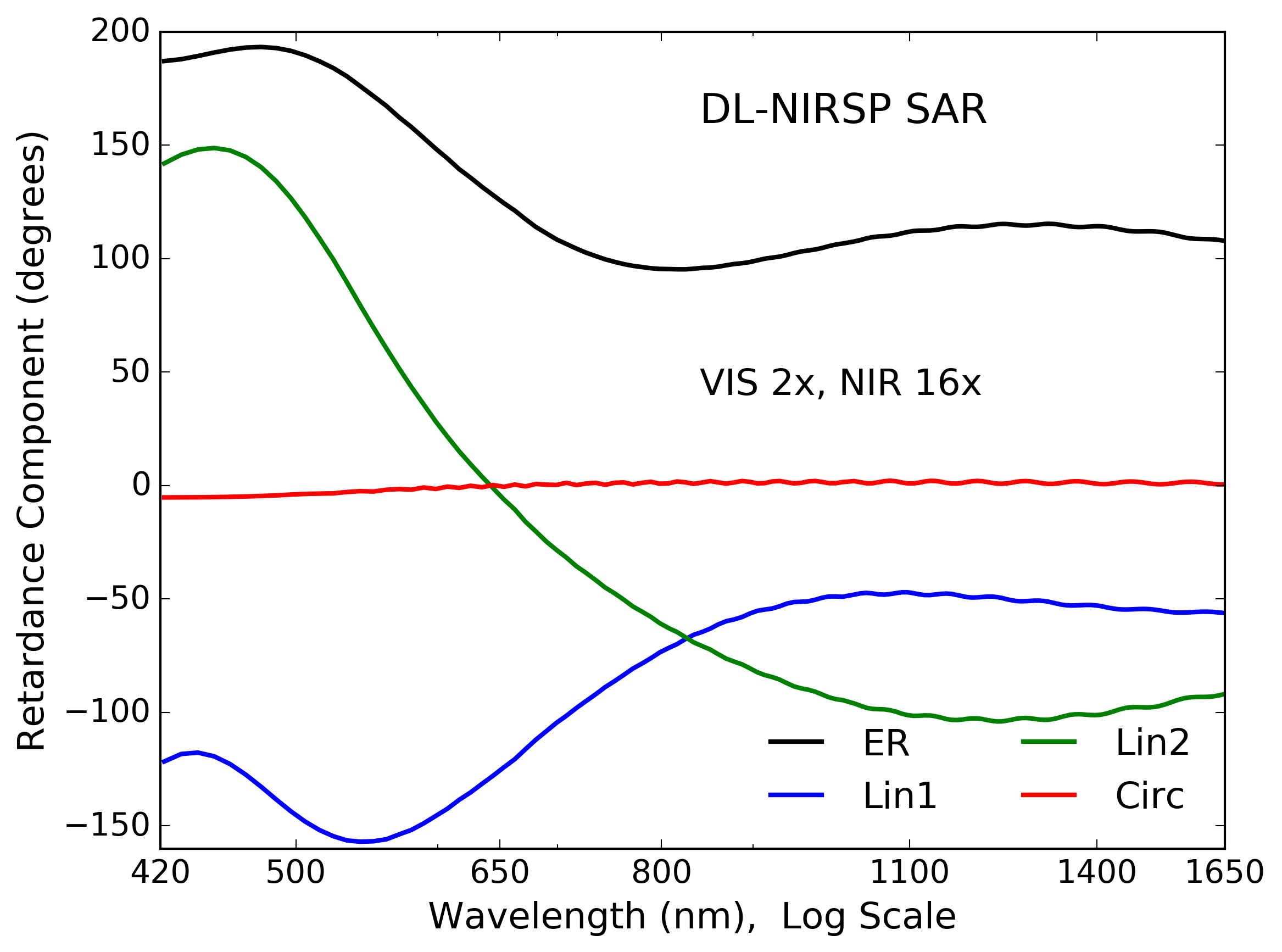}
}
\end{tabular}
\caption[] 
{ \label{fig:dl_sar_nlsp_er}  The elliptical retarder model fit to the NLSP Mueller matrix measurements for the DL-NIRSP calibration retarder. Black shows the total elliptical retardance magnitude.  Blue and green show the first and second components of linear retardance.  Red shows circular retardance.}
\vspace{-6mm}
 \end{wrapfigure}
Spatial measurements used 3 mm spatial sampling and a 3 mm beam footprint covering 108 mm diameter aperture with 1009 separate samples. Wavelength was controlled using a 660 nm interference filter with a 10 nm FWHM bandpass. The three retarder subtraction plates A-B-A were designed to be mounted at 0$^\circ$, 65$^\circ$ and 0$^\circ$ orientation respectively. Figure \ref{fig:mlo_map_dl_sar_660nm} shows the measured spatial retardance error maps. The left hand graphic shows a $\pm$3$^\circ$ elliptical retardance magnitude variation. The right hand plot shows the fast axis of linear retardance varying by just over $\pm$0.3$^\circ$ orientation.

Figure \ref{fig:dl_sar_nlsp_er} shows the elliptical retarder fits to the NLSP measurements of the optic Mueller matrix.  Similar to the ViSP PCM and DL-NIRSP PCM, this data set had the lower spectral resolving power with instrument profiles at 9 nm FWHM for the visible data set and 12 nm for the near infrared.  Crystal clocking misalignment induced oscillations are seen at low amplitude in the infrared data set. The elliptical retardance magnitude is almost entirely linear retardance. The optic meets the nominal DKIST calibration specification of 90$^\circ$ retardance $\pm$30$^\circ$ over the 650 nm to 2500 nm wavelength range.  This optic could be used for calibration at slightly shorter wavelengths, but calibrations will fail where the part is near half wave and unable to create circular retardance.

We compute the spatial variation of retardance using the eight footprints on the DL-NIRSP calibration retarder as this optic rotates during a typical calibration sequence. The center footprint remains constant for the on-axis beam and would see the same average retardance for all retarder orientations. For the footprint corresponding to the 5 arc-minute field edge, there is clear spatial separation between footprints giving rise to significant retardance variation between steps. 

\begin{figure}[htbp]
\begin{center}
\vspace{-0mm}
\hbox{
\hspace{-0.0em}
\includegraphics[height=5.6cm, angle=0]{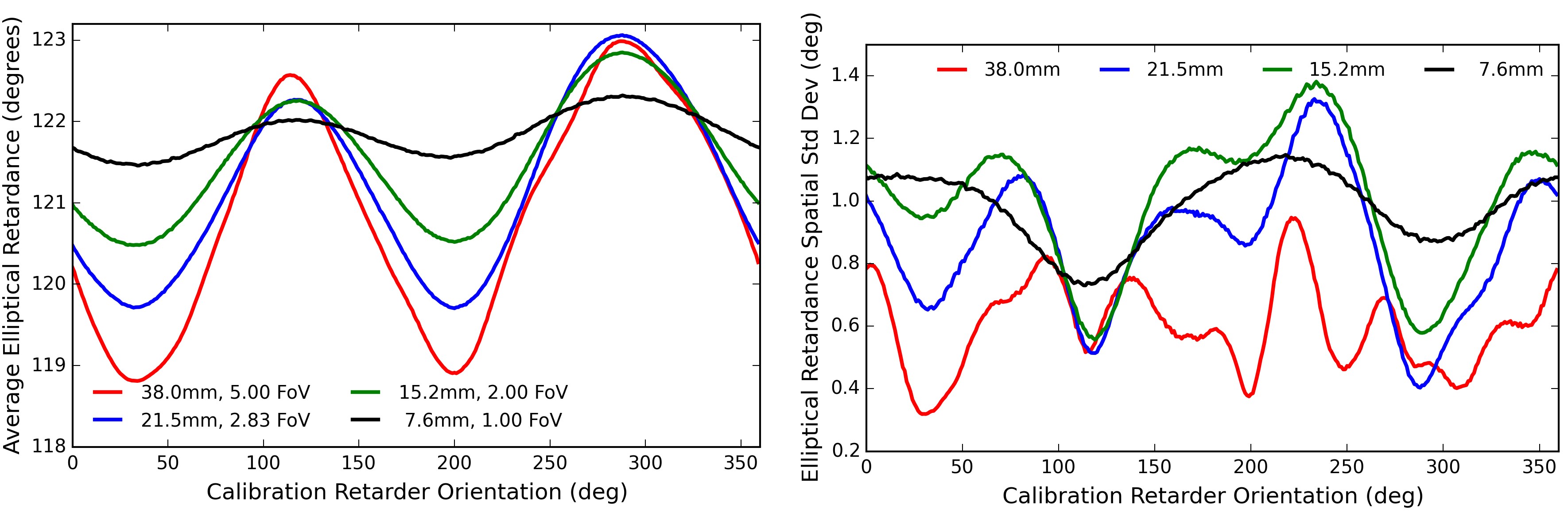}
}
\caption[] 
{\label{fig:footprint_DLSAR_660nm_field_average_and_variance} The footprint variation for the DL-NIRSP SAR during the calibration process.  The average retardance in a 26.6 mm diameter beam footprint as the optic rotates  is shown at left for a range of field angles. Black shows a 1 arc minute field of view (FoV) with a physical decenter of 7.6 mm. The standard deviation of the spatial variation across the 26.6 mm footprint is shown at right.}
\vspace{-6mm}
\end{center}
\end{figure}

We quantify the retardance variability during calibration by taking spatial averages over the footprint for varying FoV.  Figure \ref{fig:footprint_DLSAR_660nm_field_average_and_variance} shows an example of the average retardance across a footprint as the calibration retarder rotates. The various curves correspond to select field angles using the DL-NIRSP calibration retarder at 600 nm wavelength. Each individual footprint is 26.6 mm at this optical location in the converging F/ 13 beam. Black shows a 1 arc minute field of view corresponding to a beam center 7.6 mm away from the retarder center and 1$^\circ$ peak to peak spatial variation. Green shows a 2.0 arc minute field of view with 15.2 mm of decenter. Blue shows the edge of the 2.83 arc minute field of view and a 21.5 mm beam decenter. Red corresponds to the maximal DKIST field of 5.0 arc minutes and a 38.0 mm beam decenter. As wavelength increases, the magnitude of polishing spatial variation and retardance non-uniformity decreases.  The maximal retardance spatial non-uniformity is roughly 2$^\circ$ to 3$^\circ$ at for fields larger than 1 arc minute.  For fields less than this, we see significantly smaller variation.

Similar to the estimates from Sueoka for the DKIST retarders in a converging F/ 13 beam \cite{Sueoka:2016vo}, by averaging over a bundle of rays with spatially varying retardance properties, we encounter depolarization \cite{Chipman:2006iu,Anonymous:2004wl,2004OExpr..12.4941D,Chipman:2007ey,Anonymous:s4Rv1_wo,2005ApOpt..44.2490C,Chipman:2010tn,Chipman:2014ta} In the right side of Figure \ref{fig:footprint_DLSAR_660nm_field_average_and_variance} we compute the standard deviation of retardance across a zero-field footprint at varying field angles. Even though the on-axis beam sees no change in footprint location as the optic rotates, the 26.6 mm diameter beam will see (constant) spatial variation.  As the field angle increases and the beam de-centers, the spatial variation changes and varies more with retarder orientation.

Figure \ref{fig:footprint_DLSAR_660nm_field_average_and_variance} is similar to the ViSP in that we do not necessarily see the greatest spatial variation in individual footprints for the greatest field angles. The black curve corresponds to the nearly stationary beam footprint. However, the retardance spatial variation is 0.8$^\circ$ to 1.2$^\circ$ retardance. For the largest field angle with the beam de-centered by 38 mm, the spatial variation oscillates between 0.3$^\circ$ and 0.9$^\circ$, significantly less than the on-axis beam.

For this optic the peak to peak variation of each crystal pair was measured to be 0.0026 waves to 0.0081 waves retardance errors for the five spatial locations measured during acceptance metrology. The RMS values were in the range 0.0011 to 0.0033 waves. For the polishing error simulations, we apply 0.005 waves retardance error per crystal pair at 633 nm wavelength applied to every other crystal giving 3$^6$ models. Elliptical retardance fits were also done to Mueller matrix simulations where retarder orientations were given clocking errors of 0.3$^\circ$ per crystal. The circular retardance sensitivity is high in two bandpasses peaking near 400 nm and 900 nm.  The clocking error simulations show varying sensitivity across different wavelengths with circular retardance oscillations strongest around 800 nm and 2500 nm wavelength.

\subsection{Wavefront Error, Beam Deflection and Optical Quality Metrics}
\label{sec:wfe_deflection}

A major concern when initially designing these optics was the transmitted wave front flatness and the beam deflection from wedge caused by the non-parallel oil interfaces and achievable spatial uniformity. When the modulating retarders are rotating optics, any beam deflection can introduce image motion on the focal plane, creating polarization artifacts and complicating the data analysis process. We note that even liquid crystal type modulators introduce some small beam deflections when voltage switches. 

The initial DKIST retarder designs called for 10 mm thick cover windows to be used on the top and bottom of each crystal stack. As shown previously in \cite{Harrington:2017jh, Harrington:2018cx}, these windows caused issues with fringe amplitude, fringe period, mechanical vibration, exacerbated beam jitter, added heating and increased thermal gradients. We measured the beam deflection and transmitted wavefront distortion before and after window removal without observing significant differences.

The transmitted wavefront specification for the calibration retarders was 3 waves of irregularity after power removal.  Modulating retarders required 2 waves of irregularity after power removal. There was no specification for the amplitude of power allowed in the transmitted wavefront. We performed as-built metrology at Meadowlark Optics for the six crystal stacks after final assembly.  The ViSP PCM shows 0.93 peak to peak of irregularity and The Cryo-NIRSP SAR shows 1.7 peak to peak irregularity. 

We did a Zernike fit to the WFE data sets for three of the crystal stack retarders without cover windows. Power dominates with over 4 waves for the ViSP PCM and over 2 waves for the CryoNIRSP SAR.  Though there are a few waves of power in each optic, the irregularity terms are all small. Per Meadowlark, we anticipate a slight amount of cylinder-shaped error as the crystals are slightly harder in one direction than the other. The crystal retarders are cut and polished with the ordinary and extraordinary axes along the face of the polished surface. Conductivity, material strength, refractive index and expansion coefficients vary along X and Y directions.

If optics are not mounted parallel to the beam, a translation can occur.  As an example, the DL-NIRSP instrument mounts the modulator in front of the fiber bundle integral field unit.  With fiber core dimensions less than 10 microns, small tilts of the optic displace the image on the fibers introducing a stringent requirement on both beam deflection (wedge) and beam translation (parallelism during rotation). For the DKIST calibration optics, an arc minute of beam deflection introduced near Gregorian focus is enough to move the beam footprint by millimeters on the optics far down stream in the coud\'{e} laboratory. When deciding on retarder types, optical considerations can often limit design options. We use a standard beam deflection technique with a Zygo interferometer and a reference flat. As our optics have a large clear aperture and waves of power WFE, we developed a technique using multiple orientations to separate deflection caused by wedge from other sources.  By tracking the sub-aperture beam deflection as a vector and projecting only the wedge component, we estimated that the ViSP PCM and Cryo-NIRSP PCM modulators have beam deflections less than 10 arcseconds at 633 nm wavelength.  The DL-NIRSP PCM has yet to be mounted and measured.

We note that the individual crystals must be polished incredibly flat to achieve the measured 0.01 waves retardance error. Zernike polynomial fits to the WFE in transmission for three of the crystals used to create the Cryo-NIRSP SAR showed that there is minimal power and irregularity in each crystal pair. The final assembly of the six-crystal stacks involves addition of several microns thickness of oil between each interface across the 120 mm diameter optic. Given such thin crystals at high aspect ratio, some spatial variation in thickness over such a thin layer is not surprising.

\clearpage
\section{Other Telescopes \& Alternate Retarder Styles}
\label{sec:appendix_uniformity_polycarb_FLC}

We compare here alternate retarders and calibration optic locations in a variety of large solar telescopes.  The different choices and trade-offs between heating, wavelength coverage, beam deflection and uniformity will be summarized. The 1.5 m GREGOR solar telescope has had several calibration retarders located at a focus after their secondary mirror and multiple polarimeters \cite{2012AN....333..872C, Hofmann:2012ik, Gisler:2016kz, Balthasar:2011vd, Hofmann:2003kg, Volkmer:2010bk, Volkmer:2012ix, Soltau:2012hw, Schmidt:2012jm}.  Another useful comparison is The Big Bear Solar Observatory (BBSO) which operates the Goode Solar Telescope (GST) formerly named the New Solar Telescope (NST). The GST has an off-axis Gregorian design with a 1.6 meter clear aperture primary mirror.  An alternate case study with calibration performed only after several optics is the Spectro-Polarimeter for Infrared and Optical Regions (SPINOR) spectropolarimeter at the 0.76 meter Dunn Solar Telescope (DST) in Sacramento Peak, New Mexico\cite{2006SoPh..235...55S, SocasNavarro:2011gn}.

\begin{table}[htbp]
\vspace{-2mm}
\caption{Solar Telescopes \& Polarization Calibration Optic Properties}
\label{table:solar_tel_comp}
\centering
\begin{tabular}{l r r r r c r r r r l}
\hline \hline
Name	& Diam	& FoV	& Feed			& F/ 		& Power 	& Clr Ap 	& FP		& Irrad		& Num	& Optics			\\
		& m		& '		& Optics			& 		& Watts	& cm 	& mm	& Wcm$^{-2}$	& Lyr		& Upstream		\\
\hline \hline	
DKIST	& 4.0		& 5.0		& Off-axis			& 13		& 300	& 10.5	& 26.6	& 3.4			& 6		& M1, M2			\\
GREGOR	& 1.5		& 2.5	 	& On-axis			& 6		& 8		& 0.7		& 		& 24$^*$		& 5/1		& M1, M2, Filters		\\
GST		& 1.6		& 2.0		& Off-axis			& 52		& 8		& 14.0	& 		& 0.06		& 1		& M1, M2			\\
DST		& 0.76	& 2.8		& On-axis$^*$		& 72		& 3		&  4.3	& 1.8		& 0.2			& 2		& Win, Trt, Win, M1	\\
\hline \hline		
\end{tabular}
\vspace{-3mm}
\end{table}

We outline the details of the optical power, location of polarization calibration optics and some relevant optical details in Table \ref{table:solar_tel_comp}.  Details will be provided in paragraphs below about each telescope, instrument and the various retarders in use.  The second column shows the primary mirror diameter in meters.  The third column shows the field of view in arc minutes at the calibration optic. The fourth column describes the main optical path as either on-axis or off-axis.  The asterisk for the DST is due to turret mirrors described below. The beam focal ratio (F/) and power are shown in the fifth and sixth columns.  The seventh column shows the illuminated clear aperture of the calibration optic. Column eight shows the footprint for a single field point on the calibration optic.  Column nine shows the irradiance in Watts per square centimeter at the calibration optic. The tenth column lists the number of individual retarders that are used to make the calibration retarder. A single crystal or polycarbonate layer is indicated as 1, a compound retarder is listed as 2 and the DKIST six crystal achromats are indicated as 6. In some cases multiple optics can be used. The eleventh column lists optical elements ahead of the calibration optic. In some cases, multiple windows (Win) and a turret mirror pair (Trt) are used.

At GREGOR, the second focus (F2) is near the calibration optics and is stated to receive about 8 Watts of power and a 2.5 arc minute field of view \cite{Soltau:2012hw}. This can be compared to 300 Watts at DKIST covering 5 arc minutes field of view. We compute 10 Watts when scaling 300 W by (1.5/4)$^2$ and (2.5/5)$^2$. The GREGOR beam at the second focus is F/ 6 with an image scale of 23 arc seconds field angle per millimeter across the retarder \cite{Soltau:2012hw}. The power density was stated as about 24 Watts per cm$^2$ and with the GREGOR plate scale with their second focus at only 6.5 mm in diameter. The DKIST optics were 105 mm clear aperture and 120 mm diameter and see a similar irradiance of roughly 3.4 Watts per cm$^2$. From Hofmann et al. 2012 \cite{Hofmann:2012ik} the GREGOR polarization calibration unit has two superachromatic retarders consisting each of five layers of zero-order retarders. The polymethyl-metacrylat (PMMA) retarder layers were designed by Astropribor with an angular acceptance of $\pm$7$^\circ$, a temperature range of -20$^\circ$C to 50$^\circ$C and a damage threshold of 500 W per cm$^2$. These two separate retarders cover 380 nm to 800 nm wavelength range for the first and 750 nm to 1800 nm for the second.  These bandpasses are roughly similar to the DKIST calibration retarders for ViSP and DL-NIRSP respectively. 

In addition to the nominal GREGOR Polarimetric Calibration Unit (PCU, Hofmann et al. 2012), a second PCU was installed \cite{Collados:2012hi}. The retardance of this near-infrared retarder is approximated in Collados et al. 2012\cite{Collados:2012hi} as the Equation $\delta$ = 793.2$\lambda^{-1}$-415 where $\delta$ is in degrees and the wavelength $\lambda$ is given micrometers. This part is a single crystal of quartz cut as retarder with thickness estimated around 0.23 mm covering the 1000 nm to 1800 nm wavelength range (Collados, private communication). The calibration retarder is mounted at the focal plane, maximizing the heat load and field variation but minimizing beam size.  Given the single crystal type retarder and the small footprints of the GREGOR beam, this calibration unit should see substantially lower errors due to spatial non-uniformity as the polishing errors are typically a strong inverse power law of spatial frequency. DKIST observed that polishing errors are dominated by spatial scales larger than 10 mm for the DKIST retarders while the entire GREGOR field of view is contained inside a 6.5 mm footprint.  Additionally, there are heat rejection filters mounted ahead of the calibration optics that limit the wavelengths incident on the calibration optics, further reducing the heat loads.

At the GST, the first generation instrument with polarimetric capability was the Infrared Imaging Magnetograph (IRIM) \cite{Goode:2011wi, Cao:2011ti, Wang:2004ky}.The calibration unit is similar to DKIST in that both units contain a linear polarizer ahead of a $\sim$ quarter wave linear retarder mounted near Gregorian focus. The polarization modulator for IRIM is also mounted up stream of the GST tertiary mirror, after Gregorian focus. 
 
We estimate the flux delivered at the GST by comparing with DKIST.  The focal ratio of the GST aluminum coated M1 is F/ 2.4 with a protected silver coated M2 delivering an F/ 52 beam at Gregorian focus. The 2 arc minute circular heat stop defines the flux of the GST. We calculated 300 Watts for a 5 arc minute DKIST field where we have included aluminum and an enhanced protected silver coating and Haleakala atmospheric transparency. For a 1.6 meter aperture and 2 arc minute field we would see 6.25x less flux collected by M1 and the same factor for the reduced field. This gives roughly 8 Watts of heat at their Gregorian focus with a plate scale is 2.48 arc seconds per mm. The modulator is located away from focus with a 140 mm clear aperture. We compute an irradiance of less than 0.06 Watts per cm$^2$ if the full aperture is illuminated compared to 24 W per cm$^2$for GREGOR and 3.4 W per cm$^2$ for DKIST.  The UV flux load would be highly dependent on the protection and properties of the silver coating on the GST M2. 

Meadowlark Optics built 2 large polycarbonate calibration retarder assemblies at 5.5" diameter, 5" clear aperture for Big Bear Solar Observatory.  A single layer of stretched polymer was cemented between two 1" thick glass windows. MLO tested TWE at another lab to cover the large aperture. TWE was measured as 0.388 waves PV at 632.8 nm and 0.068 waves RMS over the 5” diameter aperture. The beam deviation was 46 arc seconds.  MLO built two optics to ensure successful yield of one optic. The second optic had a TWE of 0.587 waves PV and 0.140 RMS. Power was not removed in the TWE measurements.  Windows used were one inch thick and likely flat to about 0.1 wave. The TWE error on large optics is likely the result of variations in the index of refraction of the UV cure adhesive used to laminate the polycarbonate to the windows. The AR coat is a V-type coating design that measured to be centered at 1580 nm and has R=0.03\% at that wavelength and 0.04\% at the nominal 1564.8 nm operating wavelength.  Retardance was measured on the delivered optic at 5 points distributed across the clear aperture with a 1550.3nm CWL filter Dec 21st 2009. The values 0.3514, 0.3493, 0.3454, 0.3452, 0.3453 give a mean and standard deviation of 0.34732 +/- 0.0026 or in degrees  125.0 +/- 0.9.

\begin{figure}[htbp]
\begin{center}
\vspace{-2mm}
\hbox{
\includegraphics[height=6.7cm, angle=0]{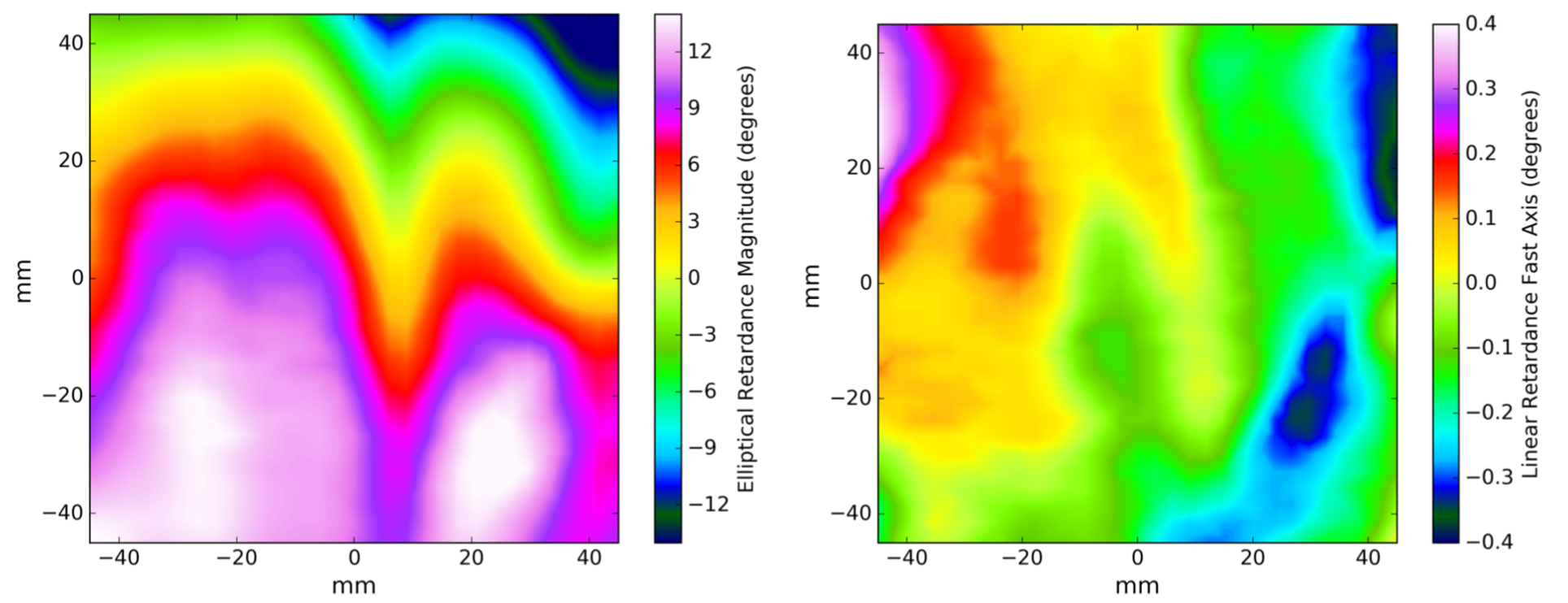}
}
\caption[] 
{\label{fig:polycarb_retardance_map_large_square} The spatial non-uniformity of a single sheet of polycarbonate stretched to be a zero order retarder over a large area. The left plot shows a spatial variation of the elliptical retardance magnitude scaled to $\pm$3.3$^\circ$. The right plot shows the fast axis of linear retardance scaled to $\pm$0.4$^\circ$. }
\vspace{-5mm}
\end{center}
\end{figure}

In pursuing this material, Meadowlark Optics made further measurements on a large stretched polycarbonate sheet near one wave retardance magnitude.  Figure \ref{fig:polycarb_retardance_map_large_square} shows the spatial variation of retardance magnitude as well as the fast axis of linear retardance. In several DKIST instruments, polycarbonate presents an opportunity to make alternate modulating retarders.  Should we need to replace the ViSP PCM with a multi-layer polycarbonate solution, the 90 mm clear aperture shown in Figure \ref{fig:polycarb_retardance_map_large_square} would be required. The spatial variation of five retarder layers would certainly become significant. However, as we showed in the main text, the main impact would be a change of modulation matrix elements leading to a possible slight loss (or increase) in modulation efficiency.

At the GST, the modulator for IRIM was also a birefringent polymer sandwiched between a pair of BK7 windows. The retarder has a 5.5 inch clear aperture which is significantly larger than the DKIST retarders at 4.1 inches\cite{Goode:2011wi, Cao:2011ti}. The optic size was nominally constrained by a UV flux threshold to avoid damaging the polymer. This zero order polymer optic achieves a retardation of $\delta$ = 0.3525 waves at 1564.85 nm wavelength, a wavefront distortion of less than a quarter wave and beam deviation was reduced to less than 2 arc minutes at these infrared wavelengths \cite{Goode:2011wi, Cao:2011ti, Wang:2004ky}.  

The second generation GST polarimetric instrument is NIRIS (Near Infrared Imaging Spectropolarimeter) which uses the adaptive optics system and is capable of measurement in the 1083 nm and 1565 nm spectral lines \cite{Cao:ti}. The modulator for NIRIS is placed near the camera using a two inch diameter optic with 80\% clear aperture. The retarder is multiple layers of birefringent polymer cemented between BK7 windows designed to be roughly 0.35 waves retardance magnitude at both operating wavelengths. This modulator is mounted near the end of a long optical path after the AO system, dichroic beam splitter, a field stop and both the Fabry-Perot and narrow band interference filters.  Heat and UV flux loads are negligible. The modulator was specified to have beam deflection less than 1 arc minute and the rotating optic creates roughly $\pm$2 pixels image motion on the focal plane"  (Kwangsu Ahn, personal communication).

SPINOR at the DST covers from 400 nm to 1600 nm and can be operated with up to four separate cameras covering nm-scale bandpasses of interest. For SPINOR, there are three retarders for the instrument. There is a 3/8 wave sapphire and quartz bi-crystalline achromat that is used as the modulator. For our observing runs used in a polarization fringe analysis, we placed this modulator in the F/ 36 beam near the slit \cite{Harrington:2017jh, Derks:2018jc}. As the SPINOR slit is scanned, the instrument samples different spatial regions of the retarder, introducing noticeable changes in system performance \cite{Derks:2018jc}. However, the modulator nominal location is in a collimated beam in the fore-optics near a pupil plane. As this optic is in a pupil plane, spatial polishing errors of the two crystals only mildly depolarize the beam and introduce a small incidence angle variation from the pupil demagnification. The modulator is mounted 190 mm down stream of a 21.652 mm pupil plane giving a demagnification of 35. For an arc minute of solar field angle the incidence angle varies by roughly half a degree on the modulator. There is a 1/4 wave retarder used for calibration near the F/ 72 entrance port 132 mm from the focus. The plate scale on the filter wheel near the calibration unit is 3.76 arcseconds per mm.  For our DST observations, we had a 140 arcsecond slit which would sample roughly 40 mm clear aperture on the 42.7 mm clear aperture calibration optic. The optical power at the calibration optic is estimated to be around 3 Watts maximum given a roughly scaling by the reduced collecting area, reduced 169.2 arcsecond field of view and a rough 10\% additional loss from the few additional reflections in the DST beam. We ignore signifiant atmospheric and wavelength dependencies compared to DKIST. The optical power is roughly 0.2 Watts per cm$^2$ computed from the 45 mm  diameter aperture stop covering the field. This optical station also has a zero-field beam footprint of 1.8 mm compared to 27 mm at DKIST calibration retarders.  Another 1/2 wave retarder is mounted after the SPINOR slit, immediately in front of the polarizing beam splitter seeing a much smaller beam diameter but insignificant heat loads.

Night time astronomical observatories do not produce significant heat loads on their optics, but polarization optics often must be mounted in converging beams near focal planes and operate in outside summit environments where temperatures change and fringes impact observations. We described some calibration impacts of fringes using the Keck 10 meter telescope and the Low Resolution Imaging Spectrograph in polarimetric mode (LRISp) \cite{Harrington:2018cx, Harrington:2015cq}. This system has a calibration polarizer and six-crystal achromatic retarder optics similar to DKIST but no independent retarder optics ahead of Cassegrain focus in front of the instrument modulator \cite{Keck:2007wf, Goodrich:2003kv, 1995AJ....110.2597T, Goodrich:1995fg,1991PASP..103.1314G, Harrington:2018cx, Harrington:2015cq}. The nominal secondary mirror gives an F/ 15 beam but after accounting for the outer edges of the hexagonal mirror segments, the beam is F/ 13.7 \cite{1995PASP..107..375O}.  This instrument sees a circularly symmetric primary and secondary mirror except for a slight off-boresight mounting location of roughly 10 arc minutes equivalent field angle \cite{Keck:2007wf,Goodrich:2003kv,1995AJ....110.2597T,Goodrich:1995fg,1991PASP..103.1314G}. The Cassegrain plate scale is 1.375 arcseconds per millimeter with a significant focal plane curvature radius of 2.18 m. With this plate scale and the mounting of the retarder some fraction of an inch behind the slit wheel, the typical seeing-limited astronomical observation would have a beam footprint of a few millimeters on the retarder. \cite{Goodrich:2003kv}  The induced polarization is low but we also found that the typical alignments and the recommended demodulation procedures can leave several percent residual cross-talk as we found in our daytime sky calibrations at LRISp \cite{Harrington:2015cq}.  
 
At other observatories, a polarization sensitive instrument may not be allowed to access upstream foci for calibration purposes. This was the case for the Air Force 4.0 meter AEOS telescope \& the HiVIS spectropolarimeter as well as for optics after primary and secondary mirrors in typical symmetric telescope designs. \cite{Harrington:2017eja,Harrington:2015dl,2014SPIE.9147E..7CH,Harrington:2011fz,Harrington:2011uj,Harrington:2010km,2008PASP..120...89H,Harrington:2006hu}  Other systems like the European Extremely Large Telescope (E-ELT) have a 39 meter diameter primary and correspondingly large beams that do not provide convenient access to calibration before a Nasmyth focus, introducing time dependent polarization from the fold mirror. \cite{DeJuanOvelar:2014bq, ovelar:2012elt, Keller:2010ig}

Calibration of the primary and secondary optics for retardance and diattenuation requires consideration for off-axis versus on-axis designs when compared to other polarization errors in the system. Goode et al. 2001\cite{Goode:2011wi} state that the polarization properties of the Gregorian focus should be benign. {\it Modeling done by Don Mickey (2009, private communication) says these off-diagonal elements should not be too large to handle, and would be much smaller than what would result if the modulation were done further downstream}. DKIST modeled a similar performance trade and concur that retardance values of the primary and secondary mirror groups are typically several degrees retardance with diattenuation below 0.5\% and mild diagonal depolarization depending significantly on the enhanced protected silver coating used on M2 \cite{Harrington:2017dj}. A significant difference between DKIST and GST is that DKIST will be operating and calibrating multiple polarimetric instruments simultaneously with up to seven cameras covering spectral channels distributed through the 380nm to 1800 nm wavelength range. Dichroic beam splitters reflect and transmit several wavelength bands for use in multiple instruments simultaneously with quite different camera frame rates times, modulation speeds and FoV scanning requirements.  A single zero order calibration retarder is not an option for DKIST, and a suite of separate retarders would compromise efficiency. However, both the field-dependent polarization caused by off-axis mirrors and the polishing non-uniformities described here are at small magnitudes over relatively small field angles, and can easily be calibrated.

\begin{figure}[htbp]
\begin{center}
\vspace{-2mm}
\hbox{
\hspace{-0.5em}
\includegraphics[height=5.8cm, angle=0]{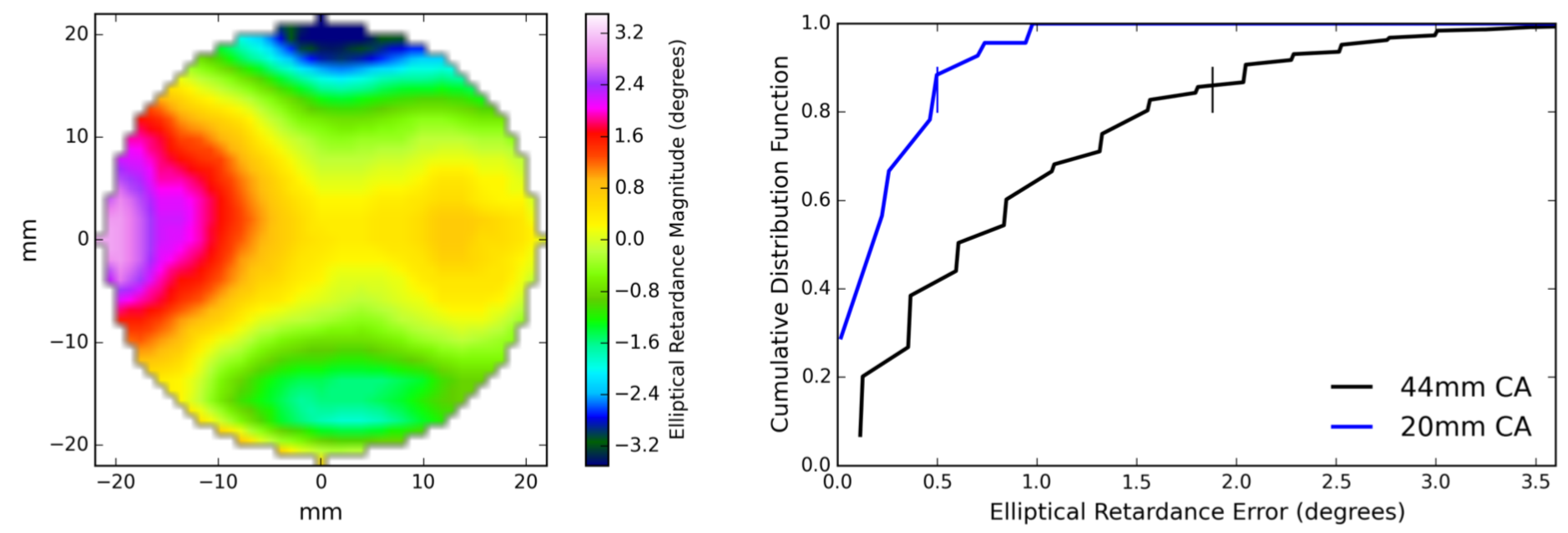}
}
\caption[] 
{\label{fig:polycarb_retardance_map} The spatial nonuniformity of a single sheet of polycarbonate stretched to be a zero order retarder as required for the VTF design. The left plot shows a spatial variation of the elliptical retardance magnitude scaled to $\pm$3.3$^\circ$. The right plot shows the cumulative distribution of errors for the full clear aperture of 44 mm in black and a reduced clear aperture of 20 mm diameter in blue. }
\vspace{-5mm}
\end{center}
\end{figure}

\subsection{Polycarbonate Spatial Uniformity \& Use with Ferro-Electric Liquid Crystals}
\label{sec:appendix_uniformity_polycarb_FLC_polysub}

Designs for achromatic retarders often include either multiple optics together as a compound achromatic retarder or in several-layer designs.  As an example, the VTF instrument on DKIST uses an achromatic modulator composed of two ferro-electric liquid crystals (FLCs) and two true zero-order polymer retarders (ZWPs). Each FLC is zero-order and is used in conjunction with a polycarbonate zero order retarder to make a compound achromat. The two pairs are then oriented and used together to make a four-optic achromatic modulator. These materials have very different spatial characteristics than polished crystals. The clear aperture for VTF is 60 mm with the optic illuminated by a diverging F/ 40 beam. The optic is close to the F/ 40 focus with a single field point having a footprint of 4.4 mm to 5.1 mm diameter on the polarization components internal to this modulator. 

One manufacturer, Astropibor, advertises five layer and seven layer super achromatic retarders broadly similar to the calibration retarder and modulator designed by Meadowlark Optics for our NSO Laboratory Spectropolarimeter. These polymer type retarders can be used either in zero-order, or in three-layer, five-layer or seven layer super-achromatic designs.  In addition, we showed some details of the Goode Solar Telescope and the IRIM spectropolarimeter calibration retarder and also the modulator. These optics are large-area polycarbonate retarders from Meadowlark Optics that would almost certainly have similar properties.  However, we show here that these optics also have uniformity issues that should be considered in instrument design.

Spatial non-uniformity can be expected at some small amplitude for the polycarbonate parts in VTF.  Meadowlark Optics used the spatial mapping equipment to map the retardance across a true zero-order single layer polycarbonate optic with a 3 mm diameter beam and 2 mm spatial sampling on a rectangular grid. There were 377 independent samples across a 44 mm diameter clear aperture.

Figure \ref{fig:polycarb_retardance_map} shows a map of the elliptical retardance magnitude errors in the left plot. The spatial variation is very smooth with $\pm$3.3$^\circ$ across the aperture. The right plot of Figure \ref{fig:polycarb_retardance_map} shows the cumulative distribution function (CDF) for two apertures. For a reduced clear aperture of 20 mm diameter shown in blue, 80\% of the aperture area is within $\pm$0.5$^\circ$ retardance error and every point in the aperture is within $\pm$1$^\circ$. For the black curve at 44 mm aperture, all points are within $\pm$3.5$^\circ$ retardance error while 80\% of the aperture is within 1.9$^\circ$ retardance error.

We note that many manufacturers say these polycarbonate devices are free of interference fringes.  We note that there always are interference fringes when plane-parallel interfaces are involved but often this is limited to just the windows. The polycarbonate layers are substantially less flat than equivalent crystals and are often either fused or cemented, minimizing internal reflections. The polycarbonate birefringence per layer is also significantly lower as they are often true zero-order at visible wavelengths as opposed to tens of waves per plate in crystal based superachromatic designs. These fringes would occur at much lower spectral periods than high resolving power astronomical instruments typically consider impactful. If the cover windows used to sandwich the polycarbonate are not anti-reflection coated, fringes will still occur at the fringe period of the flat window.

As an example, we have used five layer polycarbonate retarders at spectral resolving powers up to 50,000 in the High Resolution Visible and Infrared Spectrograph (HiVIS) at the 3.6m Advanced Electro-optical System (AEOS) on Haleakala, Maui \cite{Harrington:2017ejb,Harrington:2015dl,Harrington:2011fz,Harrington:2010km,Harrington:2009dz,2008PASP..120...89H,Harrington:2007bq}. No fringes have ever been seen in the HiVIS data.  Both VTF and HiVIS mount these optics in an F/ 40 beam near a focal plane.  As an additional fringe mitigation mechanism, the windows can be wedged on entrance and exit as is the case in several solar instruments. The VTF optical design includes a wedge on entry and exit for their modulator design, which removes concern for fringes by orders of magnitude. However, the optical design must now account for chromatic dispersion in the optical design.  Wedges pose significant problems for rotating optics but can be more easily mitigated when used with static optics.

\subsection{Ferro-Electric Liquid Crystal Spatial Uniformity With Wavelength}
\label{sec:appendix_uniformity_polycarb_FLC_flcsub}

\begin{figure}[htbp]
\begin{center}
\vspace{-0mm}
\hbox{
\includegraphics[height=4.4cm, angle=0]{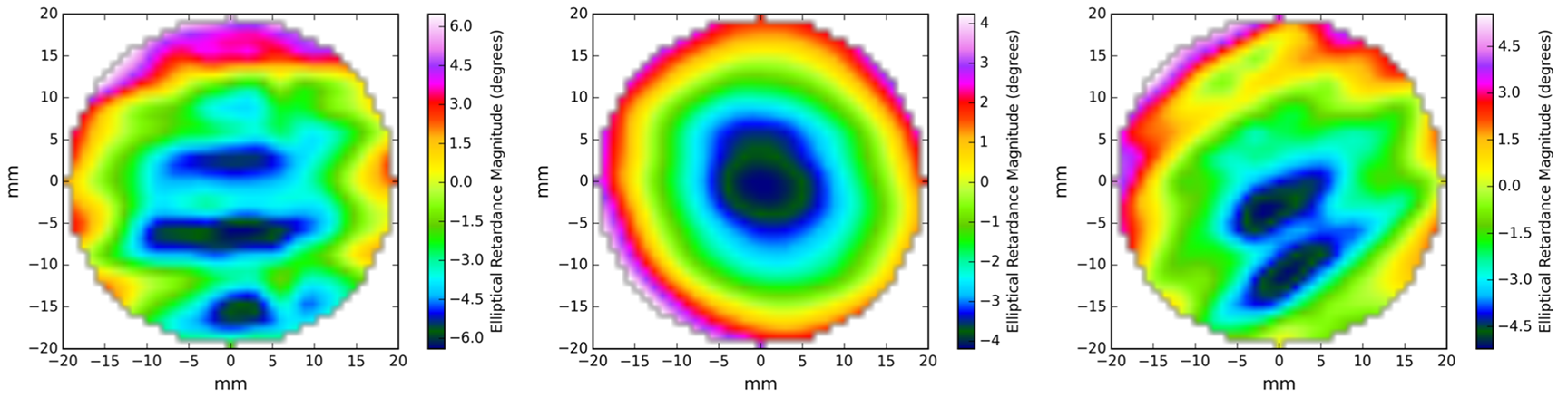}
}
\caption[] 
{\label{fig:FLC_retardance_maps} The spatial non-uniformity maps at three different wavelengths for one of the two VTF FLCs as delivered.  The left plot shows 543 nm wavelength over a range of $\pm$6$^\circ$ elliptical retardance magnitude.  The middle plot shows 650 nm wavelength over a range of $\pm$4$^\circ$ elliptical retardance magnitude. The right plot shows 870 nm wavelength over a range of $\pm$4.6$^\circ$ elliptical retardance magnitude.  These measurements were done without the $\pm$30V drive signal and do change mildly under voltage.}
\vspace{-5mm}
\end{center}
\end{figure}

The FLCs used in the DKIST VTF instrument have very large clear aperture requirements. This aperture size can cause several effects leading to strong spatial non-uniformity. The liquid crystal layers are often only a few microns thick so any small cell gap errors or polymer errors across that very large area can introduce substantial retardance impacts. Meadowlark FLC material is reasonably well aligned at zero voltage, though this material has shown retardance and fast axis switching angle that depends mildly on the voltage applied. Figure \ref{fig:FLC_retardance_maps} shows measured spatial retardance errors for three wavelengths. Wavelengths of 543 nm on the left and 870 nm on the right look considerably different than at 650 nm in the center graphic. The shorter wavelength has errors up to $\pm$6$^\circ$ but with almost similar magnitudes at 870 nm. As there are many types of liquid crystal, we show these results to encourage careful consideration of the specifications and the metrology used to verify retardance across the aperture as functions of all relevant variables.

\subsection{Polycarbonate Five Layer Superachromatic Retarder Spatial Uniformity}
\label{sec:appendix_uniformity_polycarb_FLC_poly5layer}

\begin{figure}[htbp]
\begin{center}
\vspace{-2mm}
\hbox{
\hspace{-1.5em}
\includegraphics[height=6.5cm, angle=0]{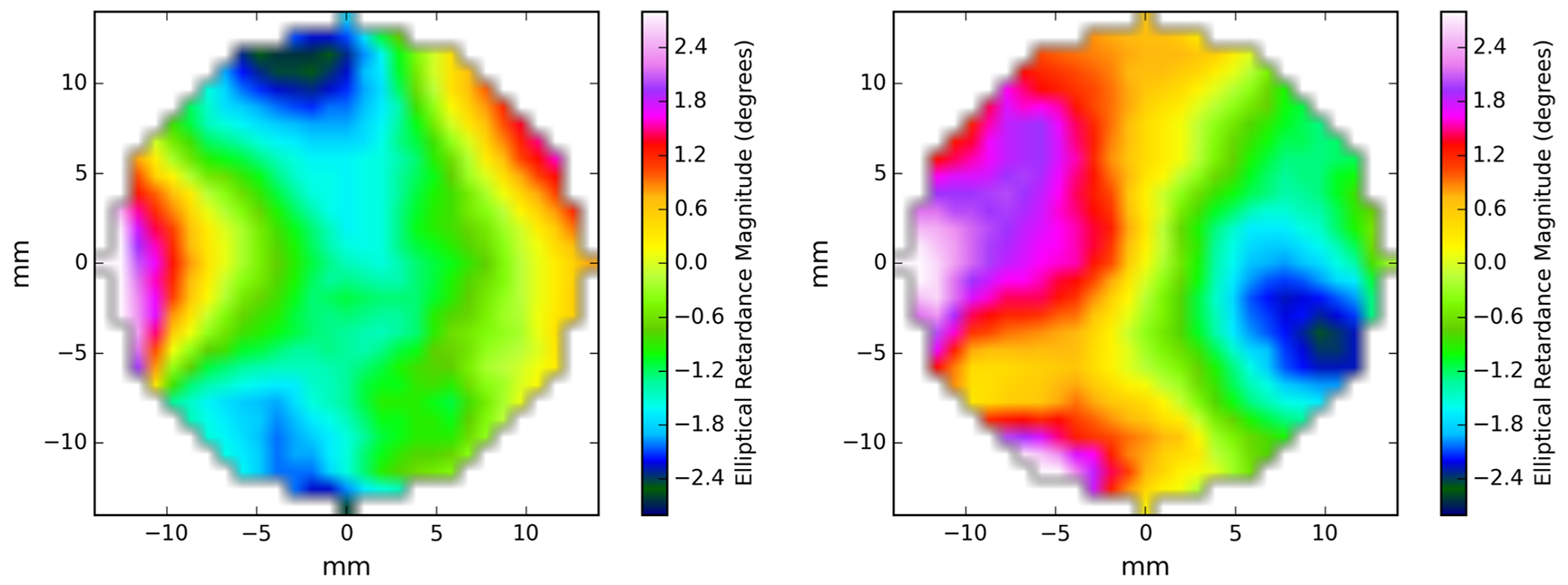}
}
\caption[] 
{\label{fig:polycarb_superachromat_retardance_map} The spatial non-uniformity of two example five-sheet polycarbonate super achromatic calibration retarders. Both test optics are scaled to $\pm$2.5$^\circ$ elliptical retardance magnitude error at 600 nm wavelength. }
\vspace{-5mm}
\end{center}
\end{figure}

Our NSO laboratory spectropolarimeter (NLSP) uses a five layer polycarbonate retarder from Meadowlark Optics. We designed this retarder to be a super-achromatic linear third-wave magnitude originally covering 380 nm to 1100 nm with reasonable efficiency. We use this optic both as a calibration retarder and as a modulator in NLSP. Figure \ref{fig:polycarb_superachromat_retardance_map} shows spatial retardance maps from Meadowlark Optics. Two example super-achromatic quarter-wave linear retarders were scanned in the same setup as the DKIST retarders. Spatial maps cover a 24 mm diameter aperture at a wavelength of 600 nm. The elliptical retardance magnitude errors are roughly $\pm$2.5$^\circ$ for these five-layer optics. Both optics have somewhat similar spatial distributions.

\begin{figure}[htbp]
\begin{center}
\vspace{-2mm}
\hbox{
\includegraphics[height=5.7cm, angle=0]{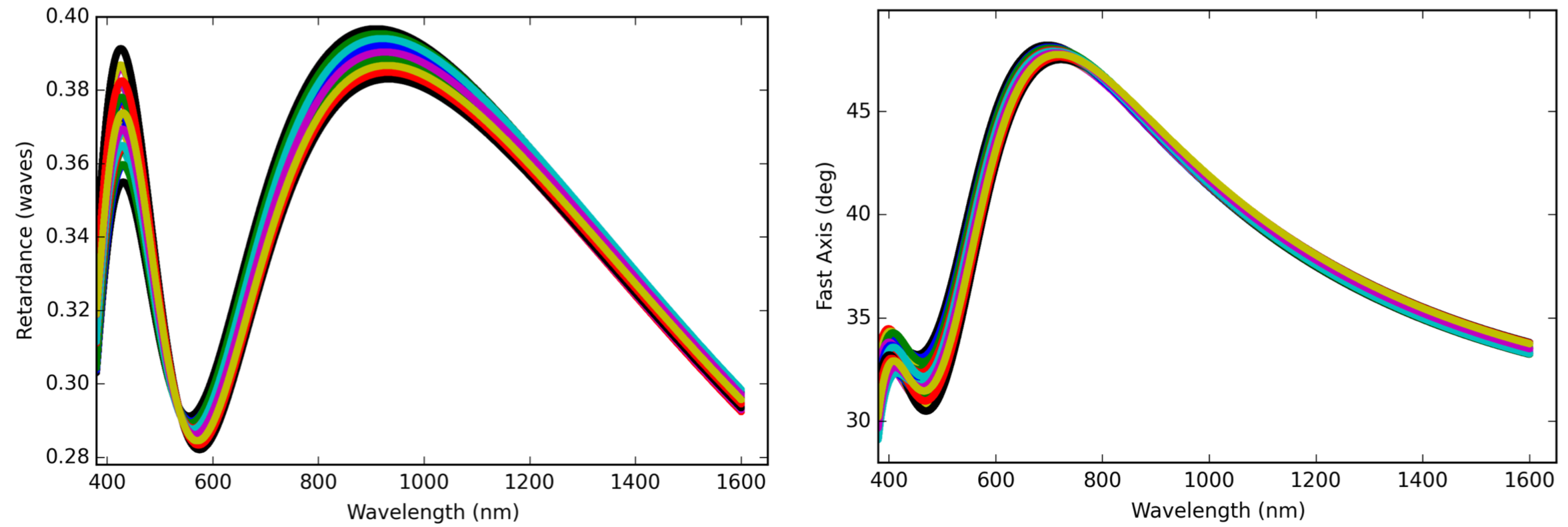}
}
\caption[] 
{\label{fig:polycarb_superachromat_retardance_MLO} The retardance design tolerance of our five-layer polycarbonate super achromatic retarder. Left plots show linear retardance magnitude.  Right plots show linear retardance fast axis orientation. }
\vspace{-5mm}
\end{center}
\end{figure}

We are also building this type of retarder as an alternate modulator for the DL-NIRSP instrument. Figure \ref{fig:polycarb_superachromat_retardance_MLO} shows a design tolerance analysis for our third wave super achromatic linear retarder for NLSP. The polycarbonate manufacturing error in magnitude is combined with a rotational uncertainty in mounting the part to produce an expected range of performance. Circular retardance will be non-zero but less than a few degrees.  Achromatic modulation efficiency is achieved by keeping the retardance magnitude around 0.33 waves across the DL-NIRSP wavelength range. For a continuously spinning modulator sampling a 180$^\circ$ rotation, the fast axis change with wavelength has no impact on the modulation efficiency. However, for our discrete, few-state NLSP sequences the calibration and modulation efficiency does require that we can sample all wavelengths with only a few orientations of the retarder.

One of these optics has been built with a 20 mm clear aperture and achieves 0.2 waves transmitted wavefront error, 6 arc seconds of beam deflection and retardance uniformity that is easily calibrated.  The DL-NIRSP fiber bundle working with the F/ 8 wide field coronal lens mode can easily work with this clear aperture. We anticipate replacing the six-crystal retarder with this five-layer polycarbonate part to assess the reduction of polarization fringes we have previously modeled. \cite{Harrington:2018cx,Harrington:2017jh}.

\clearpage
\bibliography{ms_ver04} 			
\bibliographystyle{spiebib}		

\end{document}